\newif\ifblind
\blindfalse

\ifblind
\documentclass[acmsmall,screen,review,anonymous,nonacm]{acmart}
\else
\documentclass[acmsmall,screen,nonacm]{acmart}
\fi
\AtBeginDocument{%
  }

\setcopyright{acmlicensed}
\copyrightyear{2018}
\acmYear{2018}
\acmDOI{XXXXXXX.XXXXXXX}
\acmConference[Conference acronym 'XX]{Make sure to enter the correct
conference title from your rights confirmation email}{June 03--05,
2018}{Woodstock, NY}
\acmISBN{978-1-4503-XXXX-X/2018/06}

\usepackage{amsmath}
\usepackage{amsfonts} 
\usepackage{mathtools}
\usepackage{stmaryrd}
\usepackage{amsthm}

\theoremstyle{definition}

\theoremstyle{plain}

\usepackage{graphicx}
\usepackage{tikz}
\usetikzlibrary{arrows.meta}
\usepackage{wrapfig}
\usepackage{svg}
\usepackage{caption}
\usepackage{subcaption}
\usepackage{booktabs}
\usepackage{tabularx}
\usepackage{makecell}
\usepackage{adjustbox}
\usepackage{fancyvrb} 
\usepackage{changepage}

\usepackage[ruled,vlined,linesnumbered]{algorithm2e}
\SetAlgoCaptionSeparator{.}
\SetAlgoSkip{medskip}
\SetAlCapFnt{\small\bfseries}
\SetAlCapNameFnt{\small\bfseries}
\SetAlFnt{\small}
\usepackage{textcomp}
\usepackage{comment}
\usepackage{enumitem}
\usepackage{etoolbox}
\usepackage{xcolor}
\usepackage[normalem]{ulem}
\usepackage{tcolorbox}
\usepackage{listings}
\newcommand{\tealcodefontchoice}{fira}

\newcommand{\tealcodeseries}{m}
\newbool{tealfontmatched}
\newcommand{\tealfontoption}[4]{%
  \ifdefstring{\tealcodefontchoice}{#1}{%
    \usepackage[#2]{#3}%
    \renewcommand{\tealcodeseries}{#4}%
    \booltrue{tealfontmatched}%
  }{}%
}
\tealfontoption{plex}{}{plex-mono}{m}
\tealfontoption{plex-light}{l}{plex-mono}{l}
\tealfontoption{roboto}{regular}{roboto-mono}{m}
\tealfontoption{roboto-light}{light}{roboto-mono}{l}
\tealfontoption{fira}{}{FiraMono}{m}
\tealfontoption{source-code-pro}{}{sourcecodepro}{m}
\ifdefstring{\tealcodefontchoice}{inconsolata}{\booltrue{tealfontmatched}}{}
\ifdefstring{\tealcodefontchoice}{latin-modern}{%
\booltrue{tealfontmatched}}{}
\ifdefstring{\tealcodefontchoice}{latin-modern-light}{%
  \renewcommand{\tealcodeseries}{l}%
\booltrue{tealfontmatched}}{}
\ifbool{tealfontmatched}{}{%
  \PackageError{teal-listings}{Unknown code font: \tealcodefontchoice}%
  {Choose one of the options listed above tealcodefontchoice.}%
}
\let\tealcodefamily\ttdefault
\newcommand{\tealcodefont}{\fontfamily{\tealcodefamily}\fontseries{\tealcodeseries}\fontshape{n}\selectfont}

\definecolor{tealcodekeyword}{HTML}{236B68}
\definecolor{tealcodecomment}{HTML}{66716D}
\definecolor{tealcodestring}{HTML}{875323}
\definecolor{tealcoderule}{HTML}{9DAEAA}
\definecolor{tealgridone}{HTML}{DEECE7}
\definecolor{tealgridtwo}{HTML}{F3E5D4}
\definecolor{tealgridthree}{HTML}{E0E7F1}
\definecolor{tealgridfour}{HTML}{EFDFE2}
\definecolor{tealgridfive}{HTML}{E9E2F0}
\definecolor{tealgridrule}{HTML}{A5B2AD}
\definecolor{tealgridexcluded}{HTML}{F0F0EE}
\definecolor{tealgridink}{HTML}{303B38}
\lstdefinestyle{tealcode}{
  language=C++,
  basicstyle=\fontsize{8.5}{10.5}\selectfont\tealcodefont\color{codexcolor},
  keywordstyle=\color{tealcodekeyword},
  commentstyle=\itshape\color{tealcodecomment},
  stringstyle=\color{tealcodestring},
  columns=fullflexible,
  keepspaces=true,
  showstringspaces=false,
  tabsize=2,
  breaklines=true,
  breakatwhitespace=true,
  frame=none,
  framerule=0.5pt,
  rulecolor=\color{tealcoderule},
  xleftmargin=1.4em,
  framexleftmargin=0.6em,
  aboveskip=0.8\baselineskip,
  belowskip=0.8\baselineskip,
  captionpos=b,
  numbers=left
}

\definecolor{tealcodetype}{HTML}{456783}
\definecolor{tealcodeoperator}{HTML}{3B648A}
\definecolor{tealcodetoken}{HTML}{775A78}
\lstdefinestyle{tealpaper}{
  style=tealcode,
  basicstyle=\fontsize{5.8}{6.8}\selectfont\tealcodefont\color{black},
  commentstyle=\color{tealcodecomment},
  morekeywords={mi350,kernel,out,in,Tensor,with,on,pid,as,Span,Chain,sync},
  morekeywords=[2]{LDS,VGPR,AGPR,global,_b16,_f32,_f16},
  keywordstyle=[2]\color{tealcodetype},
  morekeywords=[3]{amd_lds_tldbuf_32x16,amd_lds_tread128_32x16,amd_tmfma_f32_32x32x16_bf16_priority,amd_tmfma_f32_32x32x16_bf16,amd_tset,amd_tcast_b16,amd_tmov},
  keywordstyle=[3]\color{tealcodeoperator},
  morekeywords=[4]{ready,ra,rb,va,vb,prio,vmA,vmB,rmA,rmB,vm,rmtcgen05_completion, empty},
  keywordstyle=[4]\itshape\color{tealcodetoken},
  morecomment={[l]{\#}},
  frame=none,
  numbers=left,
  numberstyle=\fontsize{5}{6}\selectfont\color{tealcodecomment},
  numbersep=5pt,
  xleftmargin=1.5em,
  framexleftmargin=0pt,
  aboveskip=0pt,
  belowskip=0pt
}

\newsavebox{\tealgrammarbox}

\setlist[itemize]{noitemsep, topsep=0pt}

\makeatletter
\newcommand{\smalltt}[1]{{%
    \fontfamily{FiraMono-TLF}\selectfont
    \fontsize{\dimexpr\f@size pt*85/100\relax}%
    {\f@baselineskip}%
    \selectfont #1%
}}
\makeatother

\newcommand{\tengl}{Teal}
\newcommand{\name}{\tengl}

\newcommand{\operator}{operator}

\begin{document}

\ifblind
\title{Instruction-Centric Lifting for Accelerator Kernel Compilation}
\else
\title{\name: Verifiable Kernel Compilation through Instruction-Centric Lifting}
\fi

\newcommand{\rev}[1]{#1}
\newcommand{\codexrev}[1]{#1}
\definecolor{chainfixcolor}{HTML}{1B5E20}
\newcommand{\chainfix}[1]{#1}
\newenvironment{codexrevcode}{}{}
\newcommand{\tealfigureplain}[1]{#1}
\AtBeginEnvironment{figure}{\let\codexrev\tealfigureplain\let\chainfix\tealfigureplain}
\AtBeginEnvironment{figure*}{\let\codexrev\tealfigureplain\let\chainfix\tealfigureplain}
\newcommand{\agents}[1]{#1}
\definecolor{codexcolor}{HTML}{000000}
\newif\ifcodexmarkup
\codexmarkuptrue
\ifcodexmarkup\else\colorlet{codexcolor}{black}\fi
\newcommand{\codex}[1]{{\color{codexcolor}#1}}
\newcommand{\cdx}[1]{\ifcodexmarkup{\color{codexcolor}\textbf{[Codex: #1]}}\fi}

\newcommand{\langfix}[1]{#1}

\newcommand{\shriram}[1]{{\color{red}\textbf{[Shriram: #1]}}}
\newcommand{\lex}[1]{{\textbf{[Lex: #1]}}}
\definecolor{geminicolor}{HTML}{0066CC}
\definecolor{geminioldcolor}{HTML}{888888}
\newcommand{\gemini}[2]{%
  \if\relax\detokenize{#1}\relax\else{\color{geminioldcolor}(\sout{#1})} \fi%
  {\color{geminicolor}#2}%
}

\author{Shriram Chandran}
\correspondingauthor
\affiliation{%
  \institution{ETH Z\"urich}
  \city{Z\"urich}
  \country{Switzerland}
}
\email{shriram.chandran@inf.ethz.ch}

\author{Alexandru Calotoiu}
\affiliation{%
  \institution{ETH Z\"urich}
  \city{Z\"urich}
\country{Switzerland}}
\email{alexandru.calotoiu@inf.ethz.ch}

\author{Marcin Copik}
\affiliation{%
  \institution{ETH Z\"urich}
  \city{Z\"urich}
\country{Switzerland}}
\email{marcin.copik@inf.ethz.ch}

\author{Torsten Hoefler}
\affiliation{%
  \institution{ETH Z\"urich}
  \city{Z\"urich}
  \country{Switzerland}
}
\email{torsten.hoefler@inf.ethz.ch}

\renewcommand{\shortauthors}{Chandran et al.}

\begin{abstract}
  High-performance tensor kernel compilation faces a two-sided challenge as accelerator hardware diverges. Kernel authors need a way to express target-specific execution schedules without manually encoding low-level pointer addressing, memory layouts, or assembly intrinsics, while compiler engineers need a way to realize schedules on new architectures without requiring custom lowering and transformation passes for every ISA.
  We present \name, a two-sided programming model and compiler that \codexrev{treats} both \rev{the target architecture's hardware specification and the kernel's schedule as programmable inputs to a shared compilation pipeline}. On the machine side, a reusable dialect program specifies a target's processor description, memory hierarchy, instruction set, and synchronization protocols. On the client side, a kernel program composes that declared vocabulary over symbolic tensor views using generic \operator s, processor scopes, and data dependencies. Rather than recovering machine instructions from loop nests, the \name~compiler performs compositional \emph{instruction-centric lifting}: \rev{it combines the machine dialect and client schedule to automatically derive and statically verify tile loops, physical data layouts, memory and register addresses, data dependencies, and synchronization validity.} Supporting a new accelerator or instruction set therefore requires programming only its dialect specification and bindings, while reusing the lifting, verification, and composition compiler to provide a compact, statically checked kernel language for a client. We evaluate \name~across heterogeneous architectures and kernels, demonstrating that \name~can represent schedules that match or outperform hand-optimized vendor baselines while enabling high compiler reuse and rapid kernel synthesis on complex pipelines.
\end{abstract}




\maketitle

\section{Introduction}
\label{sec:intro}

High-performance tensor kernels on accelerators are defined as much by their execution \emph{schedules} \rev{-- how computations are mapped to hardware resources --} as by the computations they implement. Efficient tensor kernels select specialized vector and matrix instructions, arrange their operands in hardware-native layouts, orchestrate memory movement, assign work to distinct processor groups, and coordinate asynchronous execution~\cite{hopperdissect, hellosme, cdna4, linearlayouts, soi2025optimalsoftwarepipeliningwarp, blackwelldissect}. These decisions become increasingly architecture-specific as hardware diverges~\cite{mmus, khattak2025accuratemodelsnvidiatensor, martinez2025cambrian}. NVIDIA and AMD GPUs expose different instruction and execution mechanisms~\cite{hopperdissect, blackwelldissect, hu2025hipkittensfastfuriousamd}; Intel \textsf{AMX} and Arm \textsf{SME} introduce distinct matrix-unit interfaces~\cite{intelamx, intelamxllm, deng2025demystifyingarmsmeoptimize, hellosme}; and emerging accelerators feature custom processor arrays, local memories, and data-movement engines~\cite{xu2026maia200softwaredefined, taylor2026numericalkernelsspatialaccelerator, cerebras, ttblackhole}. Achieving high performance therefore requires expressing different schedules for different machines.

Programming systems address this complexity at different abstraction levels. Low-level languages such as CUDA and HIP~\cite{HIPdocum43, ptxdoc} provide detailed control, but entangle schedules with architecture-specific addressing, register layouts, instruction invocations, and synchronization. Frameworks such as CUTLASS, CuTe, and Composable Kernel~\cite{cutlass, cecka2026cutelayoutrepresentationalgebra, Composab19} provide structured access to these mechanisms within their respective execution models. Higher-level compilers instead rely on automated analyses or rewrite passes to derive execution schedules, shifting performance-critical decisions into opaque, target-specific heuristics where \rev{performance regressions and rejected schedules} are difficult to diagnose~\cite{triton, chou_2026, shah2024flashattention3fastaccurateattention, zadouri2026flashattention4algorithmkernelpipelining, zhang2026architectinghardwarediversitymetas}. Even systems exposing explicit execution schedules or schedule transformations~\cite{TIRxAnOp77, ye2026cakecompileragentcodesignfrontier, yadav2025taskbasedtensorcomputationsmodern} \rev{still require compiler developers to hand-write target-specific lowering passes that map those primitives to each architecture's instructions, register layouts, and synchronization mechanisms.} \agents{This tension is acute not only for human experts, but also for automated kernel synthesis agents~\cite{saba2026cutegenllmbasedagenticframework, ye2026cakecompileragentcodesignfrontier}: low-level code forces agents to coordinate brittle pointer and barrier edits across hundreds of lines, while heuristic compilers prevent them from directly controlling the schedule.} Working with agents thus requires a robust programming model that can verify schedules before starting expensive benchmarking.

\sloppy This reveals two related challenges. First, a kernel author needs a verifiable representation that captures performance-critical scheduling decisions without embedding \rev{low-level pointer arithmetic, thread indexing, and barrier code}~\cite{yadav2025taskbasedtensorcomputationsmodern,exocomp}. Second, \rev{supporting \codexrev{new hardware} requires the compiler to realize these decisions using a different ISA. In existing compilers, adding a new accelerator or instruction set entails substantial work on re-engineering custom lowering passes, layout mappings, and synchronization passes in the compiler \codexrev{backend}}~\cite{actcompiler, hong2025autocomppowerfulportablecode}\codexrev{.}
\setlength{\intextsep}{2pt}
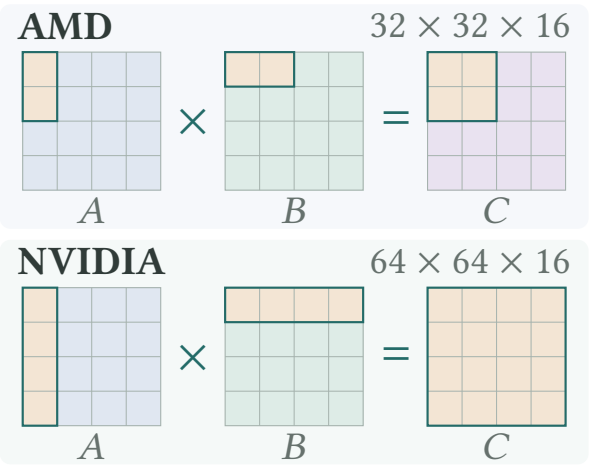
\begin{wrapfigure}{r}{0.28\linewidth}
  \centering
  \resizebox{\linewidth}{!}{%
    \begin{tikzpicture}[x=.6cm,y=.6cm,
        every node/.style={font=\fontsize{7.5}{8.4}\selectfont,
        text=tealgridink,inner sep=0pt},
        ip/title/.style={font=\bfseries\fontsize{7.5}{8.4}\selectfont},
        ip/shape/.style={font=\fontsize{7}{7.8}\selectfont,
        text=tealcodecomment},
        ip/op/.style={font=\fontsize{9}{10}\selectfont,
        text=tealcodekeyword},
      ip/selected/.style={draw=tealcodekeyword,line width=.4pt}]
      \path[use as bounding box] (0,0) rectangle (6.4,5);
      \fill[tealgridthree!30!white,rounded corners=2pt]
      (0,2.56) rectangle (6.4,5);
      \fill[tealgridone!30!white,rounded corners=2pt]
      (0,0) rectangle (6.4,2.44);
      \node[ip/title,anchor=west] at (.18,4.77) {AMD};
      \node[ip/shape,anchor=east] at (6.22,4.77) {$32\times32\times16$};
      \node[ip/title,anchor=west] at (.18,2.21) {NVIDIA};
      \node[ip/shape,anchor=east] at (6.22,2.21) {$64\times64\times16$};
      \foreach \base in {.42,2.98} {
        \fill[tealgridthree] (.25,\base) rectangle ++(1.5,1.5);
        \fill[tealgridone] (2.45,\base) rectangle ++(1.5,1.5);
        \fill[tealgridfive] (4.65,\base) rectangle ++(1.5,1.5);
        \node[ip/op] at (2.1,{\base+.75}) {$\times$};
        \node[ip/op] at (4.3,{\base+.75}) {$=$};
        \node[ip/shape] at (1,{\base-.22}) {$A$};
        \node[ip/shape] at (3.2,{\base-.22}) {$B$};
        \node[ip/shape] at (5.4,{\base-.22}) {$C$};
      }
      \foreach \base/\m/\n in {2.98/.75/.75,.42/1.5/1.5} {
        \fill[tealgridtwo] (.25,{\base+1.5-\m}) rectangle (.625,{\base+1.5});
        \fill[tealgridtwo] (2.45,{\base+1.125}) rectangle ({2.45+\n},{\base+1.5});
        \fill[tealgridtwo] (4.65,{\base+1.5-\m}) rectangle ({4.65+\n},{\base+1.5});
      }
      \foreach \base in {.42,2.98} {
        \foreach \left in {.25,2.45,4.65} {
          \draw[draw=tealgridrule,line width=.2pt]
          (\left,\base) rectangle ++(1.5,1.5);
          \foreach \offset in {.375,.75,1.125} {
            \draw[draw=tealgridrule,line width=.15pt]
            ({\left+\offset},\base) -- ++(0,1.5);
            \draw[draw=tealgridrule,line width=.15pt]
            (\left,{\base+\offset}) -- ++(1.5,0);
          }
        }
      }
      \foreach \base/\m/\n in {2.98/.75/.75,.42/1.5/1.5} {
        \draw[ip/selected] (.25,{\base+1.5-\m}) rectangle (.625,{\base+1.5});
        \draw[ip/selected] (2.45,{\base+1.125}) rectangle ({2.45+\n},{\base+1.5});
        \draw[ip/selected] (4.65,{\base+1.5-\m}) rectangle ({4.65+\n},{\base+1.5});
      }
  \end{tikzpicture}}
  \captionsetup{font=footnotesize,skip=2pt}
  \caption{First subtile in a $64\times64\times64$ GEMM on AMD and NVIDIA GPUs.}
  \label{fig:intro-placements}
\end{wrapfigure}
For instance, consider a $64\times64\times64$ bf16 tile matrix multiplication (Figure~\ref{fig:intro-placements}) using the AMD matrix-core instruction \textsf{v\_mfma\_f32\_32x32x16\_bf16} (\textsf{MFMA} in short) operating on $32\times32\times16$ fragments~\cite{cdna4}. Selecting this instruction \rev{still leaves the compiler with significant surrounding plumbing: invoking the instruction $16$ times to tile the $64\times64\times64$ problem, packing input tiles into registers distributed across the 64 threads of a wave~\cite{hu2025hipkittensfastfuriousamd, an2026simplifiedlayoutalgebragpu}, accumulating partial sums, and unpacking the distributed result back to global memory.} The exact mappings in these steps depend jointly on the instruction and the schedule's requirements. On NVIDIA Blackwell, a different schedule can instead use \textsf{tcgen05.mma.kind::f16} with $64\times64\times16$ fragments, requiring four contraction instances. Its operands can reside in shared memory, with accumulation in tensor memory and asynchronous completion coordinated through barriers~\cite{ptxdoc, blackwelldissect}. \rev{Although these two GPUs use completely different instructions, memory spaces, and synchronization primitives, both expose the same classes of compilation problems: generating the tile loops, computing per-iteration memory and register offsets, chaining data movement with accumulation, and inserting the right synchronizations~\cite{unit}.} This raises the question:

\begin{adjustwidth}{2em}{2em}
  \emph{\rev{Can a single hardware-parametric compilation algorithm take a declarative machine specification and a target-specific schedule, statically verify their compatibility, and generate high-performance native code?}}
\end{adjustwidth}

Our central insight is that \rev{dense tensor computations exhibit regular, statically analyzable coordinate and dependency structure~\cite{unit}, allowing} instruction composition over tensor domains \rev{to be separated from hardware behavior and solved together by a single compiler}. This suggests reversing conventional \emph{loop-centric} compilation, which starts from computations or iteration structures and selects instructions~\cite{feng2022tensorirabstractionautomatictensorized}. Instead, we develop a programming model that starts with selected instructions and derives their application over larger symbolic tensor domains. We call this \emph{instruction-centric lifting}: an \emph{\operator} associates a \rev{fixed-size hardware} instruction \rev{(such as a $32\times32\times16$ matrix multiply)} with a reusable construction of loops and tensor access maps, while \rev{explicit data movement, processor scopes, and synchronization compose these lifted operations into complete schedules}. \rev{Rather than mapping one hardware-independent program to many targets through transformations~\cite{CarterEdwards20143202, dace}, this structure lets one shared compiler statically verify and lower target-specific schedules across diverse machines.}

\begin{wrapfigure}{r}{0.4\linewidth}
  \centering
  \resizebox{\linewidth}{!}{%
    \begin{tikzpicture}[x=1cm,y=1.12cm,
        every node/.style={font=\fontsize{8.5}{10}\selectfont,
        text=tealgridink,align=center,inner sep=2pt},
        ov/title/.style={font=\bfseries\fontsize{9.5}{11}\selectfont},
        ov/meta/.style={font=\sffamily\fontsize{7.8}{9.2}\selectfont,
        text=tealgridink!75!white},
      ov/arrow/.style={->,draw=tealcodekeyword,line width=.8pt}]
      \path[use as bounding box] (0,0) rectangle (6,5.2);
      \fill[tealgridthree!45!white,rounded corners=2pt]
      (0,4.1) rectangle (2.8,5.2);
      \fill[tealgridone!45!white,rounded corners=2pt]
      (3.2,4.1) rectangle (6,5.2);
      \node[ov/title] at (1.4,4.94) {Define Hardware};
      \node[ov/meta] at (1.4,4.48) {Instructions, storage\\synchronization};
      \node[ov/title] at (4.6,4.94) {Write a Schedule};
      \node[ov/meta] at (4.6,4.48) {Views, operations\\and dependencies};
      \draw[draw=tealcodekeyword,line width=.8pt]
      (1.4,4.1)--(1.4,3.9)--(3,3.9);
      \draw[draw=tealcodekeyword,line width=.8pt]
      (4.6,4.1)--(4.6,3.9)--(3,3.9);
      \draw[ov/arrow] (3,3.9)--(3,3.75);
      \fill[tealgridthree!15!white,rounded corners=2pt]
      (0,.7) rectangle (6,3.7);
      \node[ov/title] at (3,3.47) {Shared \name\ Compiler};
      \fill[tealgridone!40!white,rounded corners=2pt]
      (.35,2.55) rectangle (5.65,3.2);
      \node[ov/title] at (3,2.98) {Check the Schedule};
      \node[ov/meta] at (3,2.68) {Bounds, layouts, dependencies};
      \draw[ov/arrow] (3,2.55)--(3,2.37);
      \fill[tealgridtwo!40!white,rounded corners=2pt]
      (.35,1.67) rectangle (5.65,2.32);
      \node[ov/title] at (3,2.10) {Generate Loops and Addresses};
      \node[ov/meta] at (3,1.80) {Preserve synchronization};
      \draw[ov/arrow] (3,1.67)--(3,1.49);
      \fill[tealgridfive!40!white,rounded corners=2pt]
      (.35,.79) rectangle (5.65,1.44);
      \node[ov/title] at (3,1.22) {Simplify Generated Code};
      \node[ov/meta] at (3,.92) {Fold constants, reuse calculations};
      \draw[ov/arrow] (3,.79)--(3,.55);
      \fill[tealgridthree!45!white,rounded corners=2pt]
      (0,0) rectangle (6,.5);
      \node[ov/title,text=tealcodekeyword] at (3,.25) {Target Code};
  \end{tikzpicture}}
  \captionsetup{font=footnotesize,skip=4pt}
  \caption{Hardware descriptions and client schedules feed a shared compiler.}
  \label{fig:intro-figure}
\end{wrapfigure}
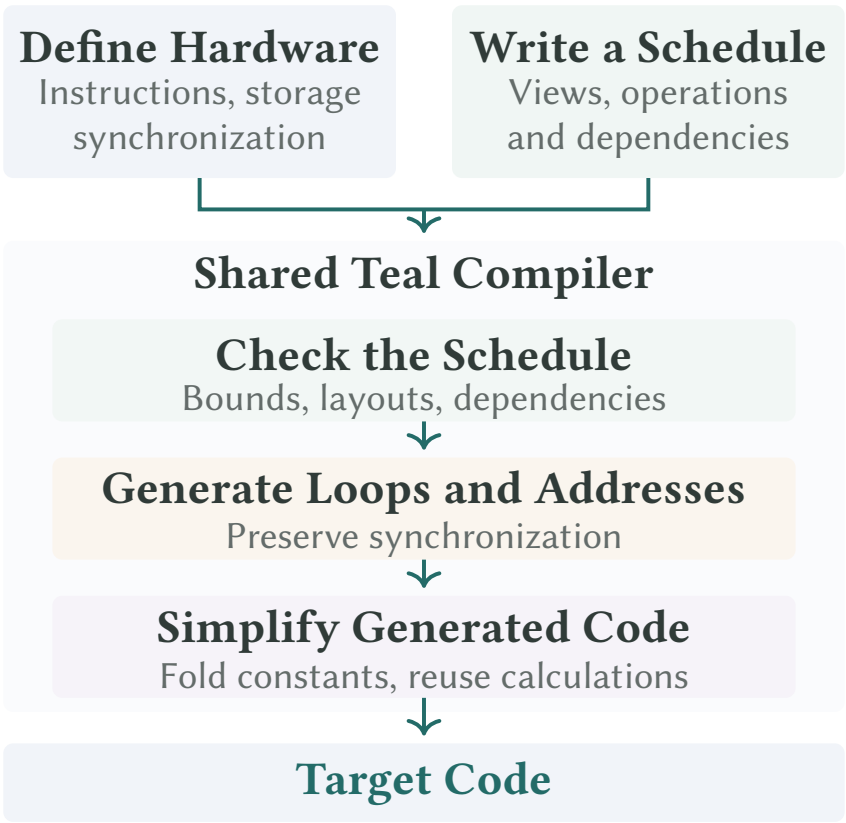

We introduce \name, a two-sided verifiable programming model and compiler built around this principle (Figure~\ref{fig:intro-figure}). \rev{On the back-facing machine side, a hardware developer \agents{(or an agent translating hardware documentation)} declares a reusable \emph{dialect} specifying the target's processor domains, state spaces, instruction signatures and layouts, and synchronization transitions. On the front-facing client side, a kernel author \agents{or agent} expresses target-specific schedules by applying generic \operator s to high-level tensor views.}
Both interfaces share a common semantic framework: new machine instructions \rev{modularly} extend the available vocabulary, while kernel authors choose how to compose that vocabulary into schedules appropriate for the target.

\setlength{\intextsep}{2pt}
We implement this model through a prototype \name~compiler. \rev{Given a dialect and a client kernel, the compiler builds a hierarchical execution graph and uses an SMT solver to statically verify before code generation that all tensor accesses stay in bounds, data layouts match across instructions, data dependencies are respected, and \codexrev{declared synchronization protocols are valid}.} The verified schedule is then lowered through a low-level representation (GLIR) to native target code. Compilation is \emph{schedule-preserving}: the client selects instructions, tensor views, processor scopes, and dependencies, while the compiler generates tiled loops, computes memory and register addresses, derives a verified instruction sequence, and inserts hardware synchronization. \rev{Consequently, supporting new hardware primarily requires specifying its machine dialect and thin native bindings rather than rewriting compiler passes.} \agents{At the same time, \name~provides a compact, statically checked language for kernel synthesis: lifting compresses hundreds of lines of error-prone indexing into a few scheduling statements, while static SMT verification catches invalid layouts and synchronization errors, giving agents actionable compile-time diagnostics without expensive on-device profiling.}

In summary, this paper makes the following contributions:

\begin{itemize}
  \item \rev{We introduce \emph{instruction-centric lifting}, a compilation paradigm that starts from fixed-size hardware instructions and automatically derives the loop nests, coordinate mappings, and address arithmetic required to apply them over larger tensors.}

  \item We develop a \emph{two-sided, architecture-parametric programming model} in which machine dialects describe native execution capabilities and client programs express target-specific schedules through a common semantic framework.

  \item \rev{We implement a retargetable, schedule-preserving compiler that statically verifies tensor bounds, data layouts, \codexrev{value consistency}, and synchronization rules via an SMT solver, fuses tile loops across chained operations, and generates low-level target code.}

  \ifblind
  \else
  \newpage
  \fi

  \item We evaluate \name~schedules across GPUs, CPUs, and spatial accelerators, competitive with or faster than hand-optimized libraries on dense GEMMs ($0.83$--$1.36\times$) and fused residual+RMSNorm ($0.97$--$1.32\times$), achieving up to $1.94\times$ native fp64 GEMM throughput via Ozaki-II int8 emulation on MI350X, and reaching up to $104.7\%$ of TT-Metalium throughput when adding Tenstorrent Blackhole as a new target accelerator.
\end{itemize}

\vspace{-0.1em}
\section{From Instructions to Schedules}

\rev{Generating low-level kernel code from chosen fixed-size hardware instructions and a schedule requires solving three problems: applying the instructions across larger tensors, coordinating data movement and synchronization, and supporting diverse hardware through a single set of compiler algorithms.} In this section, we illustrate these challenges and motivate \name's design.

\subsection{Instruction-Centric Lifting over Tensor Domains}
\label{sec:icl}

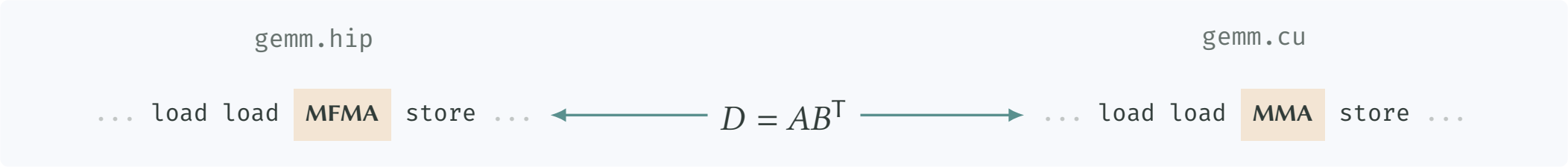
\begin{figure*}[t]
  \centering
  \captionsetup[subfigure]{font=footnotesize,skip=2pt}
  \begingroup
  \lstset{style=tealpaper,
    basicstyle=\fontsize{5.8}{6.9}\selectfont\tealcodefont\color{black},
    numberstyle=\fontsize{4.8}{5.7}\selectfont\color{tealcodecomment},
    morekeywords={nvgpu,instr,operator,Contract,actions,execute,using,layout,post},
    morekeywords=[2]{SMEM,TMEM},
    morekeywords=[3]{nv_mma_f16,nv_tmma_f16},
  aboveskip=0pt,belowskip=0pt}
  \newcommand{\liftingtraversal}[2]{%
    \par\vspace{2pt}\noindent
    \begin{tikzpicture}
      \node[draw=none,fill=tealgridone!35!white,rounded corners=2pt,
      inner sep=4pt] {
        \begin{minipage}{\dimexpr\linewidth-8pt\relax}
          \fontsize{5.5}{6.5}\selectfont\tealcodefont
          #1\par\vspace{2pt}
          {\normalfont\fontsize{6}{7}\selectfont\color{tealcodecomment}#2}
      \end{minipage}};
  \end{tikzpicture}}
  \makebox[.48\linewidth]{\bfseries\fontsize{9}{10.5}\selectfont AMD CDNA4}%
  \hfill
  \makebox[.48\linewidth]{\bfseries\fontsize{9}{10.5}\selectfont NVIDIA Blackwell}
  \par\vspace{5pt}
  \setcounter{subfigure}{0}
  \begin{subfigure}[t]{.48\linewidth}
    \begin{minipage}[t][38pt][t]{\linewidth}
\begin{lstlisting}
*(floatx16_t *)(d) =
  __builtin_amdgcn_mfma_f32_32x32x16_bf16(
    *(const bf16x8_t *)(a),
    *(const bf16x8_t *)(b),
    *(floatx16_t *)(d), 0, 0, 0);
\end{lstlisting}
    \end{minipage}
    \caption{\codexrev{Native intrinsic on per-thread register vectors.}}
    \label{fig:background-lifting-native}
  \end{subfigure}\hfill
  \setcounter{subfigure}{4}
  \begin{subfigure}[t]{.48\linewidth}
    \begin{minipage}[t][38pt][t]{\linewidth}
\begin{lstlisting}
tcgen05.mma.cta_group::1.kind::f16
  [d_tmem], a_desc, b_desc, idesc,
  {0, 0, 0, 0}, enable_input_d;
\end{lstlisting}
    \end{minipage}
    \caption{Native \textsf{PTX}: \textsf{TMEM} destination and \textsf{SMEM} descriptors.}
    \label{fig:background-blackwell-native}
  \end{subfigure}
  \par\vspace{5pt}
  \setcounter{subfigure}{1}
  \begin{subfigure}[t]{.48\linewidth}
    \begin{minipage}[t][80pt][t]{\linewidth}
\begin{lstlisting}
mi350.instr amdw_mfma_f32_32x32x16_bf16:
  out: Tensor [32,32] : layout [row,col =>
    64*(4*(row/8)+row%4)
    +32*((row%8)/4)+col] <d,_f32,AGPR>;
  in: Tensor [32,16] : layout [r,c =>
    64*(c%8)+r+32*(c/8)] <a,_b16,VGPR>;
  in: Tensor [32,16] : layout [r,c =>
    64*(c%8)+r+32*(c/8)] <b,_b16,VGPR>;
  { _amdw_mfma_f32_32x32x16_bf16(d,a,b); }
\end{lstlisting}
    \end{minipage}
    \caption{\codexrev{Instruction geometry, register spaces, and native layouts.}}
    \label{fig:background-lifting-instruction}
  \end{subfigure}\hfill
  \setcounter{subfigure}{5}
  \begin{subfigure}[t]{.48\linewidth}
    \begin{minipage}[t][80pt][t]{\linewidth}
\begin{lstlisting}[escapeinside={(*@}{@*)}]
nvgpu.instr nv_mma_f16
    [tcgen05_completion o <- _ i]:
  using T [_f16,_b16];
  out: Tensor [64,64] : layout ld
    [i,j => 65536*i+j] <d,_f32,TMEM>;
  in: Tensor [64,16] : layout la
    (*@\(\ldots\)@*) <a,T,SMEM>;
  in: Tensor [64,16] : layout lb
    (*@\(\ldots\)@*) <b,T,SMEM>;
  { tcgen05_mma_f16_issuer(d,a,b); }
\end{lstlisting}
    \end{minipage}
    \caption{\codexrev{Compact \textsf{MMA} declaration: \textsf{SMEM} inputs, \textsf{TMEM} output, and completion token.}}
    \label{fig:background-blackwell-instruction}
  \end{subfigure}
  \par\vspace{5pt}
  \setcounter{subfigure}{2}
  \begin{subfigure}[t]{.48\linewidth}
    \begin{minipage}[t][44pt][t]{\linewidth}
\begin{lstlisting}
mi350.operator amd_tmfma_f32_32x32x16_bf16
    (Contract) [empty o <- _ i]:
  actions {
    execute: amdw_mfma_f32_32x32x16_bf16 [<-];
  }
\end{lstlisting}
    \end{minipage}
    \caption{\codexrev{The Contract family lifts the instruction.}}
    \label{fig:background-lifting-operator}
  \end{subfigure}\hfill
  \setcounter{subfigure}{6}
  \begin{subfigure}[t]{.48\linewidth}
    \begin{minipage}[t][44pt][t]{\linewidth}
\begin{lstlisting}
nvgpu.operator nv_tmma_f16 (Contract)
    [tcgen05_completion o <- _ reads]:
  actions {
    execute: nv_mma_f16 [o <- _];
    post: [o.commit];
  }
\end{lstlisting}
    \end{minipage}
    \caption{\codexrev{Contract lifting with a completion action.}}
    \label{fig:background-blackwell-operator}
  \end{subfigure}
  \par\vspace{5pt}
  \setcounter{subfigure}{3}
  \begin{subfigure}[t]{.48\linewidth}
    \begin{minipage}[t][16pt][t]{\linewidth}
\begin{lstlisting}
D_acc = amd_tmfma_f32_32x32x16_bf16 A_frag B_frag;
\end{lstlisting}
    \end{minipage}
    \liftingtraversal{%
      {\color{tealcodekeyword}for} p,q,r {\color{tealcodekeyword}in} panels:\par
      \hspace*{1em}{\color{tealcodeoperator}mfma}(A[p,r], B[q,r], D[p,q])%
    }{$64\times64\times64$: $2\times2\times4=16$ calls.}
    \caption{\codexrev{Application and illustrative traversal pseudocode.}}
    \label{fig:background-lifting-application}
  \end{subfigure}\hfill
  \setcounter{subfigure}{7}
  \begin{subfigure}[t]{.48\linewidth}
    \begin{minipage}[t][16pt][t]{\linewidth}
\begin{lstlisting}
D1 = nv_tmma_f16 SA SB
  [computed <- loaded_a, loaded_b, zero];
\end{lstlisting}
    \end{minipage}
    \liftingtraversal{%
      {\color{tealcodekeyword}for} r {\color{tealcodekeyword}in} panels:\par
      \hspace*{1em}{\color{tealcodeoperator}mma}(SA[r], SB[r], D1)%
    }{$64\times64\times64$: $1\times1\times4=4$ calls.}
    \caption{\codexrev{Application and illustrative traversal pseudocode.}}
    \label{fig:background-blackwell-application}
  \end{subfigure}
  \par\vspace{5pt}
  \begin{subfigure}[t]{\linewidth}
    \centering
    \begin{tikzpicture}[x=\linewidth,y=1pt,
        every node/.style={inner sep=0pt,text=tealgridink},
        stream/.style={font=\fontsize{6}{7}\selectfont\tealcodefont},
        streamlabel/.style={font=\fontsize{6.3}{7.5}\selectfont\tealcodefont,
        text=tealcodecomment},
        connect/.style={-{Latex[length=4pt,width=3.5pt]},
          shorten <=3pt,shorten >=4pt,
      draw=tealcodekeyword!75!white,line width=.65pt}]
      \path[use as bounding box] (0,0) rectangle (1,42);
      \fill[tealgridthree!25!white,rounded corners=2pt] (0,0) rectangle (1,42);
      \node[streamlabel] at (.20,32) {gemm.hip};
      \node[streamlabel] at (.80,32) {gemm.cu};
      \node[font=\fontsize{9}{10}\selectfont] (product) at (.50,13)
      {$D=AB^{\mathsf T}$};
      \node[stream] (amdstream) at (.20,13) {%
        {\color{tealcodecomment!40!white}\ldots}\ load\ load\
        \colorbox{tealgridtwo}{\strut\bfseries \textsf{MFMA}}\ store\
      {\color{tealcodecomment!40!white}\ldots}};
      \node[stream] (nvstream) at (.80,13) {%
        {\color{tealcodecomment!40!white}\ldots}\ load\ load\
        \colorbox{tealgridtwo}{\strut\bfseries \textsf{MMA}}\ store\
      {\color{tealcodecomment!40!white}\ldots}};
      \draw[connect] (product.west)--(amdstream.east);
      \draw[connect] (product.east)--(nvstream.west);
    \end{tikzpicture}
  \end{subfigure}
  \endgroup
  \caption{\codexrev{Instruction-centric lifting on AMD and NVIDIA.
      Both dialects connect native instructions to the Contract family and
      target-specific applications. NVIDIA swizzle formulas are
  abbreviated.}}
  \label{fig:background-kernel-streams}
  \vspace{-1.5em}
\end{figure*}

\rev{A hardware instruction operates on fixed-size local operands; applying it to a larger tensor requires invoking it repeatedly across a grid of tile coordinates, which we call \emph{placements}.}
For a $64\times64\times64$ bf16 matrix multiplication $D=AB^T$, the AMD \textsf{v\_mfma} instruction on $32\times32\times16$ fragments~\cite{cdna4} requires $2\times2\times4=16$ placements: \rev{each reads a pair of input fragments and updates a $32\times32$ accumulator tile across four reduction steps.}
At the hardware level (Figure~\ref{fig:background-lifting-native}), each thread invokes a raw intrinsic on local registers with no notion of the surrounding $64\times64\times64$ problem, leaving open which tensor coordinates each call accesses and where those elements physically reside. Moreover, instructions distribute logical tile coordinates non-contiguously across execution units -- for instance, striping the $32\times32$ \textsf{MFMA} accumulator across 64 threads at 16 register slots per thread~\cite{cdna4, hu2025hipkittensfastfuriousamd}. \rev{Two identically shaped $32\times32$ tiles can therefore store elements in incompatible swizzled shared-memory addresses or register slots; passing one instruction's output directly to another is valid only if their physical element locations match, otherwise requiring explicit data movement to reshuffle the layout~\cite{ptxdoc, saba2026cutegenllmbasedagenticframework}.}

A compiler must therefore bridge three representations: high-level tensor coordinates, fixed-size instruction tiles, and physical storage layouts~\cite{linearlayouts, carlisle2026categoricalfoundationscutelayouts, cecka2026cutelayoutrepresentationalgebra}. \rev{Low-level languages (CUDA, HIP) entangle all three inside pointer arithmetic and intrinsics, while layout libraries~\cite{cecka2026cutelayoutrepresentationalgebra} describe single-tile layouts but leave loop generation, data movement, and synchronization to the user.
\name~separates these concerns cleanly through \emph{instruction-centric lifting} (\codexrev{Figure~\ref{fig:background-lifting-native}--\ref{fig:background-lifting-application}}).} First, a machine dialect declares the local instruction specification (Figure~\ref{fig:background-lifting-instruction}): two $32\times16$ bf16 vector register (\textsf{VGPR}) inputs, a $32\times32$ f32 accumulator register (\textsf{AGPR}) output, their layout formulas, and the native intrinsic. An \emph{\operator} declaration (Figure~\ref{fig:background-lifting-operator}) binds this instruction to \textsf{Contract}, a lifting schema for tensor contractions. Because dense tensor computations follow regular coordinate patterns, just six operator families (\textsf{Move}, \textsf{Unary}, \textsf{Linear}, \textsf{Reduce}, \textsf{Scan}, and \textsf{Contract}) suffice to lift the supported instruction-level tensor patterns. A client program can now apply this \operator~to high-level tensor views (Figure~\ref{fig:background-lifting-application}). From this single statement, the \name~compiler derives the $2\times2\times4$ placement grid $(p,q)\in\{0,1\}^2, r\in\{0,1,2,3\}$, reading rows $32p..32p+31$ of $A$, $32q..32q+31$ of $B$, reducing over columns $16r..16r+15$ into the same output tile, resolving all register offsets automatically.

\subsection{Different Machine Specifications, Common Compilation}
\label{sec:archs}

Hardware mechanisms vary substantially across architectures (Table~\ref{tab:overview}). NVIDIA GPUs use asynchronous tensor cores, tensor memory, and shared-memory barriers~\cite{hopperdissect, blackwelldissect, ptxdoc}; AMD GPUs use wave-distributed registers and scalar flight counters~\cite{cdna4, hu2025hipkittensfastfuriousamd}; Intel \textsf{AMX} and Arm \textsf{SME} expose tile registers~\cite{intel_2026, arm_limited_2026}; and spatial accelerators employ local \textsf{SRAMs}, \textsf{DMA} engines, and explicit processor networks~\cite{xu2026maia200softwaredefined, taylor2026numericalkernelsspatialaccelerator, ttblackhole, cerebras}, demanding target-specific schedules~\cite{mmus, khattak2025accuratemodelsnvidiatensor, martinez2025cambrian, zhang2026architectinghardwarediversitymetas}.

\begin{table*}[t]
  \caption{Diverging execution protocols across accelerator architectures.}
  \vspace{-1em}
  \label{tab:overview}
  \centering
  \footnotesize
  \setlength{\tabcolsep}{4pt}
  \renewcommand{\arraystretch}{1.18}
  \begin{tabularx}{\linewidth}{@{}>{\raggedright\arraybackslash\bfseries}p{.15\linewidth}
    *{5}{>{\raggedright\arraybackslash}X}@{}}
    \toprule
    & \makecell[c]{\textbf{NVIDIA}\\\textbf{Blackwell}}
    & \makecell[c]{\textbf{AMD}\\\textbf{CDNA4}}
    & \makecell[c]{\textbf{Arm}\\\textbf{\textsf{SME}}}
    & \makecell[c]{\textbf{Intel}\\\textbf{\textsf{AMX}}}
    & \makecell[c]{\textbf{Tenstorrent}\\\textbf{Blackhole}} \\
    \midrule
    Execution\newline organization
    & Thread, warp,\newline \textsf{CTA}, cluster
    & Thread, wave,\newline workgroup, \textsf{XCD}
    & Thread,\newline streaming mode
    & Thread
    & Tensix core \\
    \addlinespace[3pt]
    Vector ISA
    & SIMT ALU,\newline packed SIMD
    & VALU,\newline packed SIMD
    & \textsf{SVE},\newline Streaming \textsf{SVE}
    & \textsf{AVX-512}
    & SFPU \\
    \addlinespace[3pt]
    Matrix ISA
    & \smalltt{tcgen05.mma}\newline {\scriptsize (Configurable tiles)}
    & \smalltt{v\_mfma\_*}\newline {\scriptsize (Fixed tiles)}
    & \smalltt{fmopa}\newline {\scriptsize (VL-scaled tiles)}
    & \smalltt{tdpbf16ps}\newline {\scriptsize (Configurable tiles)}
    & Matrix FPU\newline {\scriptsize (Fixed tiles)} \\
    \addlinespace[3pt]
    Data Movement ISA
    & \textsf{TMA}, \smalltt{cp.async},\newline \smalltt{tcgen05.ld/st}
    & \textsf{VMEM}/\textsf{LDS}\newline buffer loads
    & \textsf{SVE}/\textsf{ZA} vector\newline load/store
    & Tile\newline load/store
    & RISC-V NoC \textsf{DMA} \\
    \addlinespace[3pt]
    Memory Hierarchy
    & \textsf{HBM}, \textsf{L1}/\textsf{L2},\newline \textsf{SMEM}, \textsf{TMEM},\newline thread registers
    & \textsf{HBM}, \textsf{L1}/\textsf{L2}/\textsf{L3},\newline \textsf{LDS}, \textsf{VGPR},\newline \textsf{AGPR}
    & \textsf{L1}/\textsf{L2}/\textsf{L3},\newline \textsf{SVE} registers,\newline \textsf{ZA} array
    & \textsf{L1}/\textsf{L2}/\textsf{L3},\newline \textsf{AVX} registers,\newline \textsf{TMM} tile registers
    & \textsf{GDDR6}, \textsf{SRAM},\newline matrix registers \\
    \addlinespace[3pt]
    Synchronization
    & \textsf{SMEM} \smalltt{mbarrier}
    & flight counters\newline \smalltt{vmcnt/lgkmcnt}
    & Acquire/release,\newline barriers
    & Fences,\newline locked atomics
    & NoC barriers,\newline semaphores \\
    \bottomrule
  \end{tabularx}
\end{table*}

\ifblind
\else
\vspace{6pt}
\fi

Despite these differences, the underlying compilation problems share a common structure~\cite{unit, Kuzma_2023}. Consider the corresponding \textsf{Contract} construction on NVIDIA Blackwell SM10x (Figure~\ref{fig:background-blackwell-native}--\ref{fig:background-blackwell-application}): unlike AMD's register-based \textsf{MFMA}, Blackwell's \textsf{tcgen05} reads shared memory (\textsf{SMEM}) and accumulates into tensor memory (\textsf{TMEM}) over a $\{32, 64, 128\}\times N\times 16$ tile, using phase barriers rather than flight counters~(\S\ref{sec:composition}). Swapping an AMD intrinsic for a Blackwell intrinsic inside the same loop nest would fail both layout and synchronization requirements.
In \name, the Blackwell dialect declares this instruction and links it to the same \textsf{Contract} family, while the client writes a Blackwell-specific schedule over \textsf{SMEM}, \textsf{TMEM}, and phase barriers. Neither the instructions nor the schedules are interchangeable, but both are verified and lowered by the same hardware-parametric compiler.

\subsection{Composing Lifted Instructions into Explicit Schedules}
\label{sec:composition}

Complete kernels must compose instructions with data movement, stores, processor specialization, and asynchronous pipelines~\cite{yadav2025taskbasedtensorcomputationsmodern, soi2025optimalsoftwarepipeliningwarp, tawa, shah2024flashattention3fastaccurateattention, wasp}. If we precede the \textsf{MFMA} contraction with a lifted load from shared memory (\textsf{LDS}) to \textsf{VGPR} as separate top-level statements, the load issues all of its tile placements before contraction begins, requiring enough registers to hold every panel simultaneously.
Wrapping the \textsf{LDS} reads and \textsf{MFMA} in a \emph{chain} (Figure~\ref{fig:background-composition}) instructs the compiler to fuse compatible outer placement loops so each register fragment is produced, consumed, and reused within a single panel iteration while the accumulator persists. \rev{Loop fusion is governed by data readiness, accumulation completeness\chainfix{, and storage lifetimes}: both the \textsf{MFMA} and final store traverse the $2\times2$ grid of $32\times32$ output tiles, but the store cannot fuse inside the reduction loop until all four reduction panels finish accumulating.}

Asynchronous transfers across global memory, scratchpads, and registers overlap with computation~\cite{1NVIDIAH98}, requiring hardware counters or barriers so consumers await completed transfers and producers do not overwrite live buffers~\cite{verifprod}. Imperative code buries these waits inside entangled loops and inline assembly, making schedules brittle to modify or synthesize -- \agents{changing one scheduling decision requires coordinated edits across addressing, layout, and synchronization code.}
\name~instead expresses asynchronous completion and buffer-reuse dependencies through explicit \emph{token wires} (Figure~\ref{fig:background-composition}): the \textsf{LDS} reads produce tokens consumed by the \textsf{MFMA}, and the compiler applies the dialect's counter rules to synthesize hardware waits while separately verifying tensor operand readiness. Additionally, processor scopes (\codexrev{\S\ref{sec:language-composition}}) decouple placement from processor participation by assigning operations to specific workgroups or waves, letting programmers and agents modify schedules without rewriting low-level address math.

\begin{figure*}[t]
  \centering
  \captionsetup[subfigure]{font=footnotesize,skip=2pt}
  \begin{subfigure}[t]{.48\linewidth}
    \vspace{0pt}
    \begin{minipage}[t]{\linewidth}
\begin{lstlisting}[style=tealpaper,
  basicstyle=\fontsize{5.5}{6.3}\selectfont\tealcodefont\color{black},
  numberstyle=\fontsize{4.6}{5.3}\selectfont\color{tealcodecomment},
  aboveskip=8pt,belowskip=8pt]
A_frag = amd_lds_tread128_32x16 A_lds;
B_frag = amd_lds_tread128_32x16 B_lds;
D_acc = amd_tmfma_f32_32x32x16_bf16
  A_frag B_frag;
\end{lstlisting}
    \end{minipage}\par\nointerlineskip
    \begin{tikzpicture}
      \node[draw=none,fill=tealgridthree!35!white,
      rounded corners=2pt,inner sep=4pt] {%
        \begin{minipage}[c][46pt][c]{\dimexpr\linewidth-8pt\relax}
          \begin{minipage}[c]{.52\linewidth}
            \fontsize{5.5}{6.3}\selectfont\tealcodefont
            {\color{tealcodekeyword}for} panel {\color{tealcodekeyword}in} panels:\par
            \hspace*{1em}{\color{tealcodeoperator}read\_A}(panel)\par
            {\color{tealcodekeyword}for} panel {\color{tealcodekeyword}in} panels:\par
            \hspace*{1em}{\color{tealcodeoperator}read\_B}(panel)\par
            {\color{tealcodekeyword}for} panel {\color{tealcodekeyword}in} panels:\par
            \hspace*{1em}{\color{tealcodeoperator}mfma}(panel)
          \end{minipage}%
          \hfill
          \begin{minipage}[c]{.44\linewidth}
            \raggedright
            \fontsize{6.5}{7.6}\selectfont\sffamily\color{tealcodecomment}
            The two reads and \textsf{MFMA} have separate traversals. No completion dependency
            is expressed between the read results and their consumer.
          \end{minipage}
      \end{minipage}};
    \end{tikzpicture}
    \caption{\codexrev{Separate applications, without chain or completion wires.}}
    \label{fig:background-composition-separate}
  \end{subfigure}%
  \hfill
  \begin{subfigure}[t]{.48\linewidth}
    \vspace{0pt}
    \begin{minipage}[t]{\linewidth}
\begin{lstlisting}[style=tealpaper,
  basicstyle=\fontsize{5.5}{6.3}\selectfont\tealcodefont\color{black},
  numberstyle=\fontsize{4.6}{5.3}\selectfont\color{tealcodecomment},
  aboveskip=0pt,belowskip=0pt,
  moredelim={**[s][\color{tealcodetoken}\itshape]{[}{]}}]
Chain(
  A_frag = amd_lds_tread128_32x16 A_lds [ra <- ];
  B_frag = amd_lds_tread128_32x16 B_lds [rb <- ];
  D_acc = amd_tmfma_f32_32x32x16_bf16
    A_frag B_frag [<- ra, rb];
);
\end{lstlisting}
    \end{minipage}\par\nointerlineskip
    \begin{tikzpicture}
      \node[draw=none,fill=tealgridone!35!white,
      rounded corners=2pt,inner sep=4pt] {%
        \begin{minipage}[c][46pt][c]{\dimexpr\linewidth-8pt\relax}
          \begin{minipage}[c]{.52\linewidth}
            \fontsize{5.5}{6.3}\selectfont\tealcodefont
            {\color{tealcodekeyword}for} panel {\color{tealcodekeyword}in} panels:\par
            \hspace*{1em}{\color{tealcodeoperator}read\_A}(panel)\par
            \hspace*{1em}{\color{tealcodeoperator}read\_B}(panel)\par
            \hspace*{1em}{\color{tealcodetoken}\itshape wait\_for\_reads(panel)}\par
            \hspace*{1em}{\color{tealcodeoperator}mfma}(panel)
          \end{minipage}%
          \hfill
          \begin{minipage}[c]{.44\linewidth}
            \raggedright
            \fontsize{6.5}{7.6}\selectfont\sffamily\color{tealcodecomment}
            Reads and \textsf{MFMA} share a panel traversal. Both reads must complete before
            \textsf{MFMA} consumes that panel's register fragments.
          \end{minipage}
      \end{minipage}};
    \end{tikzpicture}
    \caption{\codexrev{Chain composition with explicit read-to-\textsf{MFMA} completion dependencies.}}
    \label{fig:background-composition-chain}
  \end{subfigure}
  \caption{\codexrev{Composing \textsf{LDS}-to-register reads and \textsf{MFMA}. The shaded panels show intended traversal structure in illustrative pseudocode, not generated code; inner tile placements are omitted for simplicity. \textsf{LDS} inputs are already available and the accumulator is initialized.}}
  \label{fig:background-composition}
\end{figure*}

\subsection{Tensor and Machine Model}
\label{sec:model}

To support diverse architectures without building separate representations for every instruction, \name~formalizes the schedules and hardware primitives of \S\ref{sec:icl}--\S\ref{sec:composition} through two complementary abstractions: a \emph{tensor model} defining a hardware-independent coordinate system and storage layouts~\cite{cecka2026cutelayoutrepresentationalgebra}, and a \emph{machine model} characterizing processor domains, state spaces, instruction signatures, and synchronization automata~\cite{actcompiler, taidl}. 

\subsubsection{Tensor Model}
\label{sec:model-tensor}

A global input matrix, an \textsf{LDS} panel, and a \textsf{VGPR} fragment contain related matrix values in different storage. Movement transfers values between these allocations; blocking or slicing an existing tensor, in contrast, simply selects a different coordinate system without moving any data. Therefore, \name~represents a tensor as a symbolic \emph{view} -- a coordinate domain paired with a storage layout map -- decoupling logical indexing from physical storage~\cite{cecka2026cutelayoutrepresentationalgebra}. Formally, a rank-$R$ tensor is a triple $T=(\mathbf{s},\Phi,\mathcal{C})$ of symbolic extents $\mathbf{s}=(s_0,\ldots,s_{R-1})$, layout map $\Phi$, and coordinate constraints $\mathcal{C}$. $\mathcal{C}$ captures clipped domains such as partial boundary tiles, and $\Phi$ maps each coordinate to a linear storage offset (e.g., $\Phi(i,j)=Ni+j$ for a row-major $M\times N$ matrix).

This separation enables zero-copy geometric transformations: a transformation $\tau$ constructs $T'=(\mathbf{s}',\Phi',\mathcal{C}')$ from $T$ via a coordinate link $f$ with composed layout $\Phi' = \Phi \circ f$, re-indexing coordinates without moving data~\cite{lego}. For example, blocking a $64\times64$ row-major matrix \textsf{C} into $32\times32$ tiles (written \textsf{C(|32, |32)}) produces shape $(2,2,32,32)$ with link $f(p,q,u,v)=(32p+u, 32q+v)$ and layout $\Phi'(p,q,u,v)=64(32p+u)+32q+v$. Selection, slicing, permutation, tiling, ring-buffering (called \emph{spanning} in \name), and repetition follow the same composition rule \codexrev{(see the tensor model in the supplementary material)}.

To track storage identity and aliasing across views, a \emph{tagged tensor} $\widehat{T}=(T,h,a)$ associates a tensor view $T$ with a state space $h$, identifying a hardware storage class such as registers or shared memory, and an allocation $a$, identifying a particular storage object within that class, such as a register or a memory address. Two views of the same allocation alias when their layout maps select the same storage offset; views of distinct allocations are considered disjoint. For example, the matrix $C$ above contains coordinates $\{0,\ldots,63\}^2$, and coordinate $(17,5)$ has offset $64\cdot17+5=1093$.

\subsubsection{Machine Model}
\label{sec:model-machine}

Rather than fixing a universal execution hierarchy, \name~parametrizes a machine through four components that together determine how a scheduled computation can execute: its processor domain, accessible state spaces, native instructions, and synchronization transitions~\cite{actcompiler,taidl}.

\textit{Processors.}
Logical processors \codexrev{within a single device} are indexed by an abstract finite coordinate domain $\codexrev{\mathcal P} \subseteq \mathbb{N}^d$ whose dimensions represent specific hierarchy levels -- such as $(\text{workgroup}, \text{wave})$ on AMD CDNA4~\cite{HIPdocum43}, $(\text{cluster}, \textsf{CTA}, \text{warp})$ on NVIDIA Blackwell, or core coordinates in a spatial array, allowing schedules and views to select processor subsets via coordinate predicates.

\textit{State spaces.}
Each state space $h\in\mathcal{H}$ models an architectural storage class (e.g. global memory, \textsf{LDS}, \textsf{SMEM}, \textsf{TMEM}, \textsf{VGPR}, \textsf{AGPR}, etc.)~\cite{cdna4, ptxdoc}. Additionally, \emph{distributed state spaces} that partition data across cooperating execution units declare a \emph{partition extent} $P_h$ and striping level ($P_h=1$ when undistributed). For example, on MI350X, registers (\textsf{VGPR}, \textsf{AGPR}) are declared $P_h=64$ at the wave level, striping a logical tensor across 64 threads of a wave without altering its shape.

\textit{Instructions.}
Each instruction $\kappa$ has a signature $\Sigma_\kappa=(\widehat{U}_o,\widehat{U}_1,\ldots,\widehat{U}_m)$ specifying the fixed tile shape, state spaces, and physical layouts of its output $\widehat{U}_o$ and inputs $\widehat{U}_1,\ldots,\widehat{U}_m$, alongside its native invocation and synchronization actions. Operand layouts are either hardware-fixed (such as the packed $32\times16$ \textsf{VGPR} and $32\times32$ \textsf{AGPR} layouts of \textsf{v\_mfma}~\cite{cdna4, hu2025hipkittensfastfuriousamd}) or parameterized by symbolic strides inferred from the bound tensor operation (such as strided memory operations).

\textit{Tokens.}
\name~models all operations as asynchronous, with synchronous operations considered self-synchronizing. To coordinate such execution generically without hardcoding target protocols into the compiler, \name~models hardware synchronization primitives as symbolic state machines called \emph{tokens}. A token type declares integer state variables $\vec{s}$ under an invariant $I(\vec{s})$ and transition actions $\vec{s} \to \vec{s}'$, uniformly capturing flight counters (\textsf{vmcnt}, \textsf{lgkmcnt})~\cite{cdna4}, phase-and-byte barriers (\textsf{mbarrier})~\cite{ptxdoc}, and spatial \textsf{DMA} semaphores~\cite{xu2026maia200softwaredefined}.

\codexrev{Together, these components associate each processor on a device with its accessible state spaces, instructions, and token transitions.}

\begin{figure}[t]
  \centering
  \resizebox{\linewidth}{!}{%
    \begin{tikzpicture}[x=1cm,y=1cm,
        every node/.style={font=\fontsize{9}{10.5}\selectfont,
        text=tealgridink,inner sep=1pt,align=center},
        ml/title/.style={font=\bfseries\fontsize{10}{11.5}\selectfont},
        ml/meta/.style={font=\sffamily\fontsize{8.5}{10}\selectfont,
        text=tealgridink!70!white},
        ml/result/.style={font=\boldmath\fontsize{9.5}{11}\selectfont,
        text=tealcodekeyword},
      ml/arrow/.style={->,draw=tealcodekeyword,line width=1pt}]
      \path[use as bounding box] (0,0) rectangle (16,4.4);
      \node at (8,4.15)
      {Teal's canonical \textsf{MFMA} register layout~\cite{cdna4}: $\Psi_D(i,j)=64\bigl(4\lfloor i/8\rfloor+(i\bmod4)\bigr)
      +32\lfloor(i\bmod8)/4\rfloor+j$};
      \foreach \mx/\mheading/\mshade in
      {0/{Logical Tile}/tealgridthree,
        5.4/{Ownership Projection}/tealgridone,
      10.8/{Thread-Local Storage}/tealgridthree} {
        \fill[fill=\mshade!25!white,rounded corners=2pt]
        (\mx,0) rectangle ++(5.2,3.72);
        \node[ml/title] at (\mx+2.6,3.47) {\mheading};
      }
      \node[ml/meta] at (2.6,3.08) {$32\times32$ instruction tile};
      \node[ml/meta] at (8,3.08) {$(\mathrm{thread},\mathrm{slot})$};
      \node[ml/meta] at (13.4,3.08) {Thread 5, \textsf{AGPR} slots 0--15};
      \foreach \mr in {0,1,2} {
        \pgfmathtruncatemacro{\mrow}{16+\mr}
        \pgfmathtruncatemacro{\mslot}{8+\mr}
        \foreach \mc in {0,1,2} {
          \pgfmathtruncatemacro{\mcol}{4+\mc}
          \def\mleftfill{tealgridthree}
          \def\mrightfill{tealgridone}
          \ifnum\mr=1 \ifnum\mc=1
          \def\mleftfill{tealgridtwo}
          \def\mrightfill{tealgridtwo}
          \fi\fi
          \filldraw[draw=tealgridrule,fill=\mleftfill]
          (1.25+.90*\mc,2.63-.50*\mr) rectangle ++(.90,-.50);
          \node at (1.70+.90*\mc,2.38-.50*\mr) {$(\mrow,\mcol)$};
          \filldraw[draw=tealgridrule,fill=\mrightfill]
          (6.65+.90*\mc,2.63-.50*\mr) rectangle ++(.90,-.50);
          \node at (7.10+.90*\mc,2.38-.50*\mr) {$(\mcol,\mslot)$};
        }
      }
      \foreach \mx in {2.6,8} {
        \node[ml/meta] at (\mx,2.84) {$\vdots$};
        \node[ml/meta] at (\mx,1.09) {$\vdots$};
      }
      \foreach \mx in {.95,4.25,6.35,9.65}
      \node[ml/meta] at (\mx,1.88) {$\cdots$};
      \node[ml/result] at (2.6,.65) {$\Psi_D(17,5)=581$};
      \node[ml/meta] at (2.6,.25) {Packed instruction index};
      \node[ml/result] at (8,.65) {$\operatorname{location}(17,5)=(5,9)$};
      \node[font=\fontsize{8}{9.5}\selectfont] at (8,.25)
      {$581\bmod64=5,\ \lfloor581/64\rfloor=9$};
      \foreach \mr/\mlabel in {0/{\vdots},1/8,2/9,3/10,4/{\vdots}} {
        \def\mfill{tealgridthree}
        \ifnum\mr=2 \def\mfill{tealgridtwo}\fi
        \filldraw[draw=tealgridrule,fill=\mfill]
        (12.75,2.83-.38*\mr) rectangle ++(1.30,-.38);
        \node at (13.40,2.64-.38*\mr) {$\mlabel$};
      }
      \node[ml/result] at (13.4,.65) {Thread 5, slot 9};
      \node[ml/meta] at (13.4,.25) {Thread-local \textsf{AGPR} word};
      \draw[ml/arrow] (5.24,1.88)--(5.76,1.88);
      \draw[ml/arrow] (10.64,1.88)--(11.16,1.88);
  \end{tikzpicture}}
  \caption{Native \textsf{MFMA} accumulator layout $\Psi_D$ on a $32\times32$ instruction tile.
    A logical neighborhood maps through $\Psi_D$ to thread and local-slot pairs.
    The highlighted element $(17,5)$ has packed index 581 and occupies slot 9
  of thread 5's 16-word \textsf{AGPR} fragment.}
  \label{fig:mfma-ownership}
\end{figure}
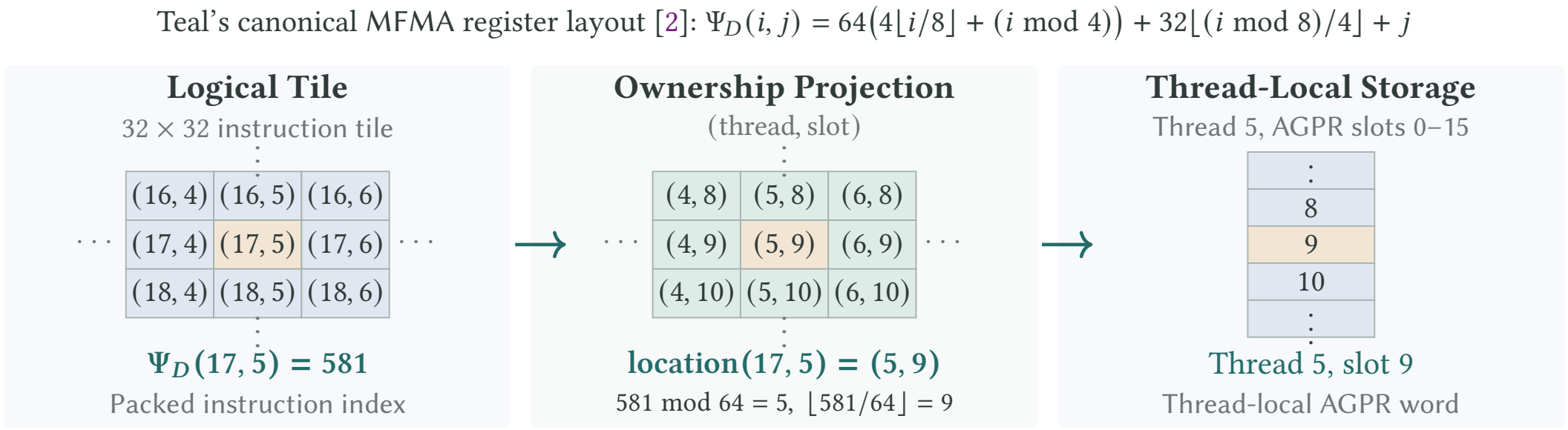

\textit{Physical layouts and materialization.}
In a state space distributed across $P_h$ owners (e.g., the 64 threads of an AMD wave), instructions impose physical layout requirements on the operands: each element must reside with a particular hardware owner and a particular local storage slot. For example, the MI350X \textsf{MFMA} accumulator element $(17,5)$ resides in thread 5, register slot 9 (Figure~\ref{fig:mfma-ownership}).

A tensor describes logical coordinates over a memory allocation, but its elements must also occupy the physical locations expected by the instruction. At each placement $\pi$, \name~verifies this compatibility by tracing the physical location of each element through its layout map. The same mapping determines the per-owner base offset $b(\pi)$ used for code generation.

\section{Programming Model}
\label{sec:language}

A \name~program constructs a target-specific schedule in three parts. First, the client declares tensors in state spaces exposed by the machine dialect, identifying the storage used by the computation. In the MI350X GEMM of Figure~\ref{fig:language-gemm}, the client declares global inputs and outputs, allocates \smalltt{SA}/\smalltt{SB} buffers in \textsf{LDS}, \smalltt{RA}/\smalltt{RB} in \textsf{VGPR}, and \smalltt{D\_acc} in \textsf{AGPR} (lines 18-23).
Then the client constructs tensor views that describe how this storage participates in computation. In the example, slicing selects each workgroup's input rows, while blocking partitions $K$ into 64-element $K$-tiles, called \emph{panels}. The \smalltt{[TK] ++} transformation introduces a zero-copy repeated dimension, allowing successive panels to reuse the same \textsf{LDS} and register allocations without copying data. Similarly, blocking \smalltt{D\_acc(|32, |32)} exposes the accumulator as a $[2,2,32,32]$ tensor of instruction-sized tiles, and repetition allows all $K$-panels to update the same allocation (lines 26–33).
Then the client applies instructions to these views and composes their execution using chains, processor scopes, and token dependencies. The compiler derives the individual placements and operand addresses instead of requiring the client to enumerate them (lines 35–57).
Figure~\ref{fig:language} illustrates the computation, AMD CDNA4 dialect declarations, and a simple bf16 GEMM schedule in which each single-wave workgroup computes a $64\times64$ output tile.

\begin{figure*}[p]
  \centering
  \captionsetup{skip=4pt}
  \captionsetup[subfigure]{font=footnotesize,skip=3pt}
  \newsavebox{\gemmOverviewBox}
  \newsavebox{\gemmDialectBox}
  \newsavebox{\gemmKernelBox}
  \begin{lrbox}{\gemmOverviewBox}
    \begin{subfigure}[t]{\dimexpr.4\linewidth-.4em\relax}
      \centering
\resizebox{\linewidth}{!}{%
  \begin{tikzpicture}[x=1cm,y=1cm,
      every node/.style={font=\fontsize{10}{11.5}\selectfont,
      text=tealgridink,inner sep=2pt,align=center},
      gs/title/.style={font=\bfseries\fontsize{11}{12.5}\selectfont},
      gs/meta/.style={font=\fontsize{9}{10.5}\selectfont},
      gs/arrow/.style={-{Latex[length=2mm,width=1.4mm]},
      draw=tealcodekeyword,line width=.9pt},
    gs/grid/.style={draw=tealgridrule,line width=.45pt}]
    \path[use as bounding box] (0,0) rectangle (9,11.2);
    \node[gs/title] at (4.5,10.95) {$C=AB^{\mathsf T}$};
    \node[gs/meta] at (4.5,10.55) {One workgroup, one wave, one $64\times64$ output tile};
    \node at (2.2,10.12) {$A$};
    \node at (6.8,10.12) {$B$};
    \foreach \gsx/\gsshade in {1.525/tealgridthree,6.125/tealgridone} {
      \fill[tealgridexcluded] (\gsx,8.55) rectangle ++(1.35,1.35);
      \ifdim\gsx pt<2pt
      \fill[\gsshade] (\gsx,9.225) rectangle ++(1.35,.3375);
      \fill[\gsshade!65!tealcodekeyword] (\gsx+.3375,9.225) rectangle ++(.3375,.3375);
      \else
      \fill[\gsshade] (\gsx,8.8875) rectangle ++(1.35,.3375);
      \fill[\gsshade!65!tealcodekeyword] (\gsx+.3375,8.8875) rectangle ++(.3375,.3375);
      \fi
      \foreach \gsi in {0,...,4} {
        \draw[gs/grid] (\gsx+.3375*\gsi,8.55)--++(0,1.35);
        \draw[gs/grid] (\gsx,8.55+.3375*\gsi)--++(1.35,0);
      }
    }
    \draw[draw=tealcodekeyword,line width=.8pt] (1.8625,9.225) rectangle ++(.3375,.3375);
    \draw[draw=tealcodekeyword,line width=.8pt] (6.4625,8.8875) rectangle ++(.3375,.3375);
    \node[gs/meta] at (2.2,8.30) {$M\times K$};
    \node[gs/meta] at (6.8,8.30) {$N\times K$};
    \node[gs/meta] at (4.5,8.75) {Select $64\times64$\\panels};
    \fill[tealgridthree!16!white,rounded corners=3pt]
    (.2,1.58) rectangle (8.8,7.98);
    \node[gs/title] at (4.5,7.70) {For each $K$-panel};
    \node[gs/meta] at (4.5,7.35) {$t=0,\ldots,K/64-1$};
    \draw[gs/arrow] (2.2,8.08)--(2.2,7.55);
    \draw[gs/arrow] (6.8,8.08)--(6.8,7.55);
    \foreach \gsx/\gsshade/\gsname in
    {2.2/tealgridthree/A,6.8/tealgridone/B} {
      \node at (\gsx,7.27) {$S_{\gsname}$};
      \foreach \gsr in {0,1} {
        \foreach \gsc in {0,...,3} {
          \filldraw[gs/grid,fill=\gsshade]
          (\gsx-.525+.2625*\gsc,7.00-.525*\gsr)
          rectangle ++(.2625,-.525);
        }
      }
      \draw[gs/arrow] (\gsx,5.95)--(\gsx,5.51);
      \foreach \gsr in {0,1} {
        \foreach \gsc in {0,...,3} {
          \filldraw[gs/grid,fill=\gsshade]
          (\gsx-.525+.2625*\gsc,5.45-.525*\gsr)
          rectangle ++(.2625,-.525);
        }
      }
      \node at (\gsx,4.17) {$R_{\gsname}$};
    }
    \node[gs/meta] at (4.5,6.48) {Load to \textsf{LDS}\\$64\times64$ (bf16)};
    \node[gs/meta] at (4.5,4.93) {Read to \textsf{VGPR}\\$64\times64$ (bf16)};
    \fill[tealgridtwo,rounded corners=2pt] (3.15,3.22) rectangle (5.85,4.00);
    \node[gs/title] at (4.5,3.75) {\textsf{MFMA}};
    \node[gs/meta] at (4.5,3.42) {$16$ calls per panel};
    \draw[gs/arrow] (2.725,4.40)--(3.20,4.06);
    \draw[gs/arrow] (6.275,4.40)--(5.80,4.06);
    \draw[gs/arrow] (4.5,3.22)--node[midway,right,gs/meta,
    inner xsep=4pt,inner ysep=0pt] {Sum across panels}(4.5,2.83);
    \foreach \gsr in {0,1} {
      \foreach \gsc in {0,1} {
        \filldraw[gs/grid,fill=tealgridfive]
        (3.975+.525*\gsc,2.77-.525*\gsr) rectangle ++(.525,-.525);
      }
    }
    \node at (2.2,2.60) {$D_{\mathrm{acc}}$};
    \node[gs/meta] at (2.2,2.20) {$64\times64$ (fp32)};
    \node[gs/meta] (gszero) at (2.2,1.84) {Zeroed once};
    \draw[gs/arrow] (gszero.east)--(3.92,2.02);
    \draw[gs/arrow] (4.5,1.72)--(4.5,1.06);
    \node[gs/meta] at (2.18,1.18) {After the last panel};
    \fill[tealgridtwo,rounded corners=2pt] (3.35,.35) rectangle (5.65,1.00);
    \node at (4.5,.675) {Cast to bf16};
    \draw[gs/arrow] (5.65,.675)--(6.71,.675);
    \node[gs/meta] at (6.18,1.03) {Store};
    \foreach \gsr in {0,1} {
      \foreach \gsc in {0,1} {
        \filldraw[gs/grid,fill=tealgridtwo]
        (6.8+.5*\gsc,1.15-.5*\gsr) rectangle ++(.5,-.5);
      }
    }
    \node[gs/meta] at (7.3,1.40) {$C_{\mathrm{tile}}$: \textsf{global}};
\end{tikzpicture}}
      \caption{One $64\times64$ GEMM tile: f32 accumulation across
      $K$-panels, followed by bf16 cast and writeback.}
      \label{fig:gemm-schedule-overview}
    \end{subfigure}
  \end{lrbox}
  \begin{lrbox}{\gemmDialectBox}
    \begin{subfigure}[t]{\dimexpr.4\linewidth-.4em\relax}
\begin{lstlisting}[style=tealpaper,
  basicstyle=\fontsize{5.5}{6.3}\selectfont\tealcodefont\color{black},
  numberstyle=\fontsize{4.6}{5.3}\selectfont\color{tealcodecomment},
  morekeywords={token,states,actions,init,commit,act,complete,nop,instr,operator,Move,execute,using,layout,base,offset}]
# Track outstanding global-memory loads.
mi350.token vmcnt : states s [0 <= s];
  actions {
    init (0 <- s) { nop; }
    commit { nop; } (s + 1 <- s)
    act (s - 1 <- s) {
      amd_wait_vmcnt<s>();
    }
    complete (0 <- s) {
      amd_wait_vmcnt<s>();
    }
  };

# Load a global-memory buffer into LDS.
mi350.instr amd_lds_ldbuf_32x16:
  using T [_f16, _b16];
  out: Tensor [32,16] : layout lds [i,j =>
    (i*16+j) ^ ((((i*16+j)%512)/256)*8)] <c,T,LDS>;
  in: Tensor [32,16] : layout la
    <a,T,global>;
  {
    _amdw_lds_ldbuf<T,~lds,la>
        (c, base a, offset a);
  }

# Lift the load over operand tensors.
mi350.operator amd_lds_tldbuf_32x16(Move)
  [vmcnt o <- _ i]:
  using T [_f16, _b16];
  actions {
    execute: amd_lds_ldbuf_32x16 [o <-];
  }
\end{lstlisting}
      \caption{\codexrev{AMD dialect examples: a global-memory completion counter, a global-to-\textsf{LDS} instruction, and its Move operator.}}
      \label{fig:dialect-examples}
    \end{subfigure}
  \end{lrbox}
  \begin{lrbox}{\gemmKernelBox}
  \begin{subfigure}[t]{\dimexpr.6\linewidth-.6em\relax}
    \vspace{0pt}
    \centering
    \begin{minipage}[t]{\linewidth}
\begin{lstlisting}[style=tealpaper,
  basicstyle=\fontsize{5.5}{6.3}\selectfont\tealcodefont\color{black},
  numberstyle=\fontsize{4.6}{5.3}\selectfont\color{tealcodecomment}]
arch amd_common;

# Launch a grid of workgroups, with one wave per 64x64 output tile.
mi350<(M / 64) * (N / 64), 1>.kernel gemm_bf16:
  # Declare bf16 inputs and bf16 output; symbolic sizes such as M, N, and K are inferred to be runtime dimensions.
  out C: global._b16[M, N];
  in A: global._b16[M, K];
  # Compute C = A B^T; both inputs store K contiguously.
  in B: global._b16[N, K];
  # Constraints support symbolic proofs and enable optimizations.
  with [M > 0, M <= 16384, 64 | M,
        N > 0, N <= 16384, 64 | N,
        K > 128, K <= 16384, 64 | K]
  {
    # Assign each workgroup its output-tile coordinates.
    on pid(0) as (tm, tn)[M / 64, N / 64] {
      # Allocate on-chip shared memory, operand registers, and accumulators.
      SA: LDS._b16[64, 64];
      SB: LDS._b16[64, 64];
      RA: VGPR._b16[64, 64];
      RB: VGPR._b16[64, 64];
      # Accumulate in fp32; convert to bf16 only at writeback.
      D_acc: AGPR._f32[64, 64];

      # Process the reduction dimension in TK panels.
      let TK = K / 64;
      # Create panel views that reuse the same physical storage.
      # [TK] ++ adds a repeated logical dimension of size TK, without copying data.
      SA_tile = [TK] ++ SA; SB_tile = [TK] ++ SB;
      RA_tile = [TK] ++ RA; RB_tile = [TK] ++ RB;
      # Block the accumulator into native 32x32 MFMA tiles.
      # |32 blocks the dimension in units of 32 elements.
      D_tile = [TK] ++ D_acc(|32, |32);

      # Initialize the accumulator before processing any K panels.
      amd_tset (0) D_acc;

      # Interleave panel loads, register reads, and compute for storage reuse.
      Chain(
          # Load this workgroup's A and B panels into LDS.
          # idx<len> selects len elements starting at index idx.
          SA_tile = amd_lds_tldbuf_32x16 (A((64 * tm)<64>, |64)) [vmA <- ];
          SB_tile = amd_lds_tldbuf_32x16 (B((64 * tn)<64>, |64)) [vmB <- ];

          # Read operands into registers, with explicit completion dependencies.
          RA_tile = amd_lds_tread128_32x16 SA_tile [rmA <- vmA];
          RB_tile = amd_lds_tread128_32x16 SB_tile [rmB <- vmB];

          # Use TK as one batch dimension (<1>), reusing the accumulator through the view.
          D_tile = amd_tmfma_f32_32x32x16_bf16 <1> RA_tile RB_tile [<- rmA, rmB];
      );

      # Interleave bf16 cast and global store to enable reuse.
      Chain(
          D_bf16 = amd_tcast_b16 (D_tile(0));
          (C((64 * tm)<64>, (64 * tn)<64>)(|32, |32)) := amd_tmov D_bf16;
      );
    };
  }
\end{lstlisting}
    \end{minipage}
    \caption{\codexrev{Naive bf16 GEMM on MI350X: asynchronous buffer loads to \textsf{LDS}, asynchronous register reads, matrix-core accumulation in fp32, and bf16 writeback.}}
    \label{fig:language-gemm}
  \end{subfigure}
  \end{lrbox}
  \begin{minipage}[t][\dimexpr\ht\gemmKernelBox+\dp\gemmKernelBox\relax][s]{\dimexpr.4\linewidth-.4em\relax}
    \vspace{0pt}
    \usebox{\gemmOverviewBox}\par
    \vfill
    \usebox{\gemmDialectBox}\par
    \vspace{0pt}
  \end{minipage}%
  \hfill
  \begin{minipage}[t]{\dimexpr.6\linewidth-.6em\relax}
    \vspace{0pt}
    \usebox{\gemmKernelBox}
  \end{minipage}
  \caption{\codexrev{\name's client and backend languages, illustrated by an AMD machine dialect and GEMM schedule. The dialect defines instruction behavior and synchronization, while the client composes their applications.}}
  \label{fig:language}
\end{figure*}

\subsection{Instruction-Centric Lifting}
\label{sec:language-lifting}

With tensors constructed, the \textsf{MFMA} operator in Figure~\ref{fig:language-gemm} receives $64\times 64$ slices of \smalltt{RA\_tile} and \smalltt{RB\_tile} as tensor operands and binds its result to one $64\times 64$ \smalltt{D\_tile}. Rather than enumerating the 16 placements of the \textsf{MFMA} and their offsets, the client applies one operator directly to the operands; \name~constructs the instruction placements.
In abbreviated form, an \operator~$\mathcal{O}=(\kappa,\Theta,\Omega,\alpha_o,\alpha_1,\ldots,\alpha_m)$ lifts an arity-$m$ instruction $\kappa$ over symbolic-size tensors. The output constructor $\Theta$ determines the result geometry, the placement family $\Omega$ produces an ordered set $\Pi$ of instruction invocations, and each access map $\alpha_j$ selects the tensor coordinate used by operand $j$ at placement $\pi\in\Pi$ and instruction-local coordinate $\vec{u}$. Thus, the operator defines both \emph{how many times} the instruction is invoked and \emph{which} tensor elements each invocation accesses.

For the operator application, specialization \smalltt{<1>} treats the leading $K$-panel axis as a shared batch dimension. At each fixed panel coordinate, the operator derives the following placements $\pi=(p,q,r)$ and access maps:
\begin{equation}
  \label{eq:language-contract-example}
  \begin{aligned}
    \Pi &= \{0,1\}^2\times\{0,\ldots,3\}, &
    \alpha_A(\pi,(u,k)) &= (32p+u,16r+k),\\
    \alpha_o(\pi,(u,v)) &= (p,q,u,v), &
    \alpha_B(\pi,(v,k)) &= (32q+v,16r+k),
  \end{aligned}
\end{equation}
for $0\leq u,v<32$ and $0\leq k<16$. The input maps select the corresponding $32\times16$ fragments, while $\alpha_o$ is independent of $r$: all four reduction contributions update the same $32\times32$ accumulator tile. Thus, combined with the tensors and instruction layouts, these maps determine the operand tiles of each invocation without client-written address calculations.

\name~provides six machine-independent \operator~schemas that construct $(\Theta,\Omega,\alpha_j)$ from an instruction signature. \textsf{Linear} and \textsf{Unary} apply instructions over independent tiles; \textsf{Move} transfers data between spaces or layouts; \textsf{Reduce} accumulates along one input dimension; \textsf{Contract} combines two inputs over a shared reduction dimension; and \textsf{Scan} performs ordered prefix updates. These schemas also expose output-availability and contribution-order requirements used when composing operations. Consequently, a new native instruction can reuse the same lifting and scheduling machinery through its dialect declaration.

Finally, the client controls how an operator's result is represented. \emph{Bind} (\textsf{=}) adopts the canonical result constructed by $\Theta$, allocating storage or rebinding a view over an existing allocation, as with \smalltt{D\_bf16}. \emph{Write} (\textsf{:=}) instead preserves a selected destination view and requires the operator's output accesses to conform to it. The final \smalltt{amd\_tmov} therefore writes into the row-major layout of the global \smalltt{C} tensor rather than adopting the \textsf{MFMA} register layout.

\subsection{Structured Composition}
\label{sec:language-composition}

\name's client program expresses decisions on operator interleaving, processor assignment and synchronizations independently through processor scopes, chains, and token wires.

\paragraph{Processor Scopes.}
Processor scopes assign computations to hardware resources without changing their operator placements. In Figure~\ref{fig:language-gemm}, \smalltt{on pid(0)} assigns each workgroup coordinates \smalltt{(tm,tn)}, selecting 64 rows of each input and a $64\times64$ output tile; each workgroup contains one wave. These processor coordinates are distinct from the local $(p,q,r)$ instruction placements within that tile. Formally, a nesting path of scope predicates $\delta=(\varphi_1,\ldots,\varphi_s)$ restricts the device's logical processor domain $\mathcal P$ to
\begin{equation}
  \label{eq:language-scope}
  \mathcal P_{\mathrm{active}}(\delta)=
  \{\vec p\in\mathcal P\mid \bigwedge_{i=1}^{s}\varphi_i(\vec p)\}.
\end{equation}
The same mechanism supports more specialized processor roles, including warp specialization~\cite{soi2025optimalsoftwarepipeliningwarp}.

\paragraph{Chains.}
The first GEMM chain illustrates why separately lifted operations must be composed: a GPU has limited on-chip memory, and the \smalltt{[TK] ++} views deliberately map successive $K$-panels onto the same \textsf{LDS} and register buffers for reuse. Executing all loads, then all register reads, and finally all contractions as separate complete traversals would overwrite intermediate values before their consumers use them. Instead, the \textsf{chain} requests a shared traversal: for each $K$-panel, load \smalltt{A} and \smalltt{B} into \textsf{LDS}, read their register fragments, perform the \textsf{MFMA} contractions, and then reuse the buffers for the next panel.

Formally, a chain of $n$ ordered applications requests that compatible outer placement dimensions be shared. For a candidate shared prefix $Q_d$ of depth $d$, with application-local suffixes $S_{k,d}(q)$, its execution order is
\begin{equation}
  \label{eq:language-chain-order}
  \mathcal{S}_d =
  \underset{q\in Q_d}{\operatorname{seq}}\;
  \underset{k=0}{\overset{n-1}{\operatorname{seq}}}\;
  \underset{t\in S_{k,d}(q)}{\operatorname{seq}}(k,q,t).
\end{equation}
Here, $d=0$ executes complete applications sequentially, whereas $d>0$ interleaves them at each shared coordinate. In Figure~\ref{fig:language-gemm}, $Q_1$ is the outer $K$-panel traversal ($d=1$); each operation retains its own internal local placements, including the sixteen \textsf{MFMA} invocations per panel. The compiler greedily admits the deepest prefix whose interleaving passes value-preservation checks (\S\ref{sec:compilation}). The second chain similarly interleaves bf16 conversion and global writeback over $32\times32$ output tiles, avoiding fully materializing \smalltt{D\_bf16}. Thus, clients request fusion structurally while the compiler establishes its legality and realizes the corresponding loops and storage reuse.

\paragraph{Synchronization and Token Wires.}
\name~expresses completion dependencies through typed token wires, written \smalltt{[out <- in]}, while the machine dialect defines their hardware realization. In Figure~\ref{fig:language-gemm}, global-to-\textsf{LDS} loads produce \smalltt{vmA} and \texttt{vmB}, which the register reads consume. Those reads produce \smalltt{rmA} and \smalltt{rmB}, and the \textsf{MFMA} consumes their completion dependencies through \smalltt{[<- rmA, rmB]}. The client specifies these relationships without having to manually encode the corresponding hardware wait instructions.

Each wire connects instruction slots to a token resource $r\in\mathcal R$ with state domain $K_{\tau(r)}$. The connected resources form a product state $\sigma\in\prod_{r\in\mathcal R}K_{\tau(r)}$, updated by the dialect's token actions. The compiler verifies the declared state transitions and completion protocol across the composed placements. Separating completion dependencies from placement order and processor assignment lets clients modify one scheduling decision without reconstructing the others or their low-level address arithmetic.

\subsection{Static Legality and Language Scope}
\label{sec:language-legality}

The same structure that makes schedules concise also makes their legality verifiable. Tensors retain symbolic coordinates and storage identities; lifted operators expose exact instruction accesses; chains describe candidate execution orders; processor scopes identify participants; and token wires expose declared completion dependencies. The compiler can therefore verify these properties before generating target code.
For the GEMM in Figure~\ref{fig:language-gemm}, the assumptions $64\mid M$, $64\mid N$, and $64\mid K$ ensure complete workgroup and $K$-panel tiles, enabling constant simplification optimizations, while the native \textsf{MFMA} retains its fixed $32\times32\times16$ geometry. The access maps in Equation~\ref{eq:language-contract-example} must stay within their tensor domains, and their physical materializations must match the \textsf{MFMA}'s required thread and register-slot layouts (\S\ref{sec:model-machine}). Composition checks that repeated buffers are not overwritten before their values are consumed, that reduction contributions occur in the required order, and that final outputs observe their intended values. Token-state checks separately validate the declared asynchronous completion protocol. Violations can be rejected at compile time rather than discovered through erroneous device execution or undefined runtime behaviour.

This verification depends on a deliberate language boundary: instruction placements, tensor accesses, and composition must be statically analyzable under symbolic kernel parameters. \name~supports tiled computations with explicit data movement and structured synchronization, but excludes unrestricted data-dependent control flow and arbitrary pointer-based accesses. Within this scope, clients retain explicit control over target-specific instruction schedules while the compiler derives their concrete loops, operand addresses, and synchronization realization from the same representation it verifies.

\section{Compilation}
\label{sec:compilation}
\tikzset{
  cf/box/.style={draw=tealgridrule,rounded corners=1.5pt,
  minimum height=5mm,inner sep=3pt,align=center},
  cf/panel/.style={cf/box,anchor=north west,align=left,
  font=\fontsize{5.8}{7}\selectfont},
  cf/arrow/.style={->,draw=tealcodekeyword,line width=.6pt},
  cf/order/.style={->,draw=tealgridrule,dashed,line width=.4pt},
  cf/small/.style={font=\fontsize{6}{7}\selectfont}
}
\newenvironment{compilationstrip}[2]{%
  \def\cfcaption{#1}\def\cflabel{#2}%
  \begin{figure*}[tbp]\centering
    \begin{tikzpicture}[x=.0625\linewidth,y=.0625\linewidth,
      every node/.style={font=\fontsize{7}{8}\selectfont,text=tealgridink}]
      \path[use as bounding box] (0,0) rectangle (16,5.2);
    }{%
    \end{tikzpicture}
    \caption{\codexrev{\cfcaption}}\label{\cflabel}
  \end{figure*}
}
\newcommand{\compilationgraphfigure}{%
  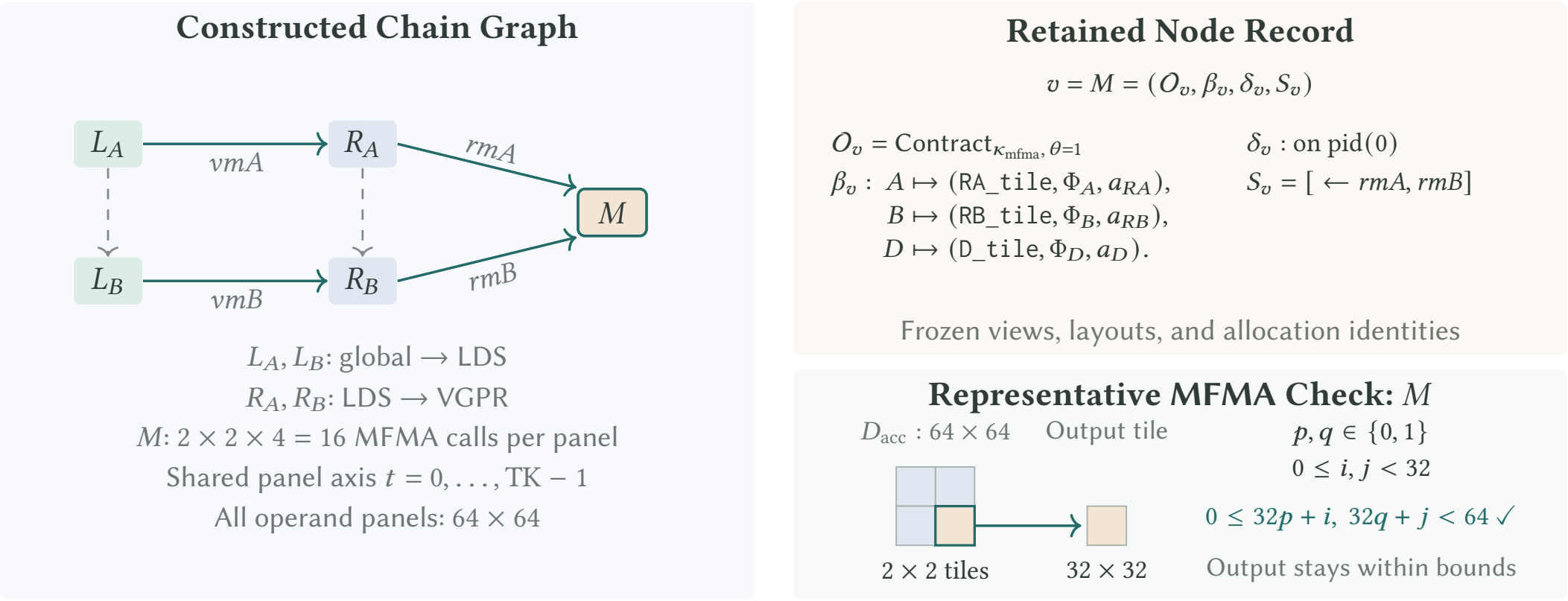
\begin{figure*}[tbp]
    \centering
    \resizebox{\linewidth}{!}{%
      \begin{tikzpicture}[x=1cm,y=1cm,
          every node/.style={font=\fontsize{9.5}{11}\selectfont,
          text=tealgridink,inner sep=2pt,align=center},
          gc/title/.style={font=\bfseries\fontsize{10}{11.5}\selectfont},
          gc/meta/.style={font=\sffamily\fontsize{8.5}{10}\selectfont,
          text=tealgridink!70!white},
          gc/math/.style={font=\fontsize{7.8}{9}\selectfont},
          gc/key/.style={font=\sffamily\fontsize{7.8}{9}\selectfont,
          text=tealgridink!70!white},
          gc/node/.style={rounded corners=2pt,minimum width=.70cm,
          minimum height=.48cm,inner sep=3pt},
          gc/token/.style={->,draw=tealcodekeyword,line width=.75pt},
          gc/order/.style={->,draw=tealgridink!60!white,dashed,line width=.6pt},
          gc/carried/.style={->,draw=tealcodekeyword,densely dotted,line width=.8pt},
        gc/result/.style={text=tealcodekeyword}]
        \path[use as bounding box] (0,0) rectangle (16,6.1);
        \fill[tealgridthree!25!white,rounded corners=2pt] (0,0) rectangle (7.7,6.1);
        \fill[tealgridtwo!25!white,rounded corners=2pt] (8.1,2.5) rectangle (16,6.1);
        \fill[tealgridone!25!white,rounded corners=2pt] (8.1,0) rectangle (16,2.35);
        \node[gc/title] at (3.85,5.82) {Constructed Chain Graph};
        \node[gc/title] at (12.05,5.82) {Retained Node Record};
        \node[gc/node,fill=tealgridone] (gla) at (1.10,4.65) {$L_A$};
        \node[gc/node,fill=tealgridone] (glb) at (1.10,3.25) {$L_B$};
        \node[gc/node,fill=tealgridthree] (gra) at (3.70,4.65) {$R_A$};
        \node[gc/node,fill=tealgridthree] (grb) at (3.70,3.25) {$R_B$};
        \node[gc/node,fill=tealgridtwo,draw=tealcodekeyword,line width=.8pt]
        (gm) at (6.25,3.95) {$M$};
        \draw[gc/token] (gla.east)--node[below,gc/meta] {$\mathit{vmA}$}(gra.west);
        \draw[gc/token] (glb.east)--node[below,gc/meta] {$\mathit{vmB}$}(grb.west);
        \draw[gc/token] (gra.east)--node[above,sloped,gc/meta] {$\mathit{rmA}$}(gm.north west);
        \draw[gc/token] (grb.east)--node[below,sloped,gc/meta] {$\mathit{rmB}$}(gm.south west);
        \draw[gc/order] (gla.south)--(glb.north);
        \draw[gc/order] (gra.south)--(grb.north);
        \node[gc/meta] at (3.85,2.45) {$L_A,L_B$: global $\rightarrow$ \textsf{LDS}};
        \node[gc/meta] at (3.85,2.04) {$R_A,R_B$: \textsf{LDS} $\rightarrow$ \textsf{VGPR}};
        \node[gc/meta] at (3.85,1.63) {$M$: $2\times2\times4=16$ \textsf{MFMA} calls per panel};
        \node[gc/meta] at (3.85,1.22) {Shared panel axis $t=0,\ldots,\mathrm{TK}-1$};
        \node[gc/meta] at (3.85,.81) {All operand panels: $64\times64$};
        \node[gc/math] (grecord) at (12.05,5.25)
        {$v=M=(\mathcal O_v,\beta_v,\delta_v,S_v)$};
        \node[gc/math,anchor=north west] at (8.40,4.85)
        {$\mathcal O_v=\mathsf{Contract}_{\kappa_{\mathrm{mfma}},\,\theta=1}$};
        \node[gc/math,anchor=north west] at (8.40,4.45)
        {\setlength{\jot}{0pt}$
          \begin{aligned}
            \beta_v:\ A&\mapsto(\mathtt{RA\_tile},\Phi_A,a_{RA}),\\
            B&\mapsto(\mathtt{RB\_tile},\Phi_B,a_{RB}),\\
            D&\mapsto(\mathtt{D\_tile},\Phi_D,a_D).
        \end{aligned}$};
        \node[gc/math,anchor=north west] at (12.65,4.85)
        {\setlength{\jot}{0pt}$
          \begin{aligned}
            \delta_v:\ &\mathrm{on}\ \mathrm{pid}(0)
        \end{aligned}$};
        \node[gc/math,anchor=north west] at (12.65,4.45)
        {\setlength{\jot}{0pt}$
          \begin{aligned}
            S_v={}&[\ \leftarrow\mathit{rmA},\mathit{rmB}]
        \end{aligned}$};
        \node[gc/meta] at (12.05,2.72) {Frozen views, layouts, and allocation identities};
        \node[gc/title] at (12.05,2.07) {Representative \textsf{MFMA} Check: $M$};
        \node[gc/key] at (9.55,1.70) {$D_{\mathrm{acc}}:64\times64$};
        \foreach \gr in {0,1} {
          \foreach \gcol in {0,1} {
            \def\gshade{tealgridthree}
            \ifnum\gr=1 \ifnum\gcol=1 \def\gshade{tealgridtwo}\fi\fi
            \filldraw[draw=tealgridrule,fill=\gshade]
            (9.15+.40*\gcol,1.35-.40*\gr) rectangle ++(.40,-.40);
          }
        }
        \draw[draw=tealcodekeyword,line width=.7pt]
        (9.55,.95) rectangle ++(.40,-.40);
        \draw[gc/token] (9.95,.75)--(11.04,.75);
        \node[gc/key] at (11.30,1.70) {Output tile};
        \filldraw[draw=tealgridrule,fill=tealgridtwo]
        (11.10,.95) rectangle ++(.40,-.40);
        \node[gc/math] at (11.30,.30) {$32\times32$};
        \node[gc/math] at (9.55,.30) {$2\times2$ tiles};
        \node[gc/math] at (13.90,1.70)
        {$p,q\in\{0,1\}$};
        \node[gc/math] at (13.90,1.32)
        {$0\le i,j<32$};
        \node[gc/math,gc/result] at (13.90,.83)
        {$0\le32p+i,\ 32q+j<64\ \checkmark$};
        \node[gc/key] at (13.90,.30) {Output stays within bounds};
    \end{tikzpicture}}
    \caption{Constructing and verifying the first chain in Figure~\ref{fig:language-gemm}. $L_A,L_B$ load the panels into \textsf{LDS}, $R_A,R_B$ read them into registers, and $M$ accumulates their product. The $\mathit{vmA}$ and $\mathit{vmB}$ wires connect each load to its register read; $\mathit{rmA}$ and $\mathit{rmB}$ connect the reads to separate \textsf{MFMA} input slots. Each vertex in the graph retains its operator, frozen tensor bindings, processor scope, and token wires. The representative bounds check shows that each $32\times32$ \textsf{MFMA} output tile lies within the $64\times64$ accumulator -- the compiler discharges this requirement automatically using Z3.}
    \label{fig:compile-graph-sketch}
  \end{figure*}
}

\newcommand{\compilationloweringfigure}{%
  \begin{figure*}[tbp]
    \centering
    \resizebox{\linewidth}{!}{%
      \begin{tikzpicture}[x=1cm,y=1cm,
          every node/.style={font=\fontsize{9.5}{11}\selectfont,
          text=tealgridink,inner sep=1pt,align=center},
          lc/arrow/.style={->,draw=tealcodekeyword,line width=1.1pt},
          lc/small/.style={font=\fontsize{8.8}{10.2}\selectfont},
          lc/meta/.style={font=\sffamily\fontsize{8.8}{10.2}\selectfont,
          text=tealgridink!70!white},
          lc/emphasis/.style={font=\sffamily\bfseries\fontsize{9}{10.5}\selectfont},
          lc/result/.style={font=\boldmath\fontsize{10}{11.5}\selectfont,
          text=tealcodekeyword},
        lc/title/.style={font=\bfseries\fontsize{10}{11.5}\selectfont}]
        \path[use as bounding box] (0,0) rectangle (16,5.4);
        \foreach \xx/\heading/\lcbackground in
        {0/{Bound Operands}/tealgridthree,4.1/{Placement Lattice}/tealgridone,
        8.2/{Output Subtile}/tealgridtwo,12.3/{Physical Realization}/tealgridthree} {
          \fill[fill=\lcbackground!25!white,rounded corners=2pt]
          (\xx,0) rectangle ++(3.7,5.4);
          \node[lc/title] at (\xx+1.85,5.175) {\heading};
        }
        \node[lc/meta] at (1.85,4.62) {One wave, one $K$-panel};
        \foreach \xx/\lcrows/\lccols/\lcselected/\lcshade/\lcname/\lcsize/\lcsubtile/\lclocalsize in
        {.37/2/4/1/tealgridthree/{\mathtt{RA}}/{64\times64}/{A_{1,1}}/{32\times16},
          1.57/2/4/1/tealgridone/{\mathtt{RB}}/{64\times64}/{B_{1,1}}/{32\times16},
        2.77/2/2/1/tealgridfive/{D_{\mathrm{acc}}}/{64\times64}/{D_{1,1}}/{32\times32}} {
          \node at (\xx+.28,4.18) {$\lcname$};
          \pgfmathtruncatemacro{\lcmaxrow}{\lcrows-1}
          \pgfmathtruncatemacro{\lcmaxcol}{\lccols-1}
          \foreach \rr in {0,...,\lcmaxrow} {
            \foreach \cc in {0,...,\lcmaxcol} {
              \def\lccolor{\lcshade}
              \ifnum\rr=\lcselected \ifnum\cc=1 \def\lccolor{tealgridtwo}\fi\fi
              \filldraw[draw=tealgridrule,fill=\lccolor]
              (\xx+.56/\lccols*\cc,3.02-.34*\rr) rectangle ++(.56/\lccols,-.34);
            }
          }
          \node[lc/meta] at (\xx+.28,1.63) {$\lcsize$};
          \node at (\xx+.28,.77) {$\lcsubtile$};
          \node[lc/meta] at (\xx+.28,.32) {$\lclocalsize$};
        }
        \node[text=tealgridink!70!white] at (1.85,3.72)
        {$(p,k)\times(q,k)\rightarrow(p,q)$};
        \node[lc/emphasis] at (1.85,1.15) {Selected subtiles};
        \node at (5.95,4.60) {At $K$-panel $t$};
        \node[lc/meta] at (5.95,4.22) {Each point is a placement};
        \begin{scope}[shift={(.15,.31)},
          x={(.99863cm,-.052336cm)},y={(0cm,1cm)}]
          \foreach \lp in {0,1} {
            \foreach \lq in {0,1}
            \draw[draw=tealgridrule]
            (4.68+.38*\lq,2.32+.42*\lq+.90*\lp)--++(1.86,0);
            \foreach \lk in {0,...,3}
            \draw[draw=tealgridrule]
            (4.68+.62*\lk,2.32+.90*\lp)--++(.38,.42);
          }
          \foreach \lk in {0,...,3} {
            \foreach \lq in {0,1} {
              \draw[draw=tealgridrule]
              (4.68+.62*\lk+.38*\lq,2.32+.42*\lq)--++(0,.90);
              \foreach \lp in {0,1}
              \fill[tealcodekeyword]
              (4.68+.62*\lk+.38*\lq,2.32+.42*\lq+.90*\lp) circle (1.6pt);
            }
          }
          \filldraw[fill=tealgridtwo,draw=tealcodekeyword,line width=.7pt]
          (5.68,3.64) circle (3.2pt);
          \draw[lc/arrow,line width=.65pt] (5.68,3.95)--(5.68,3.77);
          \draw[lc/arrow,line width=.65pt] (4.85,2.10)--(6.50,2.10) node[right] {$k$};
          \draw[lc/arrow,line width=.65pt] (6.77,2.32)--++(.32,.36) node[above] {$q$};
          \draw[lc/arrow,line width=.65pt] (4.40,2.42)--(4.40,3.35) node[above] {$p$};
        \end{scope}
        \node[text=tealcodekeyword] at (5.95,1.39) {$\pi=(t,1,1,1)$};
        \node[lc/meta] at (5.95,.86) {$t$: runtime loop};
        \node[lc/meta] at (5.95,.36)
        {$(p,q,k)$ specialized:\\$2\times2\times4=16$ instances};
        \node at (10.05,4.60) {$D_{1,1}$: $32\times32$};
        \foreach \rr in {0,...,3} {
          \foreach \cc in {0,...,3} {
            \def\lccolor{tealgridfive}
            \ifnum\rr=2 \ifnum\cc=0 \def\lccolor{tealgridtwo}\fi\fi
            \filldraw[draw=tealgridrule,fill=\lccolor]
            (9.25+.40*\cc,3.96-.40*\rr) rectangle ++(.40,-.40);
          }
        }
        \fill[tealcodekeyword] (9.525,3.085) circle (2pt);
        \draw[lc/arrow] (9.525,3.085)--(10.91,3.085);
        \node[lc/small,anchor=west] at (10.95,3.085) {$(17,5)$};
        \node[lc/small] at (10.05,2.08) {$\vec{u}=(17,5)$};
        \node[lc/small] at (10.05,1.49) {Accumulator coordinate};
        \draw[lc/arrow] (10.05,1.19)--(10.05,.89);
        \node[lc/result] at (10.05,.62) {$(49,37)$};
        \node[lc/meta] at (10.05,.23) {Accumulator coordinates};
        \node[lc/small] at (14.15,4.50)
        {$
          \begin{aligned}
            \operatorname{owner}(\vec{u})&=5,\\
            \operatorname{slot}(\vec{u})&=9.
        \end{aligned}$};
        \node at (14.15,3.85) {$b_D(\pi)=48$};
        \draw[lc/arrow] (14.15,3.60)--(14.15,3.40);
        \node[lc/small] at (14.15,3.16)
        {$\text{location }(5,48+9)$};
        \node[lc/meta] at (14.15,2.63) {Register slot};
        \node[lc/meta,rotate=90] at (12.57,1.77) {Thread};
        \foreach \rr in {0,1,2} {
          \pgfmathtruncatemacro{\lclane}{4+\rr}
          \node[lc/meta,anchor=east] at (13.00,2.00-.36*\rr) {$\lclane$};
          \foreach \cc in {0,1,2} {
            \def\lccolor{tealgridone}
            \ifnum\rr=1 \ifnum\cc=1 \def\lccolor{tealgridtwo}\fi\fi
            \filldraw[draw=tealgridrule,fill=\lccolor]
            (13.10+.70*\cc,2.18-.36*\rr) rectangle ++(.70,-.36);
            \ifnum\rr=1 \ifnum\cc=1
            \node at (13.45+.70*\cc,2.00-.36*\rr) {$\bullet$};
            \fi\fi
          }
        }
        \foreach \cc in {0,1,2} {
          \pgfmathtruncatemacro{\lcslot}{56+\cc}
          \node[lc/meta] at (13.45+.70*\cc,2.38) {$\lcslot$};
        }
        \node[lc/result] at (14.15,.62) {$\text{location }(5,57)$};
        \foreach \xx in {3.7,7.8,11.9}
        \draw[lc/arrow,line width=1.8pt] (\xx+.04,2.65)--(\xx+.36,2.65);
    \end{tikzpicture}}
    \caption{Compositional lowering of one GEMM \textsf{MFMA} application. Indexed operand subtiles select a point in the placement lattice; the local axes are specialized while the $K$-panel axis remains runtime. The selected output subtile progressively maps $\vec{u}=(17,5)$ to accumulator coordinate $(49,37)$ and thread $5$, register $57$.}
    \label{fig:compile-placement-sketch}
  \end{figure*}
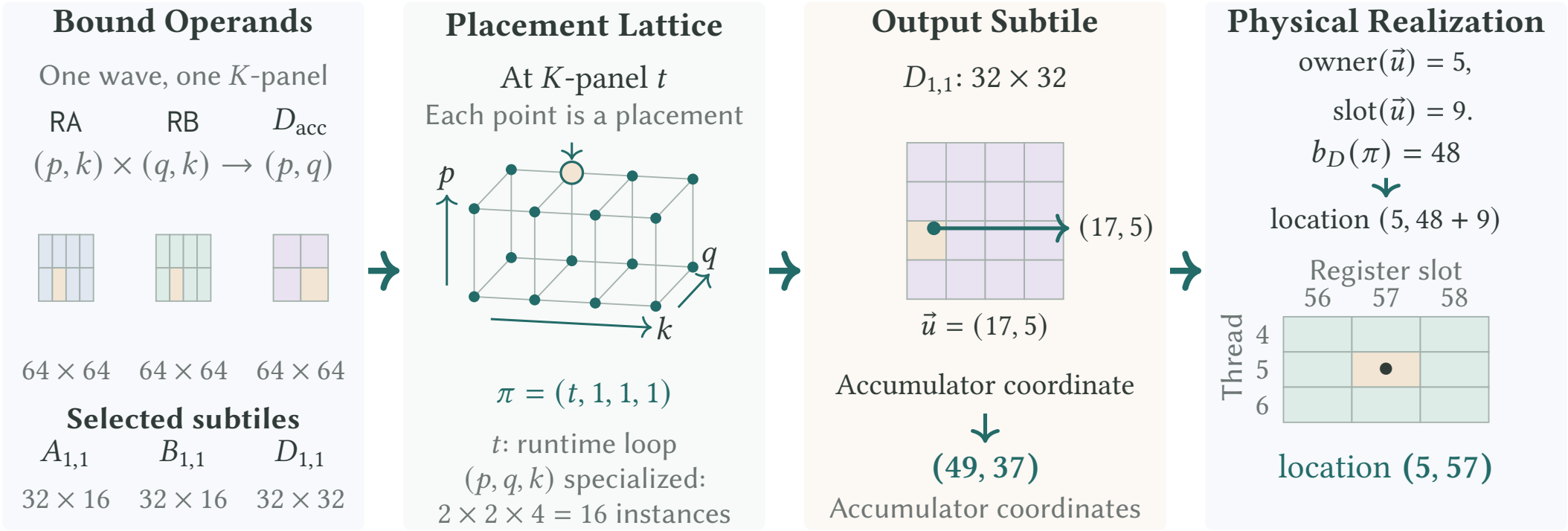
}

The \name~compiler realizes programs through a placement-indexed representation of computation and synchronization. \name's principal compilation challenge arises from the modularity of instruction-centric lifting: each \operator~describes a local computation over symbolic tensor geometry, while executable kernels require compatible physical representations, coordinated loops, and synchronization across operations. \rev{The compiler resolves and statically verifies each operation, fuses loops and compresses token synchronization across composed operations, and then \codexrev{generates} target code}, lowering them first into a verified \emph{hierarchical execution graph} and then into a \emph{generic low-level imperative representation} (GLIR).

\subsection{Constructing and Verifying the Scheduled Graph}
\label{sec:compilation-dag}

As statements introduce tensors and apply \operator s, the compiler resolves each application to a concrete instance of an \operator~$\mathcal{O}_\theta$ (instantiating specialization parameters $\theta$ against operand shapes) and constructs a canonical output tensor, operand bindings, placement domain, and access maps. In Figure~\ref{fig:language-gemm}, \langfix{the} \textsf{Contract} application resolves to a single symbolic \textsf{MFMA} vertex with a \langfix{$2\times2\times4$ local placement domain rather than sixteen separate} instructions.

Each accepted application forms a vertex $v=(\mathcal{O}_v,\beta_v,\delta_v,S_v)$ recording its resolved \operator~$\mathcal{O}_v$, immutable binding snapshot $\beta_v$, processor path $\delta_v$, and token interface $S_v$ (illustrated for \langfix{the \textsf{MFMA} node $M$} in Figure~\ref{fig:compile-graph-sketch}). The snapshot $\beta_v$ freezes the input and output views, layouts, and allocation identities at statement acceptance so later rebindings of a tensor name do not alter earlier accesses.
Vertices assemble into a hierarchical execution graph $G=(V,E_{\mathrm{fwd}},E_{\mathrm{bwd}},E_{\mathrm{dist}})$, where $V$ comprises \operator~vertices, chain vertices (each owning an internal subgraph $G_v$ to preserve scope boundaries for loop fusion and buffer planning), and local token source/sink vertices. Processor scopes are retained as nesting paths $\delta_v$ linked via $E_{\mathrm{dist}}$ rather than premature control-flow guards.
\compilationgraphfigure
Directed edges connect token slots across vertices: $E_{\mathrm{fwd}}$ encodes forward dependencies, while $E_{\mathrm{bwd}}$ encodes placement-carried dependencies within chains. Vertices are scheduled via a stable topological sort of $E_{\mathrm{fwd}}$ and explicit synchronization precedences.

\subsubsection{Symbolic Obligations}

Constructing a vertex requires proving that all symbolic instruction instances stay in bounds: for each lifted application, \name~uses Z3 to prove this symbolically for every problem size satisfying the kernel assumptions. Figure~\ref{fig:compile-graph-sketch} illustrates this bounds check for the $32\times32$ \textsf{MFMA} output tiles within the $64\times64$ accumulator \smalltt{D\_acc}. For operands in distributed state spaces, the compiler also verifies compatibility with the native instruction layout. It derives a base offset $b(\pi)$ from one valid operand coordinate at each placement, then verifies that every element maps to the required owner and local storage slot. For storage partitioned across larger processor groups, it additionally checks that accesses remain within their assigned partitions. This guarantees that all operands exist and reside in the exact locations expected by the instruction, rejecting mismatched distributions that require explicit reshuffling.

\subsubsection{Token-State Verification}

The compiler then verifies the execution graph against the dialect's token automata. A kernel's token resources form a product state $\sigma$, with each connected action inducing a transition $\sigma\rightarrow\sigma'$. Because lifted \operator s execute across placement grids, token states can depend on placement ordinals -- for example, a completion counter accumulating outstanding loads across multiple placements before an \textsf{MFMA} consumer drains them. \name~encodes these placement-indexed transitions and loop-carried recurrences symbolically and \codexrev{verifies via Z3 the symbolic state summaries and required states at action uses}.

\subsection{Compositional Placement Lowering}
\label{sec:compilation-lowering}

From the \codexrev{resolved} graph $G$, lowering constructs concrete instruction operands while coordinating loop nests, storage reuse, and synchronization across composed operators. For each admitted placement domain, the compiler \codexrev{realizes each axis as a runtime loop or static specialization in the order retained by the operator's placement plan}. In the example GEMM, specializing the three local axes emits \langfix{$16$ \textsf{MFMA} calls per panel} with constant operand offsets, while the shared $K$-panel axis remains a runtime loop.
For each instruction placement, lowering combines the operator's access maps with the frozen tensor views and layouts to derive the required operand locations, including distributed storage offsets. Parameterized instructions, such as strided loads, receive the inferred address and stride expressions, whereas fixed-layout instructions, such as \textsf{MFMA}, require only a verified per-thread base offset $b(\pi)$ (Figure~\ref{fig:compile-placement-sketch}).

\compilationloweringfigure

\subsubsection{Spatial Composition and Value Preservation}

\codexrev{Because a kernel composes independently lifted operators over aliased views, ring-buffered scratchpads, and multi-step accumulators, the compiler must ensure that physical storage reuse and loop interleaving preserve the logical values each instruction expects. \name~verifies \emph{value preservation} compositionally across both standalone applications and chains. Locally, it resolves each read's intended definition through source bindings and logical-view correspondence. Using the access maps, bound views, and execution order -- while accounting for collective tile coverage across cooperating processors -- local proofs verify contribution and state-update order (ensuring, for instance, that all reduction steps of a \textsf{Contract} or \textsf{Reduce} complete before a consumer reads the result), storage lifetimes (preventing intervening writes from clobbering live values), and final output observations; asynchronous transfer completion is checked separately via tokens.}

\codexrev{Each verified application or chain produces a region certificate recording its required incoming values and certified exports. Across the whole kernel graph $G$, the compiler composes these certificates by verifying that each consumer region satisfies its producer's proof assumptions and reads the certified value from matching physical storage, and that exported values survive producer exit and any intervening writes until consumed.}

\begin{figure*}[!t]
  \centering
  \resizebox{\linewidth}{!}{%
    \begin{tikzpicture}[x=1cm,y=1cm,
        every node/.style={font=\fontsize{10}{11.5}\selectfont,
        text=tealgridink,inner sep=2pt,align=center},
        vf/title/.style={font=\bfseries\fontsize{10.5}{12}\selectfont},
        vf/meta/.style={font=\fontsize{9}{10.5}\selectfont},
        vf/node/.style={rounded corners=1.5pt,minimum width=.70cm,
        minimum height=.45cm,inner sep=3pt},
        vf/value/.style={-{Latex[length=1.7mm,width=1.2mm]},
        draw=tealcodekeyword,line width=.8pt},
        vf/order/.style={-{Latex[length=1.6mm,width=1.1mm]},
        draw=tealgridink!60!white,line width=.7pt},
      vf/result/.style={text=tealcodekeyword}]
      \path[use as bounding box] (0,0) rectangle (20,5);
      \foreach \vfxx/\vftitle/\vfshade in
      {0/{Operator Graph of Chain}/tealgridthree,
        5.1/{\textsf{MFMA} Inputs and Output}/tealgridtwo,
        10.2/{$K$-Panel Fusion}/tealgridone,
      15.3/{32-Row Fusion}/tealgridfour} {
        \fill[\vfshade!25!white,rounded corners=2pt]
        (\vfxx,.45) rectangle ++(4.7,4.55);
        \node[vf/title,anchor=west] at (\vfxx+.15,4.73) {\vftitle};
      }
      \begin{scope}
        \node[vf/node,fill=tealgridone] (vla) at (.62,3.65) {$L_A$};
        \node[vf/node,fill=tealgridone] (vlb) at (.62,2.75) {$L_B$};
        \node[vf/node,fill=tealgridthree] (vra) at (2.18,3.65) {$R_A$};
        \node[vf/node,fill=tealgridthree] (vrb) at (2.18,2.75) {$R_B$};
        \node[vf/node,fill=tealgridtwo] (vm) at (3.98,3.20) {$M$};
        \draw[vf/value] (vla.east)--(vra.west);
        \draw[vf/value] (vlb.east)--(vrb.west);
        \draw[vf/value] (vra.east)--(vm.north west);
        \draw[vf/value] (vrb.east)--(vm.south west);
        \node[vf/meta] at (2.35,2.10) {$L_A,L_B$: global $\rightarrow$ \textsf{LDS}};
        \node[vf/meta] at (2.35,1.68) {$R_A,R_B$: \textsf{LDS} $\rightarrow$ \textsf{VGPR}};
        \node[vf/meta] at (2.35,1.26) {$M$: $2\times2\times4$ \textsf{MFMA} calls};
        \node[vf/meta] at (2.35,.72) {$K$-panel: 64 reduction elements};
      \end{scope}
      \begin{scope}[shift={(5.1,0)}]
        \foreach \vfbase/\vfshade/\vfname/\vfselrow in
        {.45/tealgridthree/RA/0,2.95/tealgridone/RB/1} {
          \node[vf/meta] at (\vfbase+.65,4.15) {$\mathtt{\vfname}:64\times64$};
          \foreach \vfr in {0,1} {
            \foreach \vfc in {0,...,3} {
              \def\vfcell{\vfshade}
              \ifnum\vfr=\vfselrow\def\vfcell{tealgridtwo}\fi
              \filldraw[draw=tealgridrule,fill=\vfcell,line width=.4pt]
              (\vfbase+.325*\vfc,3.85-.65*\vfr) rectangle ++(.325,-.65);
            }
          }
          \draw[draw=tealcodekeyword,line width=.8pt]
          (\vfbase,3.85-.65*\vfselrow) rectangle ++(1.3,-.65);
        }
        \node at (2.35,3.20) {$\times$};
        \node[vf/meta,anchor=north] (vagroup) at (1.10,2.40) {$A_{0,0:3}$};
        \node[vf/meta,anchor=north] (vbgroup) at (3.60,2.40) {$B_{1,0:3}$};
        \node[vf/node,fill=tealgridtwo,inner sep=1.5pt] (vcall) at (2.35,1.63)
        {$4\times$\textsf{MFMA}};
        \draw[vf/value] (vagroup.south)--(vcall.north west);
        \draw[vf/value] (vbgroup.south)--(vcall.north east);
        \filldraw[draw=tealcodekeyword,fill=tealgridtwo,line width=.7pt]
        (2.025,.50) rectangle ++(.65,.65);
        \node[inner sep=.5pt] at (2.35,.825) {$D_{0,1}$};
        \draw[vf/value] (vcall.south)--(2.35,1.21);
        \node[vf/meta] at (1.00,.825) {$32\times16$\\[-1pt]inputs};
        \node[vf/meta] at (3.45,.825) {$32\times32$\\[-1pt]output};
      \end{scope}
      \foreach \vfcase/\vfxx in {0/10.2,1/15.3} {
        \begin{scope}[shift={(\vfxx,0)}]
          \filldraw[fill=white,draw=tealgridrule,rounded corners=2pt,line width=.5pt]
          (.20,.92) rectangle (4.50,4.45);
          \node[vf/meta] at (2.35,4.20) {\textsf{for} $t=0,\ldots,K/64-1$};
          \ifnum\vfcase=0
          \node[vf/result] at (4.35,4.73) {$\checkmark$};
          \def\vfrows{64}
          \def\vfcalls{16}
          \else
          \node[text=tealgridfour!45!black] at (4.35,4.73) {$\times$};
          \filldraw[fill=tealgridfour!12!white,draw=tealgridfour!45!black,
          rounded corners=2pt,line width=.5pt]
          (.30,1.00) rectangle (4.40,3.95);
          \node[vf/meta] at (2.35,3.70) {\textsf{for} $p=0,1$ (32 rows)};
          \def\vfrows{32}
          \def\vfcalls{8}
          \fi
          \foreach \vfx/\vfop/\vfshade in
          {.68/L_A/tealgridone,1.515/L_B/tealgridone,
          2.35/R_A/tealgridthree,3.185/R_B/tealgridthree,4.02/M/tealgridtwo} {
            \ifnum\vfcase=0
            \node[vf/node,fill=\vfshade,minimum width=.60cm,inner sep=1.5pt,
            font=\fontsize{9.3}{10.5}\selectfont]
            (vfop\vfcase\vfop) at (\vfx,3.20) {$\vfop$};
            \else
            \node[vf/node,fill=\vfshade,minimum width=.60cm,inner sep=1.5pt,
            font=\fontsize{9.3}{10.5}\selectfont]
            (vfop\vfcase\vfop) at (\vfx,3.20) {$\vfop^{p}$};
            \fi
          }
          \draw[vf/order] (vfop\vfcase L_A.east)--(vfop\vfcase L_B.west);
          \draw[vf/order] (vfop\vfcase L_B.east)--(vfop\vfcase R_A.west);
          \draw[vf/order] (vfop\vfcase R_A.east)--(vfop\vfcase R_B.west);
          \draw[vf/order] (vfop\vfcase R_B.east)--(vfop\vfcase M.west);
          \node[font=\fontsize{8.3}{9.5}\selectfont,inner sep=1pt]
          at (2.35,2.72)
          {$\vfrows\ \text{rows} \rightarrow \vfrows\ \text{rows} \rightarrow \vfcalls\ \text{calls}$};
          \foreach \vfbase/\vfshade/\vfname/\vfcols in
          {.45/tealgridthree/RA/4,1.85/tealgridone/RB/4,3.25/tealgridfive/D/2} {
            \node[vf/meta] at (\vfbase+.5,2.33) {$\mathtt{\vfname}$};
            \pgfmathtruncatemacro{\vfmaxcol}{\vfcols-1}
            \foreach \vfr in {0,1} {
              \foreach \vfc in {0,...,\vfmaxcol} {
                \def\vfcell{\vfshade}
                \ifnum\vfcase=1\ifnum\vfr=1\def\vfcell{tealgridexcluded}\fi\fi
                \filldraw[draw=tealgridrule,fill=\vfcell,line width=.4pt]
                (\vfbase+\vfc/\vfcols,2.05-.50*\vfr)
                rectangle ++(1/\vfcols,-.50);
              }
            }
          }
          \foreach \vfk in {0,...,3} {
            \filldraw[draw=tealgridrule,fill=tealgridtwo,line width=.4pt]
            (.45+.25*\vfk,2.05) rectangle ++(.25,-.50);
            \ifnum\vfcase=0
            \filldraw[draw=tealgridrule,fill=tealgridtwo,line width=.4pt]
            (1.85+.25*\vfk,1.55) rectangle ++(.25,-.50);
            \else
            \filldraw[draw=tealgridrule,fill=tealgridfour,line width=.4pt]
            (1.85+.25*\vfk,1.55) rectangle ++(.25,-.50);
            \fi
          }
          \draw[draw=tealcodekeyword,line width=.7pt]
          (.45,2.05) rectangle ++(1,-.50);
          \ifnum\vfcase=0
          \draw[draw=tealcodekeyword,line width=.7pt]
          (1.85,1.55) rectangle ++(1,-.50);
          \filldraw[draw=tealcodekeyword,fill=tealgridtwo,line width=.7pt]
          (3.75,2.05) rectangle ++(.50,-.50);
          \node[vf/meta,vf/result] at (2.35,.68) {All required inputs read.};
          \else
          \draw[draw=tealgridfour!45!black,dashed,line width=.8pt]
          (1.85,1.55) rectangle ++(1,-.50);
          \filldraw[draw=tealgridfour!45!black,fill=tealgridfour,
          dashed,line width=.8pt]
          (3.75,2.05) rectangle ++(.50,-.50);
          \node[text=tealgridfour!45!black] at (4.00,1.80) {$\times$};
          \node[vf/meta,text=tealgridfour!45!black] at (2.35,.68)
          {$p=0$: $B_{1,0:3}$ unread.};
          \fi
        \end{scope}
      }
      \draw[vf/value] (17.65,.24)--(12.55,.24);
      \node[vf/meta,fill=white,inner xsep=3pt] at (15.10,.24)
      {retain $K$-panel fusion};
  \end{tikzpicture}}
  \vspace{-2em}
  \caption{Value preservation selects the fusion depth of the first chain. Both candidate loop nests contain the five operators in order. Sharing only the $K$-panel loop completes both operand reads; also sharing a 32-row loop reaches the \textsf{MFMA} calls for $D_{0,1}$ while $B_{1,0:3}$ is unread. \name~thus retains only the $K$-panel fusion.}
  \label{fig:value-preservation-fusion-draft}
\end{figure*}
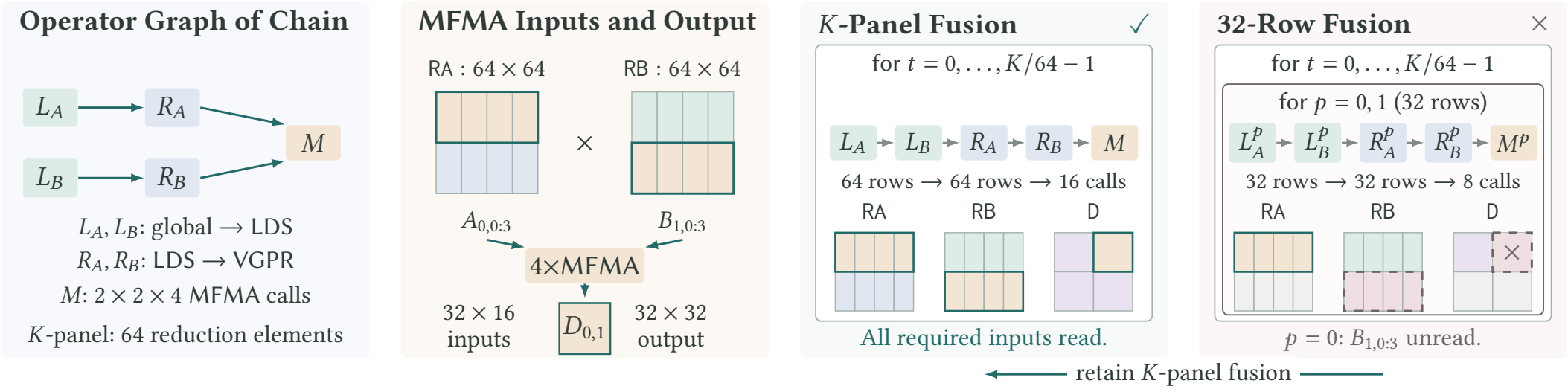

\codexrev{Within a chain, the planner uses this value-preservation check to construct a shared outer placement prefix, greedily extending it as much as possible while compatibility and value preservation hold, retaining the remaining dimensions as operator-local suffixes (which nested chains can fuse further). In \langfix{Figure~\ref{fig:compile-placement-sketch}}, the shared $K$-panel loop encloses the specialized $2\times2\times4$ local \textsf{MFMA} suffix. We illustrate the fusion itself in Figure~\ref{fig:value-preservation-fusion-draft}: $K$-panel fusion reads all 64 rows of both operands before computation, and fusing the $32$-row loop instead reaches $D_{0,1}$ before reading rows 32--63 of $B$, which supply its output columns; \name~correctly rejects this deeper fusion. When the shared placement covers an intermediate tensor's outer dimensions and the local suffixes cover each tile, the compiler materializes that intermediate in ephemeral, tile-sized register or scratchpad buffers. Once concrete scratch allocations are assigned, it re-verifies value preservation on the realized bit footprints and emits code directly from those certified bindings.}

\subsubsection{Temporal Composition}

Chain composition also aggregates token transitions across placements. Using the verified token automata (\S\ref{sec:compilation-dag}), the compiler summarizes repeated transitions across all placements (e.g., compressing $N$ increments $c\mapsto c+1$ into $c\mapsto c+N$) so an operator-scoped wait body executes once while accounting for all $N$ covered instruction requests, whereas instruction-scoped actions execute at each instruction site. At chain boundaries, token states pass directly between steps without redundant intermediate waits; final completion waits are emitted only when a wire is consumed or exits the kernel. Finally, consecutive vertices with identical processor paths $\delta_v$ are grouped into shared GLIR execution regions for target realization, yielding the physical instruction instances, shared placement traversals, and synchronization actions ready for imperative simplification.

\subsubsection{\codexrev{Verified-Schedule Conformance}} \codexrev{Together, these checks establish end-to-end value consistency for the resolved kernel graph $G$ and its concrete storage bindings across every admissible parameter set $\rho$ satisfying the kernel's assumptions. Let $\operatorname{Obs}_{\mathrm{ref}}(G,\rho,\mu_0)$ denote the reference values defined by $G$'s logical definitions and declared state-update trajectories from valid initial contents of the device memory $\mu_0$. Under faithful instruction access and effect specifications, every completed execution $\zeta$ that respects the verified ordering and token completion semantics satisfies
  \[
    \operatorname{Obs}_{\mathrm{reads,final}}(\zeta)
    =\operatorname{Obs}_{\mathrm{ref}}(G,\rho,\mu_0).
  \]
That is, every modeled instruction read and final kernel output observes its exact reference value -- guaranteeing that loop fusion, data striping, and aggressive buffer reuse preserve the schedule's logical semantics, a guarantee inherited by the generated low-level target code under lowering correspondence (derived formally in the supplementary material).}

\subsection{GLIR Optimization and Backend Integration}
\label{sec:compilation-glir}

Compositional lowering produces an imperative realization of the schedule mixed with redundant address arithmetic, temporary bindings, and control-flow scaffolding. \name~models this in GLIR, a compact imperative IR comprising constants (\textsf{const\_set}), loops (\textsf{place\_loop}), pointer bindings (\textsf{assign\_addr}), instruction calls (\textsf{inst\_call}), storage actions (\textsf{space\_action}), token actions (\textsf{token\_action}), and processor guards (\textsf{at\_entry\_point}). Because composing access maps with accelerator memory layouts often yields constant arithmetic once placement coordinates are specialized, GLIR optimization applies constant folding, dead pointer/allocation elimination, common address subexpression elimination, loop-invariant address hoisting, and full unrolling of eligible loops of up to $16$ iterations, exposing constant operands (such as the \textsf{AGPR} base address $b_D(\pi)=48$ in Figure~\ref{fig:compile-placement-sketch}) directly to the backend compiler for high-performance code generation.

\subsection{Implementation}

We implement the \name~compiler in C++. Because GLIR provides a single imperative interface across architectures alongside customizable kernel entry points and launch conventions, GLIR kernels can be embedded natively in different architectural execution models. The entire lowering, Z3 verification, materialization, chain fusion, token scheduling, and optimization pipeline is shared across targets, while machine dialects extend this core with declarative hardware specifications and lightweight C++ bindings with instruction descriptions for five hardware targets: NVIDIA GPUs (CUDA C++), AMD GPUs (HIP C++), Intel \textsf{AVX}/\textsf{AMX} and Arm \textsf{SVE}/\textsf{SME} (C++ with intrinsics), and Tenstorrent accelerators (TT-Metalium C++ APIs).
We additionally complement the compiler with a verbose mode that emits a full trace of compilation, including tensor instantiations and solver queries required for a schedule, and an ISA lookup table that enables searching for valid operators for any computation on tensors of compatible shapes, to enable iterative coding.

\section{Evaluation}

We evaluate \name~to answer four research questions:
\begin{itemize}
  \item \textbf{Q1 (Kernel Performance):} Can target-specific schedules in \name~match and/or outperform hand-tuned vendor libraries and SOTA tile compilers on standard workloads?
  \item \textbf{Q2 (Compositional Expressiveness):} Do \name's structured composition mechanisms enable high-performance schedules for complex, multi-stage pipelines?
  \item \textbf{Q3 (Hardware Extensibility \& Reuse):} Can a shared compiler support architectures with substantially different execution models through declarative machine dialects and lightweight native bindings?
  \item \textbf{Q4 (Program Verification \& Agentic Synthesis):} How can \name's language and compiler support agent-assisted dialect and kernel synthesis on new architectures?
\end{itemize}

\paragraph{Hardware and Measurement Protocol}
For AMD CDNA4, we use a single MI350X 288\,GB with an AMD EPYC 9755 host CPU, using ROCm~7.2. For NVIDIA Blackwell, we use a single B200 180\,GB with an Intel Xeon 6 host CPU, using CUDA~13.3. For Intel, we use 32 physical cores of an Intel Xeon 6975P-C 1536\,GB, using oneAPI~2026.2.0. For Arm, we use a 12-core Apple M6 16\,GB CPU, using Apple Clang 21.0.0. For Tenstorrent, we use a single Tenstorrent Blackhole P150 32\,GB with an AMD EPYC 9554 host CPU, using TT-KMD 2.4.1 and TT-Metalium v0.63.0. For each benchmark, we run 20 warmup iterations followed by 20 rounds of 20 timed iterations each. We rotate enough input/output buffers until the size of the working set exceeds $3\times$ the LLC of the device. We report the mean and sample standard deviation of the per-launch average times from 20 timed rounds. All kernels are numerically validated against a vendor-optimized baseline. Where appropriate, we also generate and sweep multiple kernels to tune parameters such as block sizes.

\begin{figure}[t]
  \centering
  \begin{subfigure}[t]{.24\linewidth}
    \centering
    \includegraphics[width=\linewidth]{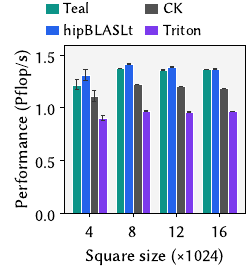}
    \caption{MI350X bf16 GEMM.}
    \label{fig:amd-bf16-gemm}
  \end{subfigure}\hfill
  \begin{subfigure}[t]{.24\linewidth}
    \centering
    \includegraphics[width=\linewidth]{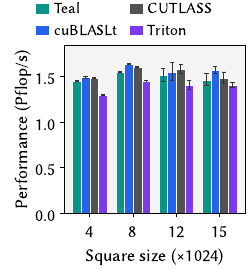}
    \caption{B200 bf16 GEMM.}
    \label{fig:nvidia-bf16-gemm}
  \end{subfigure}\hfill
  \begin{subfigure}[t]{.24\linewidth}
    \centering
    \includegraphics[width=\linewidth]{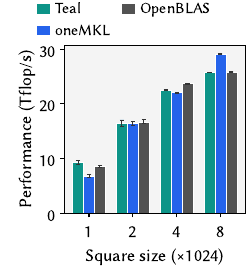}
    \caption{Xeon 6 bf16 GEMM.}
    \label{fig:intel-bf16-gemm}
  \end{subfigure}\hfill
  \begin{subfigure}[t]{.24\linewidth}
    \centering
    \includegraphics[width=\linewidth]{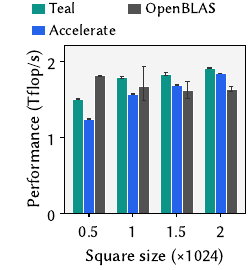}
    \caption{Apple M6 fp32 GEMM.}
    \label{fig:arm-fp32-gemm}
  \end{subfigure}
  \ifblind
  \caption{Dense GEMM throughput.}
  \else
  \caption{Performance: Dense square GEMMs in bf16, accumulating in fp32 (Apple uses fp32 throughout).}
  \fi
  \label{fig:compute-bound-gemm}
  \label{fig:kernel-performance}
\end{figure}

\paragraph{Dense GEMM}
To test whether instruction-centric lifting and GLIR lowering generate code competitive with heavily tuned vendor libraries on compute-bound workloads, we evaluate dense bf16 and fp32 GEMM schedules across GPU matrix cores and CPU tile units. On AMD MI350X, our GEMM kernel achieves $0.93$--$1.00\times$ the throughput of hipBLASLt~\cite{hipBLASL55} while consistently outperforming Composable Kernel and Triton (Figure~\ref{fig:amd-bf16-gemm}). On NVIDIA B200, \name~achieves $0.93$--$0.98\times$ cuBLASLt~\cite{cublaslt} throughput, matching CUTLASS and outperforming Triton (Figure~\ref{fig:nvidia-bf16-gemm}). On 32 Intel Xeon cores using \textsf{AMX} tile instructions, \name~achieves $0.89$--$1.36\times$ oneMKL~\cite{Develope25} and $0.95$--$1.08\times$ OpenBLAS throughput (Figure~\ref{fig:intel-bf16-gemm}). On one Apple M6 core, our SME-based fp32 GEMM achieves $1.04$--$1.21\times$ Accelerate and $0.83$--$1.17\times$ OpenBLAS throughput (Figure~\ref{fig:arm-fp32-gemm}). Answering \textbf{Q1} and \textbf{Q3}, \name's shared compilation pipeline lowers target-specific schedules across distinct GPU and CPU matrix ISAs to match (or exceed) hand-tuned vendor assembly libraries.

\begin{figure}[t]
  \centering
  \begin{subfigure}[t]{.24\linewidth}
    \centering
    \includegraphics[width=\linewidth]{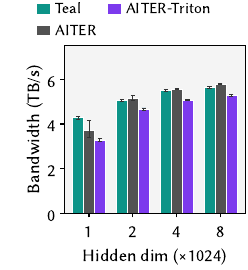}
    \caption{AMD MI350X.}
    \label{fig:amd-residual-rmsnorm}
  \end{subfigure}\hfill
  \begin{subfigure}[t]{.24\linewidth}
    \centering
    \includegraphics[width=\linewidth]{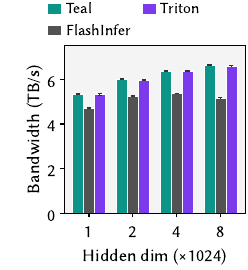}
    \caption{NVIDIA B200.}
    \label{fig:nvidia-residual-rmsnorm}
  \end{subfigure}\hfill
  \begin{subfigure}[t]{.24\linewidth}
    \centering
    \includegraphics[width=\linewidth]{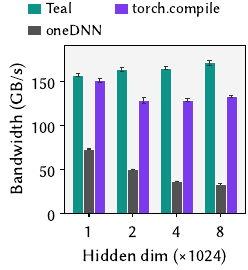}
    \caption{Intel Xeon 6 (32 cores).}
    \label{fig:intel-residual-rmsnorm}
  \end{subfigure}\hfill
  \begin{subfigure}[t]{.24\linewidth}
    \centering
    \includegraphics[width=\linewidth]{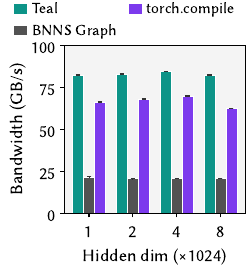}
    \caption{Apple M6 (4 cores).}
    \label{fig:arm-residual-rmsnorm}
  \end{subfigure}
  \caption{Logical bandwidth: fused residual + RMSNorm in bf16, accumulating in fp32.}
  \label{fig:residual-rmsnorm}
\end{figure}

\paragraph{Fused Residual + RMSNorm}
To evaluate how \name~composes elementwise operations (\textsf{Linear}, \textsf{Unary}) with row reductions (\textsf{Reduce}) on memory-bound workloads, we benchmark fused residual addition and RMS normalization on bf16 inputs with 8192 rows and hidden dimensions $1024$--$8192$. By chaining the residual add, sum-of-squares reduction, and normalization within a shared row placement prefix, \name~retains intermediate rows in registers rather than round-tripping to off-chip memory. On AMD MI350X, \name~sustains $0.97$--$1.14\times$ the bandwidth of hand-tuned AITER~\cite{ROCmaite73} kernels (Figure~\ref{fig:amd-residual-rmsnorm}). On NVIDIA B200, \name~outperforms FlashInfer~\cite{ye2025flashinfer} by $1.13$--$1.28\times$ (Figure~\ref{fig:nvidia-residual-rmsnorm}). On 32 Intel cores, \name~achieves $1.04$--$1.29\times$ \textsf{torch.compile} and $2.14$--$5.10\times$ oneDNN's~\cite{oneDNN2025v315} unfused baseline (Figure~\ref{fig:intel-residual-rmsnorm}). On four Apple M6 cores, \name~achieves $1.21$--$1.32\times$ \textsf{torch.compile} and $3.78$--$4.05\times$ Apple Accelerate’s BNNS Graph throughput (Figure~\ref{fig:arm-residual-rmsnorm}). Both oneDNN and BNNS Graph do not fuse the residual and RMSNorm kernels, creating an extra roundtrip to memory. Answering \textbf{Q1} and \textbf{Q2}, \name's chain fusion eliminates intermediate memory traffic across heterogeneous targets.

\begin{wrapfigure}[24]{r}{.25\linewidth}
  \centering
  \includegraphics[width=\linewidth]{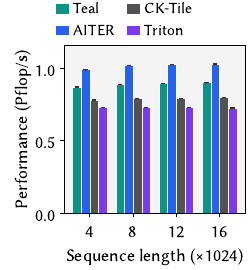}
  \caption{MI350X noncausal bf16 GQA.}
  \label{fig:amd-gqa}
  \includegraphics[width=\linewidth]{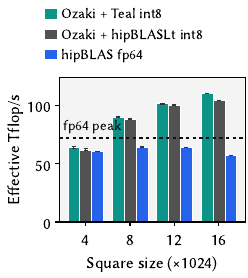}
  \caption{MI350X Ozaki-II fp64.}
  \label{fig:ozaki-gemm}
  \ifblind
  \includegraphics[width=\linewidth]{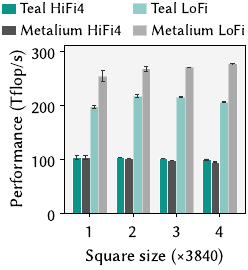}
  \caption{Blackhole bf16 GEMM.}
  \label{fig:tt-gemm}
  \fi
\end{wrapfigure}

\paragraph{Case Study: GQA Non-Causal Forward}
To stress-test multi-stage producer-consumer composition within a single fused kernel, we implement Grouped-Query Attention (GQA) forward~\cite{shah2024flashattention3fastaccurateattention}, which chains two matrix contractions ($QK^T$ and $PV$) with online softmax rescaling (\textsf{Unary}, \textsf{Linear}, \textsf{Reduce}) and asynchronous KV-tile prefetching. We evaluate noncausal bf16 GQA using batch size 16, 64 query heads, 8 key/value heads, and head dimension 128. On MI350X, \name~achieves $1.19$--$1.24\times$ Triton, $1.12$--$1.13\times$ CK-Tile~\cite{CKTileCo34:online}, and $0.87$--$0.88\times$ AITER's raw assembly kernels (Figure~\ref{fig:amd-gqa}). \name’s explicitly interleaved, double-buffered schedule hides KV-load and softmax latency behind \textsf{MFMA} execution, improving utilization over the tested Triton and CK kernels. Answering \textbf{Q2}, \name~expresses complex, multi-contraction schedules with online reductions and asynchronous pipelining directly in the client language.

\paragraph{Case Study: Ozaki Emulation fp64 GEMM}
\setlength{\intextsep}{2pt}
To evaluate whether \name~enables end-to-end optimization of emerging multi-stage numerical pipelines that go beyond monolithic library calls, we implement Ozaki Scheme II~\cite{ozaki2025schemeii} on MI350X. Ozaki II emulates fp64 matrix multiplication using high-throughput int8 matrix cores via modular arithmetic, using moduli selected via the search procedure of~\cite{mazumder2026fp64}. Our \name~implementation expresses the complete pipeline: fp64-to-int8 residue encoding, batched int8 \textsf{MFMA} contractions, modular reduction epilogues, and Chinese Remainder Theorem (CRT) reconstruction. We also evaluate a hybrid baseline that pairs our \name~encoding and reconstruction kernels with hipBLASLt's int8 GEMMs. As shown in Figure~\ref{fig:ozaki-gemm}, the full \name~pipeline outperforms the hipBLASLt-backed variant across all sizes and achieves up to $1.94\times$ the throughput of native hipBLAS fp64 GEMM, surpassing the hardware's native fp64 peak from $8\text{k}$ onward. Answering \textbf{Q1} and \textbf{Q2}, \name's compositional lifting lets programmers fuse custom arithmetic epilogues directly into matrix-core accumulations, outperforming closed vendor library calls.

\paragraph{Case Study: Targeting a New Accelerator in \name}
Finally, we evaluate hardware extensibility and agent-assisted synthesis by targeting a spatial dataflow accelerator -- Tenstorrent Blackhole -- without modifying \name's core compiler. Using GPT-6.1 Sol with Codex, we authored the Blackhole machine dialect and C++ bindings from TT-Metalium's native NoC transfer, circular-buffer, semaphore synchronization, and matrix-FPU primitives. Compared against TT-Metalium's~\cite{tenstorr95} hand-optimized multicast GEMM under matched math-fidelity settings (which trade numerical precision of the matrix engine for throughput), \name~reaches parity on HiFi4 and achieves $0.74$--$0.81\times$ Metalium throughput in LoFi (Figure~\ref{fig:tt-gemm}). More aggressive LoFi schedules that decouple A and B NoC transport currently exceed Z3's per-query timeout when discharging cross-core value-preservation obligations, highlighting improved solver efficiency and efficient SMT query framing as concrete directions for future work. Answering \textbf{Q3} and \textbf{Q4}, \name's declarative dialect and statically verified schedule interfaces enable agent-assisted coding to rapidly bring up a non-GPU spatial architecture and match hand-optimized vendor kernels while reusing the entire compiler pipeline.

\ifblind
\else
\setlength{\intextsep}{2pt}
\begin{wrapfigure}[12]{r}{.25\linewidth}
  \centering
  \includegraphics[width=\linewidth]{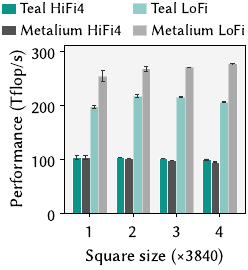}
  \caption{Blackhole bf16 GEMM.}
  \label{fig:tt-gemm}
\end{wrapfigure}
\fi

\section{Related Work}

\paragraph{Tile Languages and Layout Algebras.}
Tile languages (Triton~\cite{triton}, Pallas/Mosaic~\cite{PallasaJ14:online,mosaic_gpu}, Gluon~\cite{gluon}, Tilus~\cite{ding2025tilustilelevelgpgpuprogramming}, TileLang~\cite{wang2025tilelangcomposabletiledprogramming}, FlyDSL~\cite{FlyDSLEx5}, cuTile~\cite{nvidia_cutile_2026}) and kernel libraries (CuTe/CUTLASS~\cite{cutlass}, Composable Kernel~\cite{Composab19}, ThunderKittens/HipKittens~\cite{spector2024thunderkittenssimplefastadorable,hu2025hipkittensfastfuriousamd}) expose structured tile-level programming within target-specific execution models. Layout algebras (LEGO~\cite{lego}, Linear Layouts~\cite{linearlayouts}, TIRx~\cite{TIRxAnOp77}, TensorSeek~\cite{eisenhofer2026tensorseekslayoutformalizing}, CuTe~\cite{cecka2026cutelayoutrepresentationalgebra}) and hierarchical mapping IRs (Graphene~\cite{graphene}) formalize data distribution across threads and memory. \name~combines these structures, taking the fixed-size native instruction and its symbolic placements as the primary unit of computation, deriving loop nests, physical materialization, and synchronization from separate dialect and schedule specifications.

\paragraph{Scheduling Languages and Optimizers.}
Schedule languages (Halide~\cite{halide}, Exo~\cite{exocomp, exo2, akeley2026exogpusafeimperativeuserschedulable}, TensorIR~\cite{feng2022tensorirabstractionautomatictensorized}, Fireiron~\cite{fireiron}), polyhedral compilers~\cite{pluto, ppcg, polygeist, autosa}, tile-graph optimizers (Welder~\cite{welder}, LADDER~\cite{ladder}, Cleave~\cite{pissarra2026cleavescalingtensorprogram}, Hidet~\cite{hidet}), superoptimizers (Mirage~\cite{mirage}), and pipelining/warp-specialization compilers (Twill~\cite{soi2025optimalsoftwarepipeliningwarp}, Tawa~\cite{tawa}, WASP~\cite{wasp}) optimize schedules through transformation or search. \name~\codexrev{uses symbolic iteration domains and access maps to verify and realize} explicitly composed schedules, making it a natural target layer for such schedule-generation systems\codexrev{, which could supply instruction choices, tensor views, and scheduling structure}.

\paragraph{Task and Dataflow Representations.}
Task and dataflow systems (Sequoia~\cite{sequoia}, Legion~\cite{legion}, Regent~\cite{regent}, DaCe~\cite{dace}, Cypress~\cite{yadav2025taskbasedtensorcomputationsmodern}, Event Tensor~\cite{jin2026eventtensorunifiedabstraction}, Spatial~\cite{spatial}, TileLoom~\cite{li2026tileloomautomaticdataflowplanning}, Loom~\cite{wang2026schedulessolvablesymbolstuningfree}) synthesize physical schedules and communication from coarser task/dataflow representations with varying degrees of explicit mapping and compiler inference, whereas \name~provides a statically verified representation for explicit instruction-level schedules. While WEFT~\cite{verifprod} verifies NVIDIA GPU named-barrier synchronization, \name~symbolically verifies dialect-defined token automata across placement-indexed instruction schedules without encoding GPU semantics.

\paragraph{Architecture-Parametric and Agentic Compilation.}
LLVM/GCC provide target-independent vector operations and ISA-specific lowering; ISA description frameworks (HASCO~\cite{hasco}, TAIDL~\cite{taidl}, ACT~\cite{actcompiler}, TensorLift~\cite{gao2026tensorliftautomaticextractiontensorlevel}) describe, extract, or consume hardware instruction specifications; Arknife~\cite{ma2026retargetablecompilationtensorkernels} globally solves layouts; TIRx~\cite{TIRxAnOp77} combines layouts, scopes, instruction dispatch, and loops; UNIT~\cite{unit} unifies the detection and application of tensorized instructions by transforming loop-based tensor programs; \name~complements these by making the schedule representation itself hardware-parametric and verifiable. Finally, agentic kernel synthesis frameworks ($\mu$CUTLASS~\cite{hari2026improvingefficiencygpukernel}, Helion~\cite{HelionAH0}, CAKE~\cite{ye2026cakecompileragentcodesignfrontier}, AutoMegaKernel~\cite{jaber2026automegakernelstaticallycheckedagentharness}, CuTeGen~\cite{saba2026cutegenllmbasedagenticframework}, Ave~\cite{ave}, Zomboss~\cite{gao2026rethinkingagentickernelgeneration}, FACT~\cite{heidari2026factcompositionalkernelsynthesis}) focus on agent-assisted representation, search and evolution~\cite{ouyang2025kernelbenchllmswriteefficient}. \name~is 
orthogonal to the search \ifblind algorithm\fi: it provides a compact, statically verified target language suitable for agentic kernel synthesis.

\section{Discussion}

\paragraph{Predicated and Indirect Tensor Domains.}
\name's instruction-centric model provides a foundation for extending verified compilation beyond regular dense tensor computations. Richer bounded-integer reasoning and predicated, data-dependent indexing are natural next steps: they would broaden instruction-centric lifting to structured dynamic routing, including expert selection and token dispatch in mixture-of-experts (MoE) models, graph workloads, sparse linear algebra, and a larger set of irregular scientific computations. \name~could thus allow clients to schedule irregular data movement while the compiler derives and verifies its physical realization.

\paragraph{SMT Scalability on Mixed Layout Arithmetic.}
Universal quantification over symbolic problem sizes avoids enumerating runtime loop iterations, but combines integer quotients and remainders (from tiling and ring buffering) with bitwise XOR (from swizzled scratchpads) and alternating quantifiers in producer-resolution masks. While Z3 discharges obligations quite reliably, deeply decoupled spatial pipelines such as GQA or independent multi-axis NoC multicast on Tenstorrent Blackhole can stress general-purpose SMT heuristics. Scaling verification to larger multi-stage megakernels motivates engineering effort in efficient domain-specific algebraic simplification of symbolic expressions, reuse of resolved expressions, efficient management of program assumptions, efficient SMT query framing, and parallel discharging of independent obligations.

\paragraph{From Schedule Conformance to Functional Equivalence.}
\name~establishes \emph{verified-schedule conformance}: it proves that loop fusion, thread striping, and buffer reuse preserve the exact reference values $\operatorname{Obs}_{\mathrm{ref}}(G,\rho,\mu_0)$ defined by the schedule's lifted instructions. This cleanly factors end-to-end kernel verification: because \name~already reduces the correctness of low-level imperative schedules to a compositional functional verification problem (see the supplementary material), proving that a scheduled kernel implements a high-level specification (e.g., $C = AB^T$) reduces to reasoning about computation independently of physical layouts, loop nests, and hardware waits. Exporting \name's region certificates to interactive theorem provers such as Lean could enable end-to-end proof-producing compilation from high-level tensor equations to native machine code.

\section{Conclusion}
\label{sec:conclusion}

We presented \name, a two-sided verifiable programming model and compiler that reframes accelerator code generation around compositional \emph{instruction-centric lifting}. Rather than recovering machine instructions from loop nests through custom backend passes, \name~treats both the target's architectural specification and the kernel's execution schedule as programmable inputs to a shared compiler. From these inputs, \name~automatically derives tile loops, memory offsets, fused loop traversals, and hardware synchronization, statically verifying tensor bounds, layout compatibility, value preservation, and synchronization protocol validity before emitting target code.

More broadly, \name~shows that even as hardware diverges across GPUs, CPU tile units, and spatial dataflow arrays, the mechanics of composing and verifying instruction schedules can be unified within a single hardware-parametric framework. This separation establishes a new division of labor across the accelerator stack: hardware architects can expose emerging silicon through declarative dialects rather than bespoke compiler backends, while human experts and coding agents can explore target-specific schedules with precise control and immediate, formal compile-time feedback.


\section*{AI Use Statement}

Generative AI tools, including GPT-6.1 Sol, GPT-6 Astra, GPT-5.6 Sol, and Gemini 4 Argon, through Gemini, ChatGPT, and Codex, were used extensively in implementing and debugging the compiler, developing correctness tests and benchmarking infrastructure, research formalization, and manuscript preparation.
%
The authors directed the research, reviewed AI-generated artifacts, and assessed their correctness and performance through compilation, testing, and hardware measurements, and take full responsibility for the correctness, integrity, and reproducibility of this paper, including all AI-assisted research, code, experiments, and writing.



\ifblind
\section*{Data-Availability Statement}

Our working implementation of \name~ includes the compiler, machine dialects, native code-generation bindings, kernel programs, and correctness and benchmarking infrastructure. We intend to submit an anonymized artifact for PLDI Artifact Evaluation, containing the implementation and supporting materials needed to reproduce the reported experiments. Some performance experiments require specific accelerator hardware and vendor toolchains, which may limit reproduction on other systems. We will provide documentation of these requirements and the available validation procedures. Upon publication, we will open-source the source code, and publish the full formal specification of Teal, provided in the supplementary material, on arXiv.
\else
\section*{Source Code}

The source code and supplementary material will be made publicly available in a subsequent revision of this preprint.

\section*{Acknowledgements}

This work was supported by the Horizon Europe project DGPT, grant number 101334478. This work was also supported in part by Advanced Micro Devices, Inc. under the AMD University Program’s AI \& HPC Cluster. We thank the team at Daisytuner for providing access to the Tenstorrent hardware used in our experiments. We also thank the Swiss National Supercomputing Center (CSCS) for providing computational resources for this work. 


\fi

\bibliographystyle{ACM-Reference-Format}
\bibliography{sample-base, sc26, iclr27}

@inproceedings{shah2024flashattention3fastaccurateattention,
  author    = {Shah, Jay and Bikshandi, Ganesh and Zhang, Ying and Thakkar, Vijay and Ramani, Pradeep and Dao, Tri},
  title     = {FlashAttention-3: fast and accurate attention with asynchrony and low-precision},
  year      = {2024},
  isbn      = {9798331314385},
  publisher = {Curran Associates Inc.},
  address   = {Red Hook, NY, USA},
  booktitle = {Proceedings of the 38th International Conference on Neural Information Processing Systems},
  articleno = {2193},
  numpages  = {28},
  location  = {Vancouver, BC, Canada},
  series    = {{\codexrev{NeurIPS '24}}}
}

@misc{soi2025optimalsoftwarepipeliningwarp,
  title         = {Optimal Software Pipelining and Warp Specialization for Tensor Core GPUs},
  author        = {Rupanshu Soi and Rohan Yadav and Fredrik Kjolstad and Alex Aiken and Maryam Mehri Dehnavi and Michael Garland and Michael Bauer},
  year          = {2025},
  eprint        = {2512.18134},
  archiveprefix = {arXiv},
  primaryclass  = {cs.PL},
  url           = {https://arxiv.org/abs/2512.18134}
}

@inproceedings{hopperdissect,
  author    = {Luo, Weile and Fan, Ruibo and Li, Zeyu and Du, Dayou and Wang, Qiang and Chu, Xiaowen},
  booktitle = {2024 IEEE International Parallel and Distributed Processing Symposium (IPDPS)},
  title     = {Benchmarking and Dissecting the Nvidia Hopper GPU Architecture},
  year      = {2024},
  volume    = {},
  number    = {},
  pages     = {656-667},
  doi       = {10.1109/IPDPS57955.2024.00064}
}

@inproceedings{hellosme,
  author    = {Remke, Stefan and Breuer, Alexander},
  title     = {Hello SME! Generating Fast Matrix Multiplication Kernels Using the Scalable Matrix Extension},
  year      = {2025},
  isbn      = {9798350355543},
  publisher = {IEEE Press},
  url       = {https://doi.org/10.1109/SCW63240.2024.00185},
  doi       = {10.1109/SCW63240.2024.00185},
  booktitle = {Proceedings of the SC '24 Workshops of the International Conference on High Performance Computing, Network, Storage, and Analysis},
  pages     = {1443–1454},
  numpages  = {12},
  location  = {Atlanta, GA, USA},
  series    = {SC-W '24}
}

@misc{cdna4,
  author = {AMD},
  title  = {CDNA-4 {\codexrev{Instruction}} Set Architecture},
  url    = {https://www.amd.com/content/dam/amd/en/documents/instinct-tech-docs/instruction-set-architectures/amd-instinct-cdna4-instruction-set-architecture.pdf},
  month  = {},
  year   = {2024},
  note   = {[Online]}
}

@inproceedings{linearlayouts,
  author    = {Zhou, Keren and Lezcano-Casado, Mario and Goucher, Adam P. and Rakhmati, Akhmed and Niu, Jeff and Lebar, Justin and Szczerbuk, Pawel and Bell, Peter and Tillet, Phil and Raoux, Thomas and Moudallal, Zahi},
  title     = {Linear Layouts: Robust Code Generation of Efficient Tensor Computation Using $\mathbb{F}_2$},
  year      = {2025},
  isbn      = {9798400721656},
  publisher = {Association for Computing Machinery},
  address   = {New York, NY, USA},
  url       = {https://doi.org/10.1145/3760250.3762221},
  doi       = {10.1145/3760250.3762221},
  booktitle = {Proceedings of the 31st ACM International Conference on Architectural Support for Programming Languages and Operating Systems, Volume 1},
  pages     = {132–146},
  numpages  = {15},
  location  = {Pittsburgh, PA, USA},
  series    = {ASPLOS '26}
}

@inproceedings{blackwelldissect,
  author    = {Jarmusch, Aaron and Graddon, Nathan and Chandrasekaran, Sunita},
  booktitle = {2025 IEEE 32nd International Conference on High Performance Computing, Data and Analytics Workshop (HiPCW)},
  title     = {Dissecting the NVIDIA RTX Blackwell Architecture with Microbenchmarks},
  year      = {2025},
  volume    = {},
  number    = {},
  pages     = {309-310},
  doi       = {10.1109/HiPCW66559.2025.00109}
}

@misc{hu2025hipkittensfastfuriousamd,
  title         = {HipKittens: Fast and Furious AMD Kernels},
  author        = {William Hu and Drew Wadsworth and Sean Siddens and Stanley Winata and Daniel Y. Fu and Ryann Swann and Muhammad Osama and Christopher Ré and Simran Arora},
  year          = {2025},
  eprint        = {2511.08083},
  archiveprefix = {arXiv},
  primaryclass  = {cs.LG},
  url           = {https://arxiv.org/abs/2511.08083}
}

@misc{deng2025demystifyingarmsmeoptimize,
  title         = {Demystifying ARM SME to Optimize General Matrix Multiplications},
  author        = {Chencheng Deng and Weiling Yang and Jianbin Fang and Dezun Dong},
  year          = {2025},
  eprint        = {2512.21473},
  archiveprefix = {arXiv},
  primaryclass  = {cs.DC},
  url           = {https://arxiv.org/abs/2512.21473}
}

@article{intelamxllm,
  author   = {Kim, Hyungyo and Ye, Gaohan and Wang, Nachuan and Yazdanbakhsh, Amir and Kim, Nam Sung},
  journal  = {IEEE Computer Architecture Letters},
  title    = {Exploiting Intel Advanced Matrix Extensions (AMX) for Large Language Model Inference},
  year     = {2024},
  volume   = {23},
  number   = {1},
  pages    = {117-120},
  doi      = {10.1109/LCA.2024.3397747}
}

@inproceedings{intelamx,
  author    = {Endo, Yusuke and Ohshima, Satoshi and Nanri, Takeshi},
  title     = {Optimization of a GEMM Implementation using Intel AMX},
  year      = {2026},
  isbn      = {9798400720673},
  publisher = {Association for Computing Machinery},
  address   = {New York, NY, USA},
  url       = {https://doi.org/10.1145/3773656.3773660},
  doi       = {10.1145/3773656.3773660},
  booktitle = {Proceedings of the Supercomputing Asia and International Conference on High Performance Computing in Asia Pacific Region},
  pages     = {81–90},
  numpages  = {10},
  location  = {Osaka, Japan},
  series    = {SCA/HPCAsia '26}
}

@misc{taylor2026numericalkernelsspatialaccelerator,
  title         = {Numerical Kernels on a Spatial Accelerator: A Study of Tenstorrent Wormhole},
  author        = {Maya Taylor and Carl Pearson and Luc Berger-Vergiat and Giovanni Long and Jan Ciesko},
  year          = {2026},
  eprint        = {2603.23343},
  archiveprefix = {arXiv},
  primaryclass  = {cs.PF},
  url           = {https://arxiv.org/abs/2603.23343}
}

@misc{xu2026maia200softwaredefined,
  title         = {Maia 200: A Software Defined Dataflow System for Large-scale AI Acceleration},
  author        = {Sherry Xu and Marco Heddes and Jackson Peng and Tom Savell and Monica Tang and Prashant Ranjan and Jesse Benson and Ofer Dekel and Saurabh Dighe and Anupama Kurpad and Artour Levin and Matthew Mattina and George Petre and Cheng Tang and Yuan Yu and Li Zhang and Torsten Hoefler},
  year          = {2026},
  eprint        = {2608.24664},
  archiveprefix = {arXiv},
  primaryclass  = {cs.AR},
  url           = {https://arxiv.org/abs/2608.24664}
}

@misc{HIPdocum43,
  author = {AMD},
  title  = {HIP documentation — HIP 7.16.0 Documentation},
  url    = {https://rocm.docs.amd.com/projects/HIP/en/latest/},
  month  = {},
  year   = {2026},
  note   = {[Online]}
}

@misc{ptxdoc,
  author = {NVIDIA},
  title  = {1. Introduction — PTX ISA 9.4 documentation},
  url    = {https://docs.nvidia.com/cuda/parallel-thread-execution/},
  month  = {},
  year   = {2026},
  note   = {[Online]}
}

@misc{Composab19,
  author = {AMD},
  title  = {Composable Kernel User Guide — Composable Kernel 1.3.0 Documentation},
  url    = {https://rocm.docs.amd.com/projects/composable_kernel/en/latest/},
  month  = {},
  year   = {2026},
  note   = {[Online]}
}

@inproceedings{exocomp,
  author    = {Ikarashi, Yuka and Bernstein, Gilbert Louis and Reinking, Alex and Genc, Hasan and Ragan-Kelley, Jonathan},
  title     = {Exocompilation for productive programming of hardware accelerators},
  year      = {2022},
  isbn      = {9781450392655},
  publisher = {Association for Computing Machinery},
  address   = {New York, NY, USA},
  url       = {https://doi.org/10.1145/3519939.3523446},
  doi       = {10.1145/3519939.3523446},
  booktitle = {Proceedings of the 43rd ACM SIGPLAN International Conference on Programming Language Design and Implementation},
  pages     = {703–718},
  numpages  = {16},
  location  = {San Diego, CA, USA},
  series    = {PLDI 2022}
}

@misc{akeley2026exogpusafeimperativeuserschedulable,
  title         = {Exo-GPU: Safe, Imperative, User-schedulable Programming for Tensor Cores},
  author        = {David Zhao Akeley and Yuka Ikarashi and Jonathan Ragan-Kelley},
  year          = {2026},
  eprint        = {2609.16389},
  archiveprefix = {arXiv},
  primaryclass  = {cs.PL},
  url           = {https://arxiv.org/abs/2609.16389}
}

@misc{ye2026cakecompileragentcodesignfrontier,
  title         = {CAKE: Compiler-Agent Co-Design for Frontier Kernel Evolution},
  author        = {Zihao Ye and Yingyi Huang and Hongyi Jin and Bohan Hou and Junru Shao and Zhongming Yu and Jinqi Chen and Meghan Cowan and Shiyi Cao and Shanli Xing and Hanfeng Chen and Vinod Grover and Tianqi Chen and Luis Ceze},
  year          = {2026},
  eprint        = {2608.12629},
  archiveprefix = {arXiv},
  primaryclass  = {cs.LG},
  url           = {https://arxiv.org/abs/2608.12629}
}

@article{actcompiler,
  author  = {Jain, Devansh and Pardeshi, Akash and Frigo, Marco and Khulbe, Kaustubh and Patel, Krut and Lochan, Saatvik and Arora, Jai and Mendis, Charith},
  year    = {2026},
  month   = oct,
  pages   = {288-319},
  title   = {Automatically Generating ML Compiler Backends from Tensor Accelerator ISA Descriptions},
  volume  = {10},
  number  = {OOPSLA2},
  journal = {Proceedings of the ACM on Programming Languages},
  doi     = {10.1145/3839457}
}

@misc{feng2022tensorirabstractionautomatictensorized,
  title         = {TensorIR: An Abstraction for Automatic Tensorized Program Optimization},
  author        = {Siyuan Feng and Bohan Hou and Hongyi Jin and Wuwei Lin and Junru Shao and Ruihang Lai and Zihao Ye and Lianmin Zheng and Cody Hao Yu and Yong Yu and Tianqi Chen},
  year          = {2022},
  eprint        = {2207.04296},
  archiveprefix = {arXiv},
  primaryclass  = {cs.LG},
  url           = {https://arxiv.org/abs/2207.04296}
}

@inproceedings{ding2025tilustilelevelgpgpuprogramming,
  author    = {Ding, Yaoyao and Hou, Bohan and Zhang, Xiao and Lin, Allan and Chen, Tianqi and Yu, Cody Hao and Wang, Yida and Pekhimenko, Gennady},
  title     = {Tilus: A Tile-Level GPGPU Programming Language for Low-Precision Computation},
  year      = {2025},
  isbn      = {9798400721656},
  publisher = {Association for Computing Machinery},
  address   = {New York, NY, USA},
  url       = {https://doi.org/10.1145/3760250.3762219},
  doi       = {10.1145/3760250.3762219},
  booktitle = {Proceedings of the 31st ACM International Conference on Architectural Support for Programming Languages and Operating Systems, Volume 1},
  pages     = {281–297},
  numpages  = {17},
  location  = {Pittsburgh, PA, USA},
  series    = {ASPLOS '26}
}

@misc{saba2026cutegenllmbasedagenticframework,
  title         = {CuTeGen: An LLM-Based Agentic Framework for Generation and Optimization of High-Performance GPU Kernels using CuTe},
  author        = {Tara Saba and Zhiyang Chen and Jikai Jason Li and Anne Ouyang and Xujie Si and Fan Long},
  year          = {2026},
  eprint        = {2604.01489},
  archiveprefix = {arXiv},
  primaryclass  = {cs.LG},
  url           = {https://arxiv.org/abs/2604.01489}
}

@misc{TIRxAnOp77,
  title         = {TIRx: A Unified Tile Primitive ML Compiler Across Heterogeneous Backends}, 
  author        = {Bohan Hou and Hongyi Jin and Guanjie Wang and Shushi Hong and Jinqi Chen and Yaxing Cai and Lijie Yang and Zihao Ye and Yaoyao Ding and Ruihang Lai and Todd Mowry and Tianqi Chen},
  year          = {2026},
  eprint        = {2601.19092},
  archivePrefix = {arXiv},
  primaryClass  = {cs.DC},
  url           = {https://arxiv.org/abs/2601.19092}, 
}

@inproceedings{tawa,
  author    = {Chen, Hongzheng and Fan, Bin and Collins, Alexander and Hagedorn, Bastian and Gaburov, Evghenii and Masuda, Masahiro and Brookhart, Matthew and Sullivan, Chris and Knight, Jason and Zhang, Zhiru and Grover, Vinod},
  booktitle = {2026 IEEE/ACM International Symposium on Code Generation and Optimization (CGO)},
  title     = {Tawa: Automatic Warp Specialization for Modern GPUs with Asynchronous References},
  year      = {2026},
  volume    = {},
  number    = {},
  pages     = {255-267},
  doi       = {10.1109/CGO68049.2026.11394849}
}

@inproceedings{wasp,
  author    = {Crago, Neal C. and Damani, Sana and Sankaralingam, Karthikeyan and Keckler, Stephen W.},
  booktitle = {2024 IEEE International Symposium on High-Performance Computer Architecture (HPCA)},
  title     = {WASP: Exploiting GPU Pipeline Parallelism with Hardware-Accelerated Automatic Warp Specialization},
  year      = {2024},
  volume    = {},
  number    = {},
  pages     = {1-16},
  doi       = {10.1109/HPCA57654.2024.00086}
}

@misc{1NVIDIAH98,
  author = {NVIDIA},
  title  = {1. NVIDIA Hopper Tuning Guide — Hopper Tuning Guide 13.4 documentation},
  url    = {https://docs.nvidia.com/cuda/hopper-tuning-guide/index.html},
  month  = {},
  year   = {2026},
  note   = {[Online]}
}

@inproceedings{verifprod,
  author    = {Sharma, Rahul and Bauer, Michael and Aiken, Alex},
  title     = {Verification of producer-consumer synchronization in GPU programs},
  year      = {2015},
  isbn      = {9781450334686},
  publisher = {Association for Computing Machinery},
  address   = {New York, NY, USA},
  url       = {https://doi.org/10.1145/2737924.2737962},
  doi       = {10.1145/2737924.2737962},
  booktitle = {Proceedings of the 36th ACM SIGPLAN Conference on Programming Language Design and Implementation},
  pages     = {88–98},
  numpages  = {11},
  location  = {Portland, OR, USA},
  series    = {PLDI '15}
}

@inproceedings{mmus,
  author    = {Lu, Yuechen and Zeng, Hongwei and Casas, Marc and Liu, Weifeng},
  title     = {Characterizing Matrix Multiplication Units across General Parallel Patterns in Scientific Computing},
  year      = {2026},
  isbn      = {9798400723100},
  publisher = {Association for Computing Machinery},
  address   = {New York, NY, USA},
  url       = {https://doi.org/10.1145/3774934.3786456},
  doi       = {10.1145/3774934.3786456},
  booktitle = {Proceedings of the 31st ACM SIGPLAN Annual Symposium on Principles and Practice of Parallel Programming},
  pages     = {687–701},
  numpages  = {15},
  location  = {Sydney, NSW, Australia},
  series    = {PPoPP '26}
}

@article{Kuzma_2023,
  title     = {Fast matrix multiplication via compiler‐only layered data reorganization and intrinsic lowering},
  volume    = {53},
  issn      = {1097-024X},
  url       = {http://dx.doi.org/10.1002/spe.3214},
  doi       = {10.1002/spe.3214},
  number    = {9},
  journal   = {Software: Practice and Experience},
  publisher = {Wiley},
  author    = {Kuzma, Braedy and Korostelev, Ivan and de Carvalho, João P. L. and Moreira, José E. and Barton, Christopher and Araujo, Guido and Amaral, José Nelson},
  year      = {2023},
  month     = May,
  pages     = {1793–1814}
}

@misc{wang2026schedulessolvablesymbolstuningfree,
  title         = {Schedules Are Solvable Symbols: Tuning-Free Compilation of Tile Programs on Dataflow Architectures},
  author        = {Heru Wang and Wei Li and Zhenyu Bai and Tulika Mitra},
  year          = {2026},
  eprint        = {2609.29219},
  archiveprefix = {arXiv},
  primaryclass  = {cs.PL},
  url           = {https://arxiv.org/abs/2609.29219}
}

@misc{hari2026improvingefficiencygpukernel,
  title         = {Improving Efficiency of GPU Kernel Optimization Agents using a Domain-Specific Language and Speed-of-Light Guidance},
  author        = {Siva Kumar Sastry Hari and Vignesh Balaji and Sana Damani and Qijing Huang and Christos Kozyrakis},
  year          = {2026},
  eprint        = {2603.29010},
  archiveprefix = {arXiv},
  primaryclass  = {cs.LG},
  url           = {https://arxiv.org/abs/2603.29010}
}

@misc{HelionAH0,
  author = {PyTorch},
  title  = {Helion: A High-Level DSL for Performant and Portable ML Kernels - PyTorch},
  url    = {https://pytorch.org/blog/helion/},
  month  = {10},
  year   = {2025},
  note   = {[Online]}
}

@misc{jaber2026automegakernelstaticallycheckedagentharness,
  title         = {AutoMegaKernel: A Statically-Checked Agent Harness for Self-Retargeting Megakernel Synthesis},
  author        = {Jaber Jaber and Osama Jaber},
  year          = {2026},
  eprint        = {2606.09682},
  archiveprefix = {arXiv},
  primaryclass  = {cs.LG},
  url           = {https://arxiv.org/abs/2606.09682}
}

@inproceedings{ave,
  author    = {Mai, Haohui and Guo, Xiaoyan and Ding, Xiangyun and Li, Daifeng and Yu, Qiuchu and Guo, Chenzhun and Wang, Cong and Zhao, Jiacheng and Kozyrakis, Christos and Yuan, Binhang},
  title     = {Ave: Guiding Agentic GPU Optimization Using Data-Flow Invariants},
  year      = {2026},
  isbn      = {9798400725852},
  publisher = {Association for Computing Machinery},
  address   = {New York, NY, USA},
  url       = {https://doi.org/10.1145/3830418.3843902},
  doi       = {10.1145/3830418.3843902},
  booktitle = {Proceedings of the ACM SIGOPS 32nd Symposium on Operating Systems Principles},
  pages     = {1055–1069},
  numpages  = {15},
  location  = {Prague, Czech Republic},
  series    = {SOSP '26}
}

@inproceedings{triton,
  author    = {Tillet, Philippe and Kung, H. T. and Cox, David},
  title     = {Triton: an intermediate language and compiler for tiled neural network computations},
  year      = {2019},
  isbn      = {9781450367196},
  publisher = {Association for Computing Machinery},
  address   = {New York, NY, USA},
  url       = {https://doi.org/10.1145/3315508.3329973},
  doi       = {10.1145/3315508.3329973},
  booktitle = {Proceedings of the 3rd ACM SIGPLAN International Workshop on Machine Learning and Programming Languages},
  pages     = {10–19},
  numpages  = {10},
  location  = {Phoenix, AZ, USA},
  series    = {MAPL 2019}
}

@misc{gluon,
  author = {Triton},
  title  = {Introduction to Gluon — Triton documentation},
  url    = {https://triton-lang.org/main/getting-started/tutorials/gluon/intro.html},
  month  = {},
  year   = {2026},
  note   = {[Online]}
}

@misc{FlyDSLEx5,
  author = {Felix Li and Shijie Feng and Carlus Huang and Dewei Wang and Hongxia Yang and Peng Sun and Emad Barsoum},
  title  = {FlyDSL: Expert GPU Kernel Development with the Ease of MLIR Python Native DSL on AMD GPUs — ROCm Blogs},
  url    = {https://rocm.blogs.amd.com/software-tools-optimization/flydsl-python-native/README.html},
  month  = {2},
  year   = {2026},
  note   = {[Online]}
}

@inproceedings{exo2,
  author    = {Ikarashi, Yuka and Qian, Kevin and Droubi, Samir and Reinking, Alex and Bernstein, Gilbert Louis and Ragan-Kelley, Jonathan},
  title     = {Exo 2: Growing a Scheduling Language},
  year      = {2025},
  isbn      = {9798400706981},
  publisher = {Association for Computing Machinery},
  address   = {New York, NY, USA},
  url       = {https://doi.org/10.1145/3669940.3707218},
  doi       = {10.1145/3669940.3707218},
  booktitle = {Proceedings of the 30th ACM International Conference on Architectural Support for Programming Languages and Operating Systems, Volume 1},
  pages     = {426–444},
  numpages  = {19},
  location  = {Rotterdam, Netherlands},
  series    = {ASPLOS '25}
}

@article{halide,
  author     = {Ragan-Kelley, Jonathan and Adams, Andrew and Sharlet, Dillon and Barnes, Connelly and Paris, Sylvain and Levoy, Marc and Amarasinghe, Saman and Durand, Fr{\'e}do},
  title      = {Halide: decoupling algorithms from schedules for high-performance image processing},
  year       = {2017},
  issue_date = {January 2018},
  publisher  = {Association for Computing Machinery},
  address    = {New York, NY, USA},
  volume     = {61},
  number     = {1},
  issn       = {0001-0782},
  url        = {https://doi.org/10.1145/3150211},
  doi        = {10.1145/3150211},
  journal    = {Commun. ACM},
  month      = dec,
  pages      = {106–115},
  numpages   = {10}
}

@inproceedings{hasco,
  author    = {Xiao, Qingcheng and Zheng, Size and Wu, Bingzhe and Xu, Pengcheng and Qian, Xuehai and Liang, Yun},
  title     = {HASCO: towards agile hardware and software co-design for tensor computation},
  year      = {2021},
  isbn      = {9781450390866},
  publisher = {IEEE Press},
  url       = {https://doi.org/10.1109/ISCA52012.2021.00086},
  doi       = {10.1109/ISCA52012.2021.00086},
  booktitle = {Proceedings of the 48th Annual International Symposium on Computer Architecture},
  pages     = {1055–1068},
  numpages  = {14},
  location  = {Virtual Event, Spain},
  series    = {ISCA '21}
}

@inproceedings{legion,
  author    = {Bauer, Michael and Treichler, Sean and Slaughter, Elliott and Aiken, Alex},
  title     = {Legion: expressing locality and independence with logical regions},
  year      = {2012},
  isbn      = {9781467308045},
  publisher = {IEEE Computer Society Press},
  address   = {Washington, DC, USA},
  booktitle = {Proceedings of the International Conference on High Performance Computing, Networking, Storage and Analysis},
  articleno = {66},
  numpages  = {11},
  location  = {Salt Lake City, Utah},
  series    = {SC '12}
}

@inproceedings{sequoia,
  author    = {Fatahalian, Kayvon and Horn, Daniel Reiter and Knight, Timothy J. and Leem, Larkhoon and Houston, Mike and Park, Ji Young and Erez, Mattan and Ren, Manman and Aiken, Alex and Dally, William J. and Hanrahan, Pat},
  title     = {Sequoia: programming the memory hierarchy},
  year      = {2006},
  isbn      = {0769527000},
  publisher = {Association for Computing Machinery},
  address   = {New York, NY, USA},
  url       = {https://doi.org/10.1145/1188455.1188543},
  doi       = {10.1145/1188455.1188543},
  booktitle = {Proceedings of the 2006 ACM/IEEE Conference on Supercomputing},
  pages     = {83},
  location  = {Tampa, Florida},
  series    = {SC '06}
}

@inproceedings{dace,
  author    = {Ben-Nun, Tal and de Fine Licht, Johannes and Ziogas, Alexandros N. and Schneider, Timo and Hoefler, Torsten},
  title     = {Stateful dataflow multigraphs: a data-centric model for performance portability on heterogeneous architectures},
  year      = {2019},
  isbn      = {9781450362290},
  publisher = {Association for Computing Machinery},
  address   = {New York, NY, USA},
  url       = {https://doi.org/10.1145/3295500.3356173},
  doi       = {10.1145/3295500.3356173},
  booktitle = {Proceedings of the International Conference for High Performance Computing, Networking, Storage and Analysis},
  articleno = {81},
  numpages  = {14},
  location  = {Denver, Colorado},
  series    = {SC '19}
}

@phdthesis{regent,
  author    = {Slaughter, Elliott},
  advisor   = {Alexander Aiken, and Philip Levis, and Oyekunle Olukotun},
  title     = {Regent: A High-Productivity Programming Language for Implicit Parallelism with Logical Regions},
  year      = {2017},
  isbn      = {9798698509349},
  publisher = {Stanford University},
  address   = {Stanford, CA, USA},
  note      = {AAI28115976}
}

@misc{jin2026eventtensorunifiedabstraction,
  title         = {Event Tensor: A Unified Abstraction for Compiling Dynamic Megakernel},
  author        = {Hongyi Jin and Bohan Hou and Guanjie Wang and Ruihang Lai and Jinqi Chen and Zihao Ye and Yaxing Cai and Yixin Dong and Xinhao Cheng and Zhihao Zhang and Yilong Zhao and Yingyi Huang and Lijie Yang and Jinchen Jiang and Gabriele Oliaro and Jianan Ji and Xupeng Miao and Vinod Grover and Todd C. Mowry and Zhihao Jia and Tianqi Chen},
  year          = {2026},
  eprint        = {2604.13327},
  archiveprefix = {arXiv},
  primaryclass  = {cs.DC},
  url           = {https://arxiv.org/abs/2604.13327}
}

@inproceedings{taidl,
  author    = {Jain, Devansh and Frigo, Marco and Arora, Jai and Pardeshi, Akash and Wang, Zhihao and Patel, Krut and Mendis, Charith},
  title     = {TAIDL: Tensor Accelerator ISA Definition Language with Auto-generation of Scalable Test Oracles},
  year      = {2025},
  isbn      = {9798400715730},
  publisher = {Association for Computing Machinery},
  address   = {New York, NY, USA},
  url       = {https://doi.org/10.1145/3725843.3756075},
  doi       = {10.1145/3725843.3756075},
  booktitle = {Proceedings of the 58th IEEE/ACM International Symposium on Microarchitecture},
  pages     = {1316–1333},
  numpages  = {18},
  location  = {Seoul, Korea},
  series    = {MICRO '25}
}

@misc{gao2026tensorliftautomaticextractiontensorlevel,
  title         = {TensorLift: Automatic Extraction of Tensor-Level ISA Semantics from Accelerator RTL via MLIR Semantic Lifting},
  author        = {Ruijie Gao and Haoran Jin and Jirong Yang and Nathaniel Bleier},
  year          = {2026},
  eprint        = {2604.13523},
  archiveprefix = {arXiv},
  primaryclass  = {cs.AR},
  doi           = {https://doi.org/10.1145/3831252.3833946},
  url           = {https://arxiv.org/abs/2604.13523}
}

@misc{an2026simplifiedlayoutalgebragpu,
  title         = {A Simplified Layout Algebra for GPU Kernels},
  author        = {Nan An and Xiaotian Zhou},
  year          = {2026},
  eprint        = {2610.13679},
  archiveprefix = {arXiv},
  primaryclass  = {cs.PL},
  url           = {https://arxiv.org/abs/2610.13679}
}

@article{ppcg,
  author     = {Verdoolaege, Sven and Carlos Juega, Juan and Cohen, Albert and Ignacio G{\'o}mez, Jos{\'e} and Tenllado, Christian and Catthoor, Francky},
  title      = {Polyhedral parallel code generation for CUDA},
  year       = {2013},
  issue_date = {January 2013},
  publisher  = {Association for Computing Machinery},
  address    = {New York, NY, USA},
  volume     = {9},
  number     = {4},
  issn       = {1544-3566},
  url        = {https://doi.org/10.1145/2400682.2400713},
  doi        = {10.1145/2400682.2400713},
  journal    = {ACM Trans. Archit. Code Optim.},
  month      = jan,
  articleno  = {54},
  numpages   = {23}
}

@inproceedings{pluto,
  author    = {Bondhugula, Uday and Hartono, Albert and Ramanujam, J. and Sadayappan, P.},
  title     = {A practical automatic polyhedral parallelizer and locality optimizer},
  year      = {2008},
  isbn      = {9781595938602},
  publisher = {Association for Computing Machinery},
  address   = {New York, NY, USA},
  url       = {https://doi.org/10.1145/1375581.1375595},
  doi       = {10.1145/1375581.1375595},
  booktitle = {Proceedings of the 29th ACM SIGPLAN Conference on Programming Language Design and Implementation},
  pages     = {101–113},
  numpages  = {13},
  location  = {Tucson, AZ, USA},
  series    = {PLDI '08}
}

@inproceedings{lego,
  author    = {Tavakkoli, Amir Mohammad and Oancea, Cosmin E. and Hall, Mary},
  booktitle = {2026 IEEE/ACM International Symposium on Code Generation and Optimization (CGO)},
  title     = {LEGO: A Layout Expression Language for Code Generation of Hierarchical Mapping},
  year      = {2026},
  volume    = {},
  issn      = {},
  pages     = {228-241},
  doi       = {10.1109/CGO68049.2026.11394846},
  url       = {https://doi.ieeecomputersociety.org/10.1109/CGO68049.2026.11394846},
  publisher = {IEEE Computer Society},
  address   = {Los Alamitos, CA, USA},
  month     = Feb
}

@inproceedings{polygeist,
  author    = {Moses, William S. and Chelini, Lorenzo and Zhao, Ruizhe and Zinenko, Oleksandr},
  title     = {Polygeist: Raising C to Polyhedral MLIR},
  year      = {2021},
  isbn      = {9781665442787},
  publisher = {IEEE Press},
  url       = {https://doi.org/10.1109/PACT52795.2021.00011},
  doi       = {10.1109/PACT52795.2021.00011},
  booktitle = {Proceedings of the 30th International Conference on Parallel Architectures and Compilation Techniques},
  pages     = {45–59},
  numpages  = {15},
  location  = {Atlanta, GA, USA},
  series    = {PACT '21}
}

@inproceedings{autosa,
  author    = {Wang, Jie and Guo, Licheng and Cong, Jason},
  title     = {AutoSA: A Polyhedral Compiler for High-Performance Systolic Arrays on FPGA},
  year      = {2021},
  isbn      = {9781450382182},
  publisher = {Association for Computing Machinery},
  address   = {New York, NY, USA},
  url       = {https://doi.org/10.1145/3431920.3439292},
  doi       = {10.1145/3431920.3439292},
  booktitle = {The 2021 ACM/SIGDA International Symposium on Field-Programmable Gate Arrays},
  pages     = {93–104},
  numpages  = {12},
  location  = {Virtual Event, USA},
  series    = {FPGA '21}
}

@misc{zhang2026architectinghardwarediversitymetas,
  title         = {Architecting for Hardware Diversity: Meta's End-to-End Co-Design for High-Performance AI Training on AMD GPUs},
  author        = {Haoyu Zhang and Arshad Rahman and Adnan Aziz and Ajith Prabhudev and Amin Farmahini and Ben Rhoads and Benson Ma and Brian Shi and Cen Zhao and CK Luk and Dingming Wu and Emad El-Haraty and Eric Chen and Frank Li and Gary Huang and Gefei Zuo and Haoran Zhang and Hiwot Kassa and Hongyi Zeng and Jack Taylor and Jeremy Hadidjojo and John Donati and Jonathan Sheffield and Julio Machado Silva and Karthik Ganesan and Kent Russell and KR Kishore and Lei Tian and Liu Ke and Muhammad Osama and Peng Chen and Rakesh Hemnani and Ryan Swann and Saleel Kudchadker and Shankar Yanamandram and Shobhit Kanaujia and Shucai Xiao and Srinath Bayareddy and Srinivas Vaidyanathan and Sudharssun Subramanian and Supadchaya Puangpontip and Taylor Robie and Tianhao Huang and Tyler Hart and Viswesh Sankaran and Weijun Jiang and Wenkai Du and Xiaodong Wang and Xiaohu Guo and Yao Rong and Yinbin Ma and Yue Dong and Yuzhen Huang and Yvonne Liu and Zhan Shu and Dan Rabinovitsj and Ajit Mathews and Chunqiang Tang},
  year          = {2026},
  eprint        = {2610.22230},
  archiveprefix = {arXiv},
  primaryclass  = {cs.DC},
  url           = {https://arxiv.org/abs/2610.22230}
}

@inproceedings{welder,
  author    = {Yining Shi and Zhi Yang and Jilong Xue and Lingxiao Ma and Yuqing Xia and Ziming Miao and Yuxiao Guo and Fan Yang and Lidong Zhou},
  title     = {Welder: Scheduling Deep Learning Memory Access via Tile-graph},
  booktitle = {17th USENIX Symposium on Operating Systems Design and Implementation (OSDI 23)},
  year      = {2023},
  isbn      = {978-1-939133-34-2},
  address   = {Boston, MA},
  pages     = {701--718},
  url       = {https://www.usenix.org/conference/osdi23/presentation/shi},
  publisher = {USENIX Association},
  month     = jul
}

@misc{pissarra2026cleavescalingtensorprogram,
  title         = {Cleave: Scaling Tensor Program Optimization via Decoupled Algebraic Search and Operator Scheduling},
  author        = {David Pissarra and Jinkun Lin and Haitian Jiang and Aurojit Panda and Jinyang Li},
  year          = {2026},
  eprint        = {2610.07742},
  archiveprefix = {arXiv},
  primaryclass  = {cs.PL},
  url           = {https://arxiv.org/abs/2610.07742}
}

@misc{gao2026rethinkingagentickernelgeneration,
  title         = {Rethinking Agentic Kernel Generation for Emerging Accelerators},
  author        = {Ruijie Gao and Jirong Yang and Barry Lyu and Haoran Jin and Nathan Bleier},
  year          = {2026},
  eprint        = {2608.00894},
  archiveprefix = {arXiv},
  primaryclass  = {cs.AR},
  url           = {https://arxiv.org/abs/2608.00894}
}

@inproceedings{spatial,
  author    = {Koeplinger, David and Feldman, Matthew and Prabhakar, Raghu and Zhang, Yaqi and Hadjis, Stefan and Fiszel, Ruben and Zhao, Tian and Nardi, Luigi and Pedram, Ardavan and Kozyrakis, Christos and Olukotun, Kunle},
  title     = {Spatial: a language and compiler for application accelerators},
  year      = {2018},
  isbn      = {9781450356985},
  publisher = {Association for Computing Machinery},
  address   = {New York, NY, USA},
  url       = {https://doi.org/10.1145/3192366.3192379},
  doi       = {10.1145/3192366.3192379},
  booktitle = {Proceedings of the 39th ACM SIGPLAN Conference on Programming Language Design and Implementation},
  pages     = {296–311},
  numpages  = {16},
  location  = {Philadelphia, PA, USA},
  series    = {PLDI 2018}
}

@misc{eisenhofer2026tensorseekslayoutformalizing,
  title         = {Tensor Seeks Layout: Formalizing Layout Selection for ML Compilers},
  author        = {Clemens Eisenhofer and Yuwen Jia and Daniel Kroening and Sergey Pupyrev},
  year          = {2026},
  eprint        = {2608.21555},
  archiveprefix = {arXiv},
  primaryclass  = {cs.PL},
  url           = {https://arxiv.org/abs/2608.21555}
}

@misc{li2026tileloomautomaticdataflowplanning,
  title         = {TileLoom: Automatic Dataflow Planning for Tile-Based Languages on Spatial Dataflow Accelerators},
  author        = {Wei Li and Zhenyu Bai and Heru Wang and Pranav Dangi and Zhiqiang Zhang and Cheng Tan and Huiying Lan and Weng-Fai Wong and Tulika Mitra},
  year          = {2026},
  eprint        = {2512.22168},
  archiveprefix = {arXiv},
  primaryclass  = {cs.DC},
  url           = {https://arxiv.org/abs/2512.22168}
}

@inproceedings{ladder,
  author    = {Wang, Lei and Ma, Lingxiao and Cao, Shijie and Zhang, Quanlu and Xue, Jilong and Shi, Yining and Zheng, Ningxin and Miao, Ziming and Yang, Fan and Cao, Ting and Yang, Yuqing and Yang, Mao},
  title     = {LADDER: enabling efficient low-precision deep learning computing through hardware-aware tensor transformation},
  year      = {2024},
  isbn      = {978-1-939133-40-3},
  publisher = {USENIX Association},
  address   = {USA},
  booktitle = {Proceedings of the 18th USENIX Conference on Operating Systems Design and Implementation},
  articleno = {17},
  numpages  = {17},
  location  = {Santa Clara, CA, USA},
  series    = {OSDI'24}
}

@inproceedings{fireiron,
  author    = {Hagedorn, Bastian and Elliott, Archibald Samuel and Barthels, Henrik and Bodik, Rastislav and Grover, Vinod},
  title     = {Fireiron: A Data-Movement-Aware Scheduling Language for GPUs},
  year      = {2020},
  isbn      = {9781450380751},
  publisher = {Association for Computing Machinery},
  address   = {New York, NY, USA},
  url       = {https://doi.org/10.1145/3410463.3414632},
  doi       = {10.1145/3410463.3414632},
  booktitle = {Proceedings of the ACM International Conference on Parallel Architectures and Compilation Techniques},
  pages     = {71–82},
  numpages  = {12},
  location  = {Virtual Event, GA, USA},
  series    = {PACT '20}
}

@misc{heidari2026factcompositionalkernelsynthesis,
  title         = {FACT: Compositional Kernel Synthesis with a Three-Stage Agentic Workflow},
  author        = {Sina Heidari and Dimitrios S. Nikolopoulos},
  year          = {2026},
  eprint        = {2604.26666},
  archiveprefix = {arXiv},
  primaryclass  = {cs.DC},
  url           = {https://arxiv.org/abs/2604.26666}
}

@misc{hong2025autocomppowerfulportablecode,
  title         = {Autocomp: A Powerful and Portable Code Optimizer for Tensor Accelerators},
  author        = {Charles Hong and Sahil Bhatia and Alvin Cheung and Yakun Sophia Shao},
  year          = {2025},
  eprint        = {2505.18574},
  archiveprefix = {arXiv},
  primaryclass  = {cs.PL},
  url           = {https://arxiv.org/abs/2505.18574}
}

@misc{zadouri2026flashattention4algorithmkernelpipelining,
  title         = {FlashAttention-4: Algorithm and Kernel Pipelining Co-Design for Asymmetric Hardware Scaling},
  author        = {Ted Zadouri and Markus Hoehnerbach and Jay Shah and Timmy Liu and Vijay Thakkar and Tri Dao},
  year          = {2026},
  eprint        = {2603.05451},
  archiveprefix = {arXiv},
  primaryclass  = {cs.CL},
  url           = {https://arxiv.org/abs/2603.05451}
}

@article{cerebras,
  author   = {Lie, Sean},
  journal  = {IEEE Micro},
  title    = {Inside the Cerebras Wafer-Scale Cluster},
  year     = {2024},
  volume   = {44},
  number   = {3},
  pages    = {49-57},
  doi      = {10.1109/MM.2024.3386628}
}

@misc{ouyang2025kernelbenchllmswriteefficient,
  title         = {KernelBench: Can LLMs Write Efficient GPU Kernels?},
  author        = {Anne Ouyang and Simon Guo and Simran Arora and Alex L. Zhang and William Hu and Christopher Ré and Azalia Mirhoseini},
  year          = {2025},
  eprint        = {2502.10517},
  archiveprefix = {arXiv},
  primaryclass  = {cs.LG},
  url           = {https://arxiv.org/abs/2502.10517}
}

@inproceedings{graphene,
  author    = {Hagedorn, Bastian and Fan, Bin and Chen, Hanfeng and Cecka, Cris and Garland, Michael and Grover, Vinod},
  title     = {Graphene: An IR for Optimized Tensor Computations on GPUs},
  year      = {2023},
  isbn      = {9781450399180},
  publisher = {Association for Computing Machinery},
  address   = {New York, NY, USA},
  url       = {https://doi.org/10.1145/3582016.3582018},
  doi       = {10.1145/3582016.3582018},
  booktitle = {Proceedings of the 28th ACM International Conference on Architectural Support for Programming Languages and Operating Systems, Volume 3},
  pages     = {302–313},
  numpages  = {12},
  location  = {Vancouver, BC, Canada},
  series    = {ASPLOS 2023}
}

@inproceedings{hidet,
  author    = {Ding, Yaoyao and Yu, Cody Hao and Zheng, Bojian and Liu, Yizhi and Wang, Yida and Pekhimenko, Gennady},
  title     = {Hidet: Task-Mapping Programming Paradigm for Deep Learning Tensor Programs},
  year      = {2023},
  isbn      = {9781450399166},
  publisher = {Association for Computing Machinery},
  address   = {New York, NY, USA},
  url       = {https://doi.org/10.1145/3575693.3575702},
  doi       = {10.1145/3575693.3575702},
  booktitle = {Proceedings of the 28th ACM International Conference on Architectural Support for Programming Languages and Operating Systems, Volume 2},
  pages     = {370–384},
  numpages  = {15},
  location  = {Vancouver, BC, Canada},
  series    = {ASPLOS 2023}
}

@inproceedings{mirage,
  author    = {Wu, Mengdi and Cheng, Xinhao and Liu, Shengyu and Shi, Chunan and Ji, Jianan and Ao, Man Kit and Velliengiri, Praveen and Miao, Xupeng and Padon, Oded and Jia, Zhihao},
  title     = {Mirage: a multi-level superoptimizer for tensor programs},
  year      = {2025},
  isbn      = {978-1-939133-47-2},
  publisher = {USENIX Association},
  address   = {USA},
  booktitle = {Proceedings of the 19th USENIX Conference on Operating Systems Design and Implementation},
  articleno = {13},
  numpages  = {18},
  location  = {Boston, MA, USA},
  series    = {OSDI '25}
}

@misc{PallasaJ14:online,
  author  = {{JAX}},
  title   = {{Pallas}: a {JAX} kernel language},
  url     = {https://docs.jax.dev/en/latest/pallas/index.html},
  year    = {2026},
  urldate = {2026-10-29}
}

@misc{mosaic_gpu,
  author  = {{JAX}},
  title   = {{Pallas Mosaic GPU}},
  url     = {https://docs.jax.dev/en/latest/pallas/gpu/index.html},
  year    = {2026},
  urldate = {2026-10-29}
}

@article{CarterEdwards20143202,
  title   = {Kokkos: Enabling manycore performance portability through polymorphic memory access patterns },
  journal = {Journal of Parallel and Distributed Computing },
  volume  = {74},
  number  = {12},
  pages   = {3202 - 3216},
  year    = {2014},
  note    = {Domain-Specific Languages and High-Level Frameworks for High-Performance Computing },
  issn    = {0743-7315},
  doi     = {https://doi.org/10.1016/j.jpdc.2014.07.003},
  url     = {http://www.sciencedirect.com/science/article/pii/S0743731514001257},
  author  = {H. Carter Edwards and Christian R. Trott and Daniel Sunderland}
}

@inproceedings{unit,
  author    = {Weng, Jian and Jain, Animesh and Wang, Jie and Wang, Leyuan and Wang, Yida and Nowatzki, Tony},
  booktitle = {2021 IEEE/ACM International Symposium on Code Generation and Optimization (CGO)},
  title     = {UNIT: Unifying Tensorized Instruction Compilation},
  year      = {2021},
  volume    = {},
  number    = {},
  pages     = {77-89},
  doi       = {10.1109/CGO51591.2021.9370330}
}

@misc{hipBLASL55,
  author = {AMD},
  title  = {hipBLASLt documentation — hipBLASLt 1.4.1 Documentation},
  url    = {https://rocm.docs.amd.com/projects/hipBLASLt/en/latest/index.html},
  month  = {},
  year   = {2026},
  note   = {[Online]}
}

@misc{ROCmaite73,
  author = {AMD},
  title  = {ROCm/aiter: AI Tensor Engine for ROCm},
  url    = {https://github.com/ROCm/aiter},
  month  = {},
  year   = {2026},
  note   = {[Online]}
}

@article{ye2025flashinfer,
  title   = {FlashInfer: Efficient and Customizable Attention Engine for LLM Inference Serving},
  author  = {
             Ye, Zihao and
             Chen, Lequn and
             Lai, Ruihang and
             Lin, Wuwei and
             Zhang, Yineng and
             Wang, Stephanie and
             Chen, Tianqi and
             Kasikci, Baris and
             Grover, Vinod and
             Krishnamurthy, Arvind and
             Ceze, Luis
             },
  journal = {arXiv preprint arXiv:2501.01005},
  year    = {2025},
  url     = {https://arxiv.org/abs/2501.01005}
}

@misc{cublaslt,
  author = {NVIDIA},
  title  = {1. Introduction — cuBLAS 13.5 documentation},
  url    = {https://docs.nvidia.com/cuda/cublas/index.html#using-the-cublaslt-api},
  month  = {},
  year   = {2026},
  note   = {[Online]}
}

@software{oneDNN2025v315,
  title   = {oneAPI Deep Neural Network Library ({oneDNN})},
  author  = {{oneDNN Contributors}},
  year    = {2026},
  version = {3.15},
  url     = {https://github.com/uxlfoundation/oneDNN},
  note    = {Apache-2.0. Documentation: \url{https://uxlfoundation.github.io/oneDNN}}
}

@misc{Develope25,
  author = {Intel},
  title  = {Developer Reference for Intel® oneAPI Math Kernel Library for C},
  url    = {https://www.intel.com/content/www/us/en/docs/onemkl/developer-reference-c/2026-2/overview.html},
  month  = {},
  year   = {2026},
  note   = {[Online]}
}

@inproceedings{ttblackhole,
  author    = { Vasiljevic, Jasmina and Capalija, Davor },
  booktitle = { 2024 IEEE Hot Chips 36 Symposium (HCS) },
  title     = {{ Blackhole \& TT-Metalium: The Standalone AI Computer and its Programming Model }},
  year      = {2024},
  volume    = {},
  issn      = {},
  pages     = {1-30},
  doi       = {10.1109/HCS61935.2024.10664810},
  url       = {https://doi.ieeecomputersociety.org/10.1109/HCS61935.2024.10664810},
  publisher = {IEEE Computer Society},
  address   = {Los Alamitos, CA, USA},
  month     = Aug
}

@misc{tenstorr95,
  author = {Tenstorrent},
  title  = {tenstorrent/tt-metal: TT-NN operator library, and TT-Metalium low level kernel programming model.},
  url    = {https://github.com/tenstorrent/tt-metal/tree/main},
  month  = {},
  year   = {2026},
  note   = {[Online]}
}

@misc{CKTileCo34:online,
  author = {AMD},
  title  = {CK Tile Conceptual Documentation — Composable Kernel 1.3.0 Documentation},
  url    = {https://rocm.docs.amd.com/projects/composable_kernel/en/latest/conceptual/ck_tile/index.html},
  month  = {},
  year   = {2026},
  note   = {[Online]}
}

@misc{ma2026retargetablecompilationtensorkernels,
  title         = {Retargetable Compilation for Tensor Kernels via a Unified Layout Abstraction},
  author        = {Likai Ma and Renze Chen and Tianqi Li and Yuhao Luo and Youwei Xiao and Size Zheng and Xiuhong Li and Yun Liang},
  year          = {2026},
  eprint        = {2611.13741},
  archiveprefix = {arXiv},
  primaryclass  = {cs.PL},
  url           = {https://arxiv.org/abs/2611.13741}
}

@misc{wang2025tilelangcomposabletiledprogramming,
  title         = {TileLang: A Composable Tiled Programming Model for AI Systems},
  author        = {Lei Wang and Yu Cheng and Yining Shi and Zhengju Tang and Zhiwen Mo and Wenhao Xie and Lingxiao Ma and Yuqing Xia and Jilong Xue and Fan Yang and Zhi Yang},
  year          = {2025},
  eprint        = {2504.17577},
  archiveprefix = {arXiv},
  primaryclass  = {cs.LG},
  url           = {https://arxiv.org/abs/2504.17577}
}

@misc{cecka2026cutelayoutrepresentationalgebra,
  title         = {CuTe Layout Representation and Algebra},
  author        = {Cris Cecka},
  year          = {2026},
  eprint        = {2603.02298},
  archiveprefix = {arXiv},
  primaryclass  = {cs.MS},
  url           = {https://arxiv.org/abs/2603.02298}
}

@misc{carlisle2026categoricalfoundationscutelayouts,
  title         = {Categorical Foundations for CuTe Layouts},
  author        = {Jack Carlisle and Jay Shah and Reuben Stern and Paul VanKoughnett},
  year          = {2026},
  eprint        = {2601.05972},
  archiveprefix = {arXiv},
  primaryclass  = {cs.PL},
  url           = {https://arxiv.org/abs/2601.05972}
}

@misc{spector2024thunderkittenssimplefastadorable,
  title         = {ThunderKittens: Simple, Fast, and Adorable AI Kernels},
  author        = {Benjamin F. Spector and Simran Arora and Aaryan Singhal and Daniel Y. Fu and Christopher Ré},
  year          = {2024},
  eprint        = {2410.20399},
  archiveprefix = {arXiv},
  primaryclass  = {cs.LG},
  url           = {https://arxiv.org/abs/2410.20399}
}

@misc{cutlass,
  author  = {{NVIDIA}},
  title   = {{CUTLASS}: {CUDA} Templates and Python {DSLs} for High-Performance Linear Algebra},
  year    = 2026,
  url     = {https://github.com/NVIDIA/cutlass},
  urldate = {2026-10-29}
}

@misc{khattak2025accuratemodelsnvidiatensor,
  title         = {Accurate Models of NVIDIA Tensor Cores},
  author        = {Faizan A. Khattak and Mantas Mikaitis},
  year          = {2025},
  eprint        = {2512.07004},
  archiveprefix = {arXiv},
  primaryclass  = {cs.MS},
  url           = {https://arxiv.org/abs/2512.07004}
}

@article{martinez2025cambrian,
  title     = {The cambrian explosion of mixed-precision matrix multiplication for quantized deep learning inference},
  author    = {Mart{\'\i}nez, H{\'e}ctor and Castell{\'o}, Adri{\'a}n and Igual, Francisco D and Quintana-Ort{\'\i}, Enrique S},
  journal   = {Future Generation Computer Systems},
  pages     = {108231},
  year      = {2025},
  publisher = {Elsevier}
}

@misc{arm_limited_2026,
  title   = {{\codexrev{Arm A64 Instruction Set for A-profile architecture (DDI0602)}}},
  url     = {https://developer.arm.com/documentation/ddi0602/2025-12/?lang=en},
  journal = {Arm.com},
  author  = {Arm},
  year    = {2026}
}

@misc{intel_2026,
  author  = {Intel},
  title   = {Intel® 64 and IA-32 Architectures Software Developer Manuals},
  url     = {https://www.intel.com/content/www/us/en/developer/articles/technical/intel-sdm.html},
  journal = {Intel},
  author  = {Intel},
  year    = {2026}
}

@misc{nvidia_cutile_2026,
  title   = {GitHub - NVIDIA/cuda-tile: CUDA Tile IR is an MLIR-based intermediate representation and compiler infrastructure for CUDA kernel optimization, focusing on tile-based computation patterns and optimizations targeting NVIDIA tensor core units.},
  url     = {https://github.com/NVIDIA/cuda-tile},
  journal = {GitHub},
  author  = {NVIDIA},
  year    = {2026},
  month   = {Mar}
}

@article{yadav2025taskbasedtensorcomputationsmodern,
  author     = {Yadav, Rohan and Garland, Michael and Aiken, Alex and Bauer, Michael},
  title      = {Task-Based Tensor Computations on Modern GPUs},
  year       = {2025},
  issue_date = {June 2025},
  publisher  = {Association for Computing Machinery},
  address    = {New York, NY, USA},
  volume     = {9},
  number     = {PLDI},
  url        = {https://doi.org/10.1145/3729262},
  doi        = {10.1145/3729262},
  journal    = {Proc. ACM Program. Lang.},
  month      = jun,
  articleno  = {163},
  numpages   = {25}
}

@misc{chou_2026,
  title   = {Fast 2-Simplicial Attention: Hardware-Efficient Kernels in TLX – PyTorch},
  url     = {https://pytorch.org/blog/fast-2-simplicial-attention-hardware-efficient-kernels-in-tlx/},
  journal = {Pytorch.org},
  author  = {Chou, Timothy},
  year    = {2026}
}

@article{ozaki2025schemeii,
  title   = {{Ozaki Scheme II}: A {GEMM}-oriented emulation of floating-point matrix multiplication using an integer modular technique},
  author  = {Katsuhisa Ozaki and Yuki Uchino and Toshiyuki Imamura},
  journal = {The International Journal of High Performance Computing Applications},
  pages   = {10943420261467787},
  year    = {2026},
  doi     = {10.1177/10943420261467787},
  url     = {https://doi.org/10.1177/10943420261467787}
}

@misc{mazumder2026fp64,
  title         = {{FP64} Is All You Want, {INT8} Is All You Need, {FP4/6/8} Is All You Have},
  author        = {Pratyai Mazumder and Alexandru Calotoiu and Torsten Hoefler},
  year          = {2026},
  eprint        = {2609.37693},
  archiveprefix = {arXiv},
  primaryclass  = {cs.DC},
  url           = {https://arxiv.org/abs/2609.37693}
}

\appendix

\end{document}